\documentclass[11pt,letter]{article}
\pdfoutput=1

\usepackage{mymacros}

\title{\boldmath Circuit complexity in quantum field theory}

\author[a,b]{Ro Jefferson}
\author[a]{and Robert C. Myers}

 \affiliation[a]{Perimeter Institute for Theoretical Physics,\\31 Caroline Street North, Waterloo, Ontario N2L 2Y5, Canada}
 
\affiliation[b]{Institute of Physics, Universiteit van Amsterdam,\\Science Park 904, 1098 XH Amsterdam, the Netherlands}

\emailAdd{rjefferson@aei.mpg.de}
\emailAdd{rmyers@perimeterinstitute.ca}

\abstract{
Motivated by recent studies of holographic complexity, we examine the question of circuit complexity in quantum field theory. We provide a quantum circuit model for the preparation of Gaussian states, in particular the ground state, in a free scalar field theory for general dimensions. Applying the geometric approach of Nielsen to this quantum circuit model, the complexity of the state becomes the length of the shortest geodesic in the space of circuits. We compare the complexity of the ground state of the free scalar field to the analogous results from holographic complexity, and find some surprising similarities. 
}

\begin{document}  
\maketitle

\section{Introduction}
\begin{figure}[h!]
\centering
\includegraphics[width=0.45\textwidth]{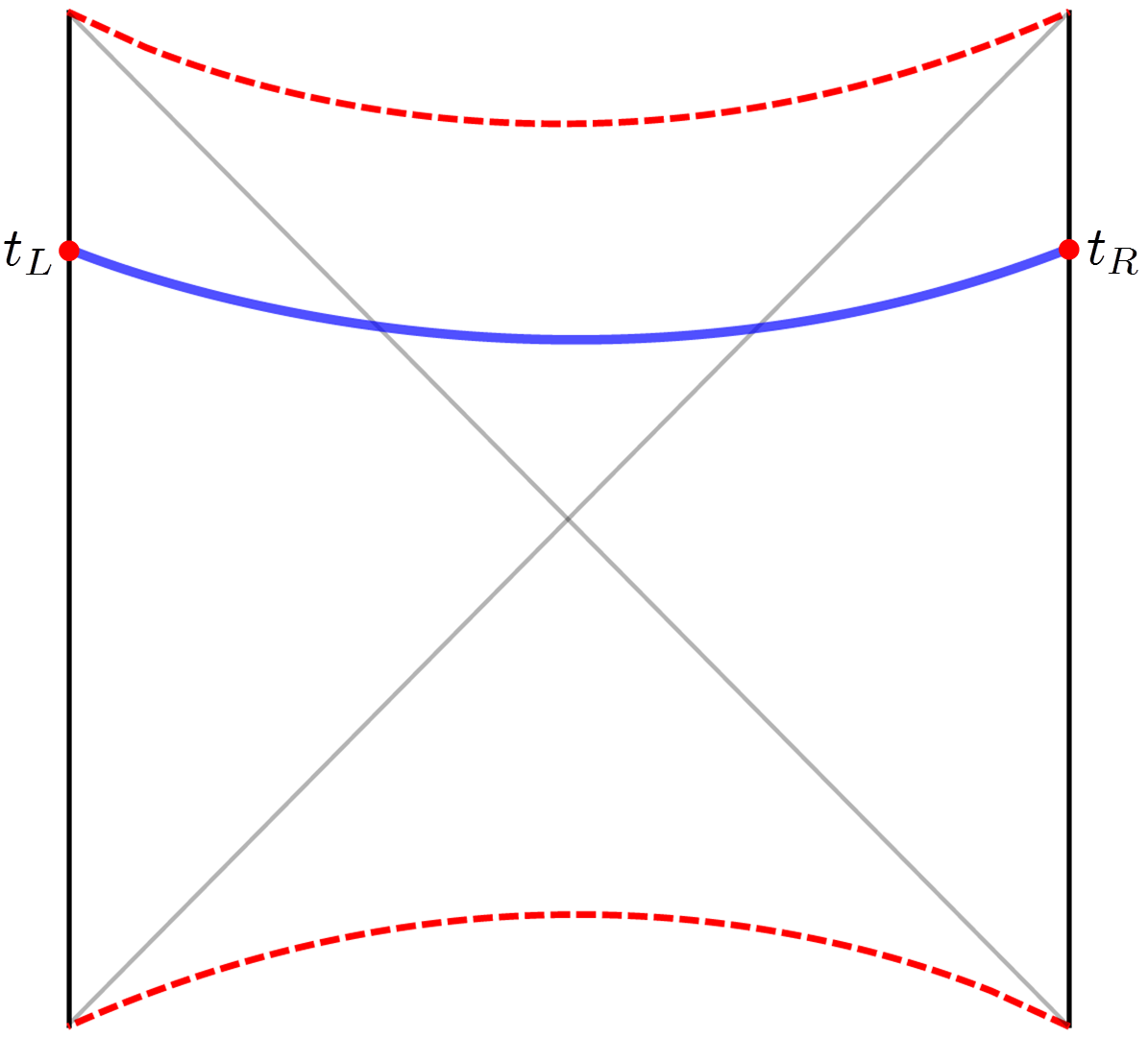}
\ \ \ \ \ 
\includegraphics[width=0.45\textwidth]{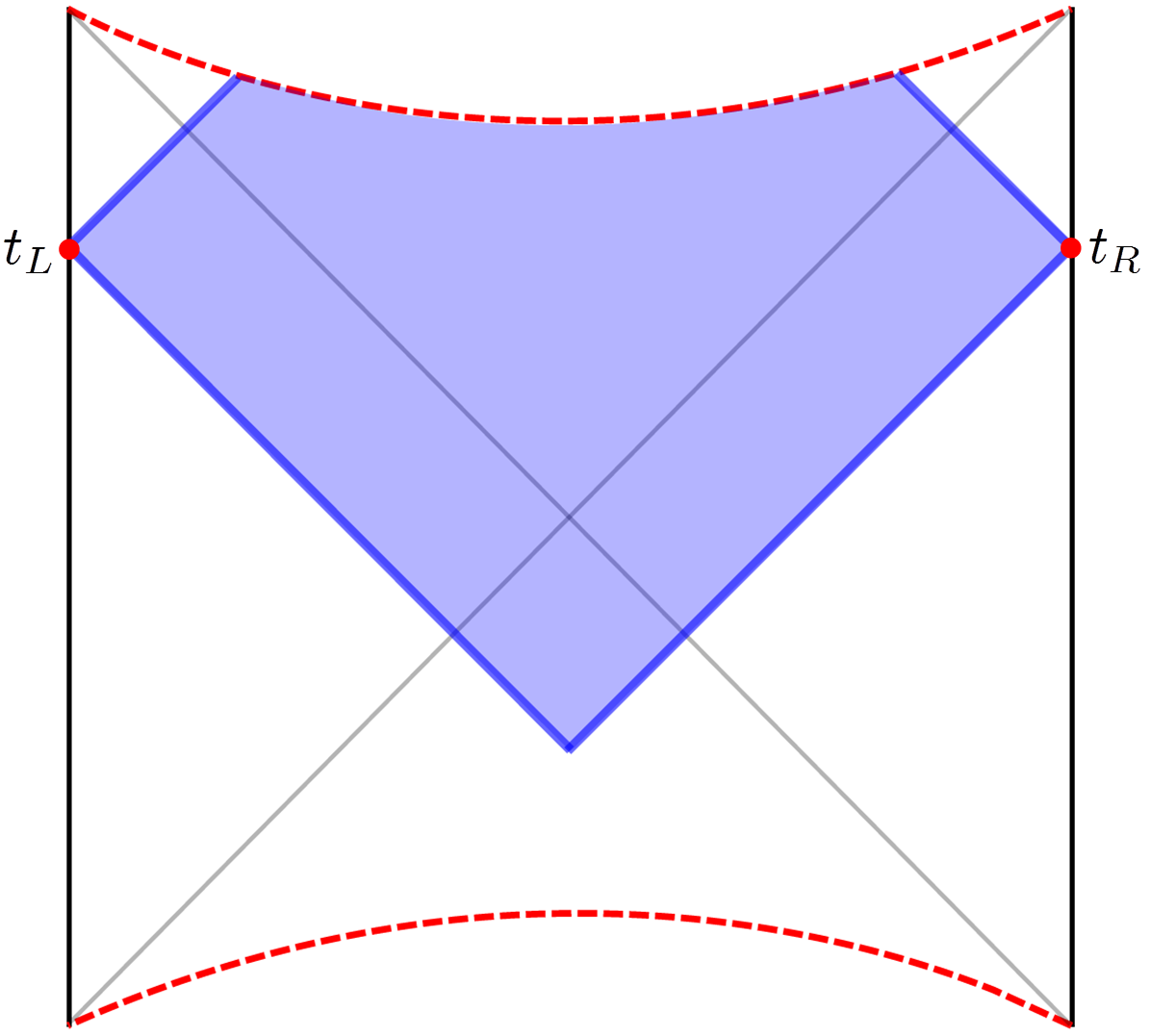}
\caption{Complexity=volume (CV, left) and complexity=action (CA, right) for the eternal AdS black hole dual to the thermofield double state \eqref{eq:TFD}. In the left panel, the blue curve represents the maximal spacelike surfaces that connects the specified time slices on the left and right boundaries. In the right image, the shaded region is the corresponding WDW patch. \label{fig:CVCA}}
\end{figure}

Recent years have seen exciting progress in understanding the connection between entanglement and geometry \cite{Swingle:2014uza,Faulkner:2013ica,Bianchi:2012ev,Lashkari:2013koa,Balasubramanian:2013lsa,VanRaamsdonk:2010pw}. However, in the context of the AdS/CFT correspondence, our ability to decipher the bulk geometry (or bulk physics, more generally) from information in the boundary CFT remains very incomplete. The challenges are most pronounced if one considers physics behind the horizon of a black hole. Consider for example the eternal AdS black hole, which is dual to the thermofield double (TFD) state \cite{Maldacena:2001kr}
\be
\ket{TFD(t_L,t_R)}=\frac{1}{\sqrt{Z_\beta}}\sum_ie^{-\beta E_i/2}\,e^{-iE_i(t_L+t_R)}\,\ket{i}_L\,\ket{i}_R~.\label{eq:TFD}
\ee
This describes an entangled state of the two copies of the CFT associated with the asymptotic boundaries (see figure \ref{fig:CVCA}), which are joined by a wormhole, \ie an Einstein-Rosen bridge (ERB), in the bulk \cite{Maldacena:2013xja}. The AdS/CFT correspondence demands that the interior region have an equivalent description in terms of the boundary field theory. But now, in addition to the usual difficulties involved in probing behind the horizon, we have another conundrum: the boundary field theory reaches thermal equilibrium very quickly, on the order of the thermalization time $1/T$, while the ERB continues to grow on much longer timescales \cite{Stanford:2014jda}. Therefore, there must be some quantity in the field theory that corresponds to this fine-grained information -- which is evidently not captured by entanglement entropy \cite{Hartman:2013qma,Liu:2013iza} -- that continues to evolve long after thermal equilibrium is reached. 

These considerations led Susskind to introduce \emph{holographic complexity} as the boundary entity whose growth corresponds to the evolution of the ERB \cite{Susskind:2014rva,Susskind:2014jwa,Susskind:2014moa}. In particular, with his collaborators, he developed two new gravitational observables, both of which successfully probe the late-time growth of the ERB. The first of these is referred to as the complexity=volume (CV) conjecture, which posits that the complexity of the boundary state is proportional to the volume of a maximal codimension-one bulk surface $\mathcal{B}$ that extends to the AdS boundary, and asymptotes to the time slice $\Sigma$ on which the boundary state is defined \cite{Stanford:2014jda,Susskind:2014rva}:
\beq
\mathcal{C}_{\text{V}}(\Sigma) =\ \mathrel{\mathop {\rm
max}_{\scriptscriptstyle{\Sigma=\partial \mathcal{B}}} {}\!\!}\left[\frac{\mathcal{V(B)}}{G_N \, \ell}\right] \, ,
\label{volver}
\eeq
where $\ell$ is some length scale associated with the bulk geometry, \eg the AdS radius or the radius of the black hole. For example, in the eternal AdS black hole, this bulk surface connects the time slices denoted $t_L$ and $t_R$ on the left and right boundaries through the ERB; see the left panel in figure \ref{fig:CVCA}. The second proposal is the complexity=action (CA) conjecture. This identifies the complexity of the boundary state with the gravitational action evaluated on a bulk region known as the Wheeler-DeWitt (WDW) patch \cite{Brown:2015bva,Brown:2015lvg}:
\be
\mathcal{C}_{\text{A}}(\Sigma)=\frac{I_\mt{WDW}}{\pi \,\hbar}\,.
\label{ccaction}
\ee
One can think of the WDW patch as the causal development of the spacelike surface $\mathcal{B}$ picked out by the CV construction. The right panel in figure \ref{fig:CVCA} illustrates the WDW patch for the example of the eternal AdS black hole, where the CFT state is again evaluated on the $t_L$ and $t_R$ slices of the left and right boundaries, respectively.

Both proposals have their merits, as well as certain shortcomings. In any case, they bring to our attention two new classes of interesting gravitational observables which should certainly be studied in further detail. In fact, various aspects of the proposals and these new observables have been examined in a number of recent papers, \eg \cite{Lehner:2016vdi,Chapman:2016hwi,Carmi:2016wjl, Reynolds:2016rvl,Ben-Ami:2016qex,Abad:2017cgl,CCMMS}. And while both the CV and CA conjectures appear to provide viable candidates for holographic complexity, this research program is still at a very preliminary stage. In particular, one would like to establish a concrete translation of the new observables in the bulk to a specific quantity in the boundary theory, \eg as was recently found for holographic entanglement entropy \cite{CHM,Lewkowycz:2013nqa,Dong:2016hjy}. However, a stumbling block to this endeavor is finding the answer to an even simpler question: what does ``complexity'' mean in the boundary CFT?

This question is the focus of the present paper. Specifically, our objective is to provide the first steps towards defining circuit complexity in quantum field theory (QFT).\footnote{We also refer the reader to ref.~\cite{Hashimoto:2017fga} for a recent complementary investigation in this direction.} A precise understanding of this quantity will not only shed light on the CV and CA proposals, but is also an interesting question deserving of study in its own right. For example, it may also provide new insights into quantum algorithms for the simulation of quantum field theories\cite{Jordan:2011ne,Jordan:2011ci, Jordan:2014tma,Jordan:2017lea}, or more generally into Hamiltonian complexity \cite{osborne2012hamiltonian, gharibian2015quantum}, or the efficient description of many-body wave functions \cite{Orus:2013kga,vidal2009entanglement}.

In computer science, the notion of computational complexity refers to the minimum number of operations necessary to implement a given task \cite{Aaronson:2016vto,watrous}. In the present context, the task of interest will be the preparation of a state in the QFT, and we will define the complexity in terms of a quantum circuit model. That is, we will begin with a simple reference state $\ket{\psi_\mt{R}}$, and construct a unitary transformation $U$ that produces the desired target state $\ket{\psi_\mt{T}}$ via
\beq
\ket{\psi_\mt{T}}=U\,\ket{\psi_\mt{R}}~.\label{circuitDef}
\eeq
The unitary $U$ will be constructed from a particular set of simple elementary or universal gates, which can be applied sequentially to the state. When working with such discrete operations, we should also introduce a tolerance $\veps$ so that even if we cannot achieve the precise equality above, we may still judge the transformation to be successful when the two states are sufficiently close to one another according to some distance measure, \ie
\beq
\big|\!\big|\, \ket{\psi_\mt{T}}-U\,\ket{\psi_\mt{R}}\,\big|\!\big|^2\le\veps\,. \label{scotch}
\eeq
Of course, there will not be a unique circuit which implements the desired transformation \reef{circuitDef}: generally there will exist infinitely many sequences of gates which produce the same target state. However, the complexity of the state $\ket{\psi_\mt{T}}$ may be defined as the minimum number of gates required to produce the transformation \eqref{circuitDef}, \ie the complexity is the number of elementary gates in the optimal or shortest circuit. The challenge then is to identify this optimal circuit from amongst the infinite number of possibilities.

Our work takes inspiration from the geometric approach of Nielsen and collaborators \cite{Nielsen:2005mn1,Nielsen:2006mn2,Nielsen:2007mn3},\footnote{See \cite{Chemissany:2016qqq} for another application of Nielsen's ideas in holography. We also refer the interested reader to ref.~\cite{alvarez1999comment}, which introduces an interesting connection between quantum algorithms and geodesics on the Fubini-Study metric.} which itself was developed using ideas from the theory of optimal quantum  control, \eg \cite{gordon1997active,shapiro2003principles,rabitz2000whither,rice2000optimal}. In Nielsen's case, the question of interest was to find the minimal size quantum circuit required to exactly implement a specified $n$-qubit unitary operation $U$ (without the use of ancilla qubits). Neilsen approaches this question as the Hamiltonian control problem of finding a time-dependent Hamiltonian $H(t)$ that synthesizes the desired $U$, 
\beq
U=\cev{\cal P} \exp\left[\int_0^1dt\ H(t) \right]\qquad{\rm where}\ \ 
H(t)=\sum_I Y^I(t)\,M_I~,\label{eq:controlY}
\eeq
where the Hamiltonian is expanded in terms of generalized Pauli matrices, denoted here as $M_I$,\footnote{Our notation diverges from that of Neilsen, in order to increase the similarity of these equations with our notation in the main text. In particular, note that we have absorbed a factor of $-i$ in $M_I$ so that these are now anti-Hermitian operators.} and the $\cev{\cal P}$ indicates a time ordering such that the Hamiltonian at earlier times is applied to the state first, \ie the circuit is built from right to left. In \cite{Nielsen:2005mn1}, the control functions $Y^I$ form a $\lp 4^n-1\rp$-dimensional vector space, and can be seen as specifying the tangent vector to a trajectory in the space of unitaries,
\beq
U(t)=\cev{\cal P} \exp\left[\int_0^t \dd\tilde t\ H(\tilde t) \right]\ .
\label{row}
\eeq
In this general space, the paths of interest satisfy the boundary conditions $U(t=0)=\mathbb{1}$ and $U(t=1)=U$. Neilsen's idea is then to define a \emph{cost} for the various possible paths
\beq
{\cal D}(U(t))=\int_0^1\dd t\ F\!\lp U(t),\dot U(t)\rp~,\label{costco}
\eeq
and to identify the optimal circuit or path by minimizing this functional. In general, the cost function $F(U,v)$ is some local functional of the position $U$ in the space of unitaries and a vector $v$ in the tangent space at this point. Neilsen further argues that for the present problem, a physically reasonable cost function must satisfy a number of desirable features: 
\vskip 0.8ex

\noindent \ \ \ \
$1.$\ \ \textit{Continuity}: $F$ should be continuous, \ie $F\in C^0$.
\vskip 0.8ex

\noindent \ \ \ \
$2.$\ \ \textit{Positivity:} $F(U,v)\ge0$ with equality if and only if $v=0$. 
\vskip 0.8ex

\noindent \ \ \ \
$3.$\ \ \textit{Positive homogeneity}: $F(U,\lambda v)= \lambda\, F(U,v)$ for any positive real number $\lambda$. 
\vskip 0.8ex

\noindent \ \ \ \
$4.$\ \ \textit{Triangle inequality}: $F(U,v+v')\le F(U,v)+F(U,v')$ for all tangent vectors $v$ and $v'$. 
\vskip 0.8ex

\noindent these four properties come very close to defining a class of geometries known as Finsler manifolds. In particular, if we replace the first condition above with
\vskip 0.8ex

\noindent \ \ \ 
$1'.$\ \ \textit{Smoothness}: $F$ should be smooth, \ie $F\in C^\infty$,
\vskip 0.8ex

\noindent then eq.~\reef{costco} defines length functional for a Finsler manifold, a particular class of differential manifolds equipped with a quasimetric structure in which the length of any curve is measured by a length functional of the form \reef{costco}, with a Finsler metric $F$ satisfying the four properties enumerated above, see \eg \cite{web1,web2}. While the familiar notion of Riemannian manifolds would fall within this definition, Finsler geometry provides a generalization to a broader class of manifolds where the norm on the tangent space is not (generally) induced by a metric tensor. Hence Neilsen has identified the problem of finding an optimal circuit with the problem of finding extremal curves, \ie geodesics, in a Finsler geometry, and the complexity is then identified with the length of the geodesic.\footnote{For future reference, when referring to \emph{general} paths or circuits, we will use ``size,'' ``length,'' ``cost,'' and ``depth'' interchangeably; however, ``complexity'' will be reserved for the length of the \emph{optimal} path or circuit.} 

Of course, this still leaves open the question of the precise form of the cost function, and various possibilities are examined in \cite{Nielsen:2005mn1}:\footnote{The functions $F_1$ and $F_p$ are not technically Finsler metrics, since both fail to meet the smoothness requirement. However, as explained in \cite{Nielsen:2005mn1}, they can be approximated arbitrarily well by metrics which \emph{are} Finsler. This subtlety will not be important for our analysis.} 
\beq
\bal
F_1(U,Y)=&\sum_I \left|Y^I\right|~,\qquad\qquad
F_p(U,Y)\!&\!=&\sum_I p_I \left|Y^I\right|~,\\
F_2(U,Y)=&\sqrt{\sum_I \lp Y^I\rp^2}~,\qquad\;\;
F_q(U,Y)\!&\!=&\sqrt{\sum_I q_I\lp Y^I\rp^2}~.
\eal\label{eq:Fmetrics}
\eeq
In the two measures on the right, $p_I$ and $q_I$ are {penalty factors} which can be chosen to favour certain directions in the circuit space over others, \ie to give a higher cost to certain classes of gates. We do not include such factors in most of our analysis, but we return to this issue in section \ref{sec:penalty}. Of course, the $F_2$ measure yields a standard Riemannian geometry --- and in fact, it will be the focus of much of our discussion. 

The preceding exposition of Nielsen's approach is of course very incomplete, and the interested reader is referred to \cite{Nielsen:2005mn1,Nielsen:2006mn2,Nielsen:2007mn3} for more details. The key feature of this approach is that it enables one to bring the full power of differential geometry to bear on the problem of constructing the optimal quantum circuit, and this provides an objective manner in which to measure the complexity as the length of extremal paths in the geometry. However, at many points our approach will necessarily differ from that of Nielsen since we are studying a different problem, namely complexity in a quantum field theory. The primary purpose of the above presentation was to provide motivation for our geometrical analysis, but we should add that the details of Finsler geometry will not play any role in the following. Rather, a simpler physics-oriented perspective is to view the problem of finding the optimal circuit as a trajectory in the space of all possible circuits, as a classical mechanics problem for the motion of particle governed by the usual Lagrangian in eq.~\reef{costco}.

This paper is organized as follows: we begin in section \ref{sec:gates} by examining complexity for a simple free scalar field theory. Following the preceding discussion, this requires identifying a simple reference state, introducing a set of elementary gates, and also identifying a family of interesting target states. However, the first step will be to regulate the theory by placing it on a lattice, which reduces the scalar field theory to a family of coupled harmonic oscillators. Hence, as a warm up problem, we consider the case of a single pair of harmonic oscillators. Then, having built up some intuition, we shall geometrize the problem in section \ref{sec:geometry}. The main ideas from Nielsen's approach are implemented here: we represent the circuit as a path-ordered exponential analogous to eq.~\reef{row}, show that our space of circuits forms a representation of $\GLtwo$, and construct the appropriate (Euclidean) metric. With this in hand, we proceed to find the geodesics, and identify the complexity of the ground state  as the geodesic length of the global minimum. In section \ref{sec:N}, we return to the field theory problem by generalizing these results to a lattice of coupled oscillators. Given the complexity for the (regulated) field theory, we then ask how our results compare to holographic complexity, and we find some surprising similarities. In section \ref{sec:penalty}, we conduct a preliminary exploration of the effects of introducing penalty factors for nonlocal gates. Finally, we close in section \ref{discuss} with a brief discussion of our results and directions for future work. Various technical details have been relegated to several appendices: we construct some explicit example circuits using the elementary gates given in section \ref{sec:gates} in appendix \ref{sec:example}, elaborate on some geometrical details in appendix \ref{sec:appxKilling}, derive the normal-mode frequencies for a one-dimensional lattice in appendix \ref{sec:omegaDeriv}, find a closed-form approximation to the circuit complexity for the $d$-dimensional lattice in appendix \ref{sec:approx}, and compute an approximation to the optimal circuit in the presence of penalty factors in appendix \ref{sec:subspace}.

\newpage
\section{Complexity for harmonic oscillators}\label{sec:gates}
As a first step towards understanding circuit complexity in QFT, we will consider for simplicity a free scalar field in $d$ spacetime dimensions. However, having identified this particular QFT, we must first regulate the theory by placing it on a lattice,\footnote{Our experience with holographic complexity suggests that we will not be able to sensibly define complexity in a QFT without a UV regulator in place \cite{Carmi:2016wjl}.} which reduces the system to an infinite family of harmonic oscillators. This in turn suggests the much simpler warm-up problem of two coupled harmonic oscillators. As it turns out, this simple model retains enough of the structure of the original problem that we will be able to learn several important lessons, which we can then carry over to the problem of circuit complexity in our scalar field theory. As in the general case, to study complexity in the two oscillator problem, we must identify a target state, a reference state, and a suitable family of elementary gates.

We begin with the Hamiltonian of a free scalar field in $d$ spacetime dimensions,
\be
H=\frac{1}{2}\int\dd^{d-1}x\left[\pi(x)^2+\vec\nabla\phi(x)^2+m^2\phi(x)^2\right]~.
\ee
As mentioned above, our first step is to regulate the theory by placing it on a (square) lattice with lattice spacing  $\delta$, in which case the Hamiltonian becomes:
\be
H=\frac{1}{2}\sum_{\vecn}\left\{\frac{p(\vecn)^2}{\delta^{d-1}}+\delta^{d-1}\left[\frac{1}{\delta^2}\sum_i\lp\phi(\vecn)-\phi(\vecn-\hat{x}_i)\rp^2+m^2\phi(\vecn)^2\right]\right\}~,\label{eq:Hlattice}
\ee
where $\hat{x}_i$ are unit vectors pointing along the spatial directions of the lattice. The resulting theory is essentially a quantum mechanical problem with an infinite family of coupled (one-dimensional) harmonic oscillators. We can make this description manifest by redefining $X(\vecn)=\delta^{d/2}\phi(\vecn)$, $P(\vecn)=p(\vecn)/\delta^{d/2}$, $M=1/\delta$, $\omega=m$ and $\Omega=1/\delta$, whereupon the Hamiltonian \reef{eq:Hlattice} takes the familiar form
\be
H=\sum_{\vecn}\left\{\frac{P(\vecn)^2}{2M}+\frac12 M \left[\omega^2 X(\vecn)^2+\Omega^2\sum_i\lp X(\vecn)-X(\vecn-\hat{x}_i)\rp^2\right]\right\}~.\label{hqm}
\ee
Hence the frequency of the individual masses is given by $\omega=m$, and the inter-mass coupling is given by $\Omega=1/\delta$.

Now, the above suggests that we begin with an even simpler warm-up problem, namely, the case of two coupled harmonic oscillators:
\be
H=\frac{1}{2}\left[ p_1^2+p_2^2+\omega^2\lp x_1^2+x_2^2\rp+\Omega^2\lp x_1-x_2\rp^2\right]~,
\label{qm1}
\ee
where $x_1,x_2$ label their spatial positions, and we have set $M_1=M_2=1$ for simplicity. Of course, to solve this system, one simply rewrites the Hamiltonian in terms of the normal modes,
\be
H=\frac{1}{2}\lp \tilde p_+^2+\tilde\omega_+^2\tilde x_+^2+\tilde p_-^2+\tilde\omega_-^2\tilde x_-^2\rp~,
\label{qm2}
\ee
where\footnote{When working in the normal-mode basis, we denote variables (\eg positions, frequencies), with a tilde to clearly distinguish from the physical basis. The utility of this convention will become apparent later. \label{footy78}}
\be
\tilde x_\pm\equiv\frac{1}{\sqrt{2}}\lp x_1\pm x_2\rp~,\qquad
\tilde\omega_+^2=\omega^2~,\qquad\tilde\omega_-^2=\omega^2+2\Omega^2~.
\label{qm3}
\ee
This recasts the problem as that of two decoupled simple harmonic oscillators, and hence it is now straightforward to solve for the eigenstates and eigen-energies of the Hamiltonian. For example, we can write the ground-state wave function as the product of the ground-state wave functions for the two individual oscillators:
\be
\psi_0(\tilde x_+,\tilde x_-)=\psi_{0+}(\tilde x_+)\psi_{0-}(\tilde x_-)
=\frac{\lp\tilde\omega_+\tilde\omega_-\rp^{1/4}}{\sqrt{\pi}}\,\mathrm{exp}\!\left[-\frac{1}{2}\lp\tilde\omega_+\tilde x_+^2+\tilde\omega_-\tilde x_-^2\rp\right]~,\label{eq:targetNorm}
\ee
where the normalization has been chosen such that $\int\! d^2x\,|\psi_0|^2=1$. We may also express this wave function in terms of the physical positions of the two masses:
\be
\psi_0(x_1,x_2)=\frac{\lp\omega_1\omega_2-\beta^2\rp^{1/4}}{\sqrt{\pi}}\mathrm{exp}\left[-\frac{\omega_1}{2}x_1^2-\frac{\omega_2}{2}x_2^2-\beta x_1x_2\right]~,\label{eq:targetPhys}
\ee
where
\be
\omega_1=\omega_2=\frac{1}{2}\lp\tilde\omega_++\tilde\omega_-\rp~,\;\;\;
\beta\equiv\frac{1}{2}\lp\tilde\omega_+-\tilde\omega_-\rp<0~.\label{eq:omega12pm}
\ee
We note in passing that our notation for the wave function in eq.~\reef{eq:targetPhys} is slightly more general than necessary; however, these Gaussian wave functions constitute an interesting family of target states for the present exercise.\footnote{For example, Gaussian states play an important role in quantum optics, and much of our analysis is closely related to ideas developed in the quantum information literature for this purpose, \eg \cite{WANG20071,weedbrook2012gaussian,adesso2014continuous}.}

The next step is to identify a simple reference state. Motivated by discussions of holographic complexity \cite{Susskind:2014rva,Susskind:2014jwa,Susskind:2014moa}, as well as cMERA \cite{Haegeman:2011uy}, we choose a reference state where the two masses are unentangled, namely a factorized Gaussian state,
\be
\psi_\mt{R}(x_1,x_2)=\sqrt{\frac{\omega_0}{\pi}}\,\mathrm{exp}\!\left[-\frac{\omega_0}{2}\lp x_1^2+x_2^2\rp\right]~.\label{eq:refPhys}
\ee
For the time being, we will simply leave $\omega_0$ as a free parameter which characterizes our reference state. We shall examine specific choices of this frequency in section \ref{sec:continuum}. 

Having chosen our reference and target states, it remains to identify a simple set of unitary gates with which to construct the desired unitary $U$, which implements $\psi_\mt{T}=U\,\psi_\mt{R}$. The natural operators appearing in the quantum mechanics problem of the two coupled oscillators are the positions $x_1, x_2$ and the momenta $p_1\!=\!-i\partial_1,\,p_2\!=\!-i\partial_2$, which satisfy the canonical commutation relations $\left[x_a,p_b\right]=i\,\delta_{ab}$.  We can use these operators to build an interesting set of elementary gates for our problem:
\be
\bal
&\quad H=e^{i\eps x_0p_0}~, \qquad J_{a}=e^{i\eps x_0p_a}~,\qquad K_{a}=e^{i\eps x_ap_0}~, \ \\
Q_{ab}&=e^{i\eps x_ap_b}\ \ ({\rm with}\ a\ne b)~, \qquad Q_{aa}=e^{\frac{i\eps}{2}\lp x_ap_a+p_ax_a\rp}=e^{\eps/2}\,e^{i\eps x_ap_a}~,\label{eq:gates}
\eal
\ee
where $x_0$ and $p_0$ are c-number constants. A key point is that we have introduced an infinitesimal parameter $\eps\ll1$ into the exponent of each one of these operators. This ensures that the action of any one of these gates only produces a small change on the wave function. The action of each of these gates can be understood with the following general examples:
\be
\bal
H\,\psi(x_1,x_2)&=e^{i\eps p_0x_0}\psi(x_1,x_2) \qquad && \mathrm{(global)~phase~change}\\
J_{1}\,\psi(x_1,x_2)&=\psi(x_1+\eps x_0,x_2) \qquad && \mathrm{shift~} x_1 \mathrm{~by~constant~}\eps x_0\\
K_{1}\,\psi(x_1,x_2)&=e^{i\eps p_0x_1}\psi(x_1,x_2) \qquad && \mathrm{shift~} p_1 \mathrm{~by~constant~}\eps p_0\\
Q_{21}\,\psi(x_1,x_2)&=\psi(x_1+\eps x_2,x_2) \qquad && \mathrm{shift~} x_1 \mathrm{~by~} \eps x_2 \quad\ \mathrm{~(entangling\ gate)}\\
Q_{11}\,\psi(x_1,x_2)&=e^{\eps/2}\psi\lp e^\eps x_1,x_2\rp \qquad && \mathrm{scale~} x_1 \to e^\eps x_1 \ \ \ \mathrm{~(scaling\ gate)}
\eal
\label{eq:gatesAct}
\ee
When working with position-space wave functions, the momentum shift produced by $K_{1}$ (or $K_{2}$) amounts to introducing a small plane-wave component in the wave function, as illustrated in \eqref{eq:gatesAct}. We refer to $Q_{11}$ and $Q_{22}$ as scaling gates, for the obvious reason that these operators scale the corresponding coordinate by a small amount. Note that they also introduce an overall normalization factor, which ensures that the norm of the wave function is preserved. The operators $Q_{21}$ and $Q_{12}$ mix the positions of the two masses, thereby increasing (or decreasing) the entanglement between the two oscillators; hence we refer to these as the entangling gates. The scaling and entangling gates will play a key role in the circuits we construct below.

Of course, one could extend the ensemble of gates introduced in eq.~\reef{eq:gates} with operators like
\beq
\exp\left[i\eps \,\frac{p_0}{x_0}\, x_1x_2\right] \quad{\rm or}\quad
\exp\left[i\eps \,\frac{x_0}{p_0}\, p_1^2\right]\,.
\label{extra}
\eeq
Furthermore, one could also introduce gates with even higher powers of $x$'s and $p$'s in the exponent. However, we know that the collection of gates in eq.~\reef{eq:gates} is sufficient to implement the unitary transformation from the specified reference state \reef{eq:refPhys} to the desired target state \reef{eq:targetPhys}. Hence for simplicity, we shall work within this subset of all possible unitary gates.

A \emph{circuit} then consists of a sequence of these gates, whose action on $\psi_\mt{R}$ produces the desired state $\psi_\mt{T}$. For example, consider the following circuit:
\beq
\psi_\mt{T}=U\psi_\mt{R}\equiv Q_{22}^{\alpha_3}\,Q_{21}^{\alpha_2}\,Q_{11}^{\alpha_1}\,\psi_\mt{R}~.\label{eq:eg1}
\eeq
Here, $Q_{11}$ acts first, and by acting with the appropriate number of times $\alpha_1$, we will increase the reference frequency $\omega_0$ appearing in front of $x_1^2$ in eq.~\reef{eq:refPhys} to the desired frequency $\omega_1$ appearing in eq.~\reef{eq:targetPhys}. Similarly, the number of times that the $Q_{21}$ and $Q_{22}$ gates are required to appear in the circuit, namely $\alpha_2$ and $\alpha_3$, are uniquely fixed by the desired $\omega_2$ and $\beta$ in the target state. The details of the corresponding calculations are given in appendix \ref{sec:example}, and the final result is 
\beqa
&&\alpha_1=\frac{1}{2\eps}\log\lp\frac{\omega_1}{\omega_0}\rp\,,
\qquad
\alpha_2=\frac{1}{\eps}\sqrt{\frac{\omega_0}{\omega_1}}\frac{\beta}{\sqrt{\omega_1\omega_2-\beta^2}}~,
\nonumber\\
&&\qquad\qquad\quad\alpha_3=\frac{1}{2\eps}\log\lp\frac{\omega_1\omega_2-\beta^2}{\omega_0\,\omega_1}\rp~.
\label{answer1}
\eeqa

We then define the \emph{circuit depth} as the total number of gates in the circuit. In the above example, we have simply
\beqa
\mathcal{D}(U)&=&|\alpha_1|+|\alpha_2|+|\alpha_3|
\nonumber\\
&=&\frac{1}{\eps}\left[\frac12\,\log\lp\frac{\omega_1\omega_2-\beta^2}{\omega_0^2}\rp+\sqrt{\frac{\omega_0}{\omega_1}}\frac{|\beta|}{\sqrt{\omega_1\omega_2-\beta^2}}\right]\,.
\label{leach}
\eeqa
Note the use of the absolute values in the first line. At a pragmatic level, this is required because $\alpha_2$ is negative in this particular example, \ie $\beta<0$. But this means that we are giving an equal complexity cost for the inverse gates $Q_{ij}^{-1}$ as for the original gates $Q_{ij}$, \ie we count the appearance of $Q_{ij}^{-1}$ as one gate in a circuit. 

We refer to the result in eq.~\reef{leach} as the circuit depth of the particular circuit $U$ given in eq.~\reef{eq:eg1}. But we must distinguish this from the complexity of the target state $\psi_\mt{T}$, which is the minimum number of gates required to produce the desired transformation. In other words, the complexity is the circuit depth of the optimal circuit. At present, we have no reason to believe that the simple circuit proposed in eq.~\reef{eq:eg1} is the optimal circuit, and in fact, our calculations below will show that it is not.

We can describe the general form of the result in eq.~\reef{leach} as being an overall factor of $1/\epsilon$, and a coefficient determined by the various physical parameters characterizing the target and reference states. More generally, the circuit depth might be given by an expansion in $\epsilon$, beginning with a $1/\epsilon$ term followed by a finite term and then potentially terms involving positive powers of $\epsilon$. However, since $\epsilon\ll1$, determining the complexity essentially requires finding the circuit which minimizes the coefficient of the leading $1/\epsilon$ term. For further discussion and additional examples, the interested reader may turn to appendix \ref{sec:example}. 

In the next section, we apply Neilsen's approach of geometrizing the circuit complexity to find the optimal circuit. Before leaving present example however, for comparison to later results it is convenient to express the circuit depth in eq.~\reef{leach} in terms of the normal-mode frequencies using eq.~\reef{eq:omega12pm}. This substitution yields
\beq
\mathcal{D}_1=\frac{1}{\eps}\left[\,\frac{1}{2}\log\lp\frac{\tom_+}{\omega_0}\rp+\frac{1}{2}\log\lp\frac{\tom_-}{\omega_0}\rp+\frac{\tom_-\!-\tom_+}{\sqrt{2\tom_+\tom_-}}\sqrt{\frac{\omega_0}{\tom_+\!+\tom_-}}\,\right]~. \label{eq:Deg1}
\eeq
Recall that $\tom_->\tom_+$ from eq.~\reef{qm3} (and implicitly, we are assuming $\tilde \omega_\pm>\omega_0$). 

\newpage

\section{Geometrizing complexity}\label{sec:geometry}

In the introduction, we discussed Neilsen's approach \cite{Nielsen:2005mn1,Nielsen:2006mn2,Nielsen:2007mn3} of geometrizing the problem of finding the optimal circuit. We now wish to apply this geometric approach to the problem of finding the optimal preparation of the ground-state of two coupled harmonic oscillators. Our first step is to represent the circuit $U$ as a path-ordered exponential,
\beq
U=\cev{\mathcal{P}}\,\mathrm{exp}\int_0^1\dd s\,Y^I\!(s)\,\op_I~,\qquad
\psi_\mt{T}\lp x_1,x_2\rp=U\psi_\mt{R}\lp x_1,x_2\rp~.\label{eq:pathPsi}
\eeq
This structure replaces the representation of the circuits as products of the discrete gates in eq.~\reef{eq:gates}. The connection with these gates comes about since we choose the operators $\op_I$ appearing in the exponential to be precisely those appearing in the scaling and entangling gates introduced previously; that is, we write 
\beq
Q_{ab} = \exp\!\left[\eps\,\cO_{ab}\right]
\qquad{\rm with}\qquad \cO_{ab}=\left(i\,x_a\,p_b+\frac12\,\delta_{ab}\right)\,.
\label{operate}
\eeq
Our notation in eq.~\reef{eq:pathPsi} is that the sum over $I$ runs over the pairs $ab$, \ie $I\in\{11,12,21,22\}$. Hence in the path-ordered exponential, we can think of $s$ as parametrizing a (continuous) product of gates, and the functions $Y^I(s)$ as indicating whether the $I$'th type of gate is turned on or off in this sequence (analogous to the control functions in Nielsen's time-dependent Hamiltonian \eqref{eq:controlY}). In the integral appearing in the exponent, the differential $\dd s$ plays a role analogous to that of the infinitesimal parameter $\eps$. Finally, the path-ordering symbol indicates that we build the circuit from right to left, \ie the operators at smaller values of $s$ act on the wave function before those at larger values of $s$. Furthermore, with this framework, we consider a particular circuit as being constructed by following a particular trajectory, specified by $Y^I(s)$, through the space of unitary circuits. Hence we begin with $U(s=0)=\mathbb{1}$, and have the family of unitaries
\beq
U(s)=\cev{\mathcal{P}}\,\mathrm{exp}\int_0^s\dd \tilde s\ Y^I\!(\tilde s)\,\op_I~.\label{path2}
\eeq
Eq.~\reef{eq:pathPsi} then specifies the final unitary at the end-point $s=1$, which corresponds to the desired circuit that generates the target state, \ie $U_\mt{fin}=U(s\!=\!1)$ with $\psi_\mt{T}=U_\mt{fin}\psi_\mt{R}$. From this perspective, $Y^I(s)$ specifies the velocity vector tangent to this trajectory, in a manner in which we will make precise below. In more geometric language which may be familiar from general relativity, we would say that the $Y^I(s)$ are the components of the velocity in a particular frame basis, rather than in a coordinate basis.  

As in the example in section \ref{sec:gates} above, the circuit depth is determined by counting the total number of gates appearing in the full sequence comprising the circuit, \cf eq.~\reef{leach}. For our path-ordered exponential \reef{eq:pathPsi}, the analogous expression becomes\footnote{Actually this expression \reef{cost1} is the continuum limit of the cost function $\mathcal{D}(U)=\sum \epsilon\,|\alpha_i|$. Including the extra factor of $\epsilon$ in the sum eliminates the $1/\epsilon$ factor, so the circuit depth remains finite in the limit $\epsilon\to0$.\label{foot55}}
\beq
\mathcal{D}(U)=
\int_0^1\dd s\sum_I\left|Y^I(s)\right|=
\int_0^1\dd s\bigg[\left|Y^{11}(s)\right|+\left|Y^{12}(s)\right|+\left|Y^{21}(s)\right|+\left|Y^{22}(s)\right|\bigg]~.
\label{cost1}
\eeq
This cost function corresponds to the $F_1$ metric in the notation of \cite{Nielsen:2005mn1} --- see eq.~\eqref{eq:Fmetrics}. Our goal of finding the optimal circuit then amounts to finding the functions $Y^I(s)$ which yield the desired unitary $U_\mt{fin}$ while minimizing this cost function. However, having also identified $Y^I(s)$ as the velocity along the trajectories $U(s)$, we can use our physical intuition to think of this as a classical mechanics problem where we aim to find the extremal trajectory given a particular set of boundary conditions and the somewhat unusual Lagrangian in eq.~\reef{cost1}.

A mentioned in the introduction, we can also make other choices for the cost function, and the analysis will go through essentially unchanged. Hence in order to develop the present problem most easily, we shall consider the $F_2$ or $F_q$ metric in eq.~\eqref{eq:Fmetrics}. That is, we replace eq.~\reef{cost1} with
\beq
\mathcal{D}(U)=\int_0^1\dd s\sqrt{G_{IJ}\,Y^I(s)\,Y^J(s)}~.\label{cost2}
\eeq
This expression should be familiar as the action of a particle moving in a curved space, and hence the optimal path corresponds to a geodesic in the corresponding (Riemannian) geometry. As we mentioned above, $Y^I(s)$ are the components of the velocity in a particular frame, for which the metric $G_{IJ}$ then defines the inner product. In our examples, $G_{IJ}$ is taken to be a purely constant (and usually diagonal) matrix. We will begin by studying the simple Euclidean metric $G_{IJ}=\delta_{IJ}$, which corresponds to the $F_2$ metric above. With this choice, motion in every direction in the space of unitaries is assigned the same cost, \ie the cost of each type of gate is the same. However, our notation is sufficiently general to allow for the assignment of penalty factors for particular gates, as in the $F_q$ metric. We shall return to this possibility in section \ref{sec:penalty}.

To proceed further, we must find a prescription to explicitly identify the functions $Y^I(s)$. Given eq.~\reef{path2}, it is straightforward to show that
\beq
Y^I(s)\,\op_I=\pd_sU(s)\,U^{-1}(s)~.\label{v12}
\eeq
However, this expression is not particularly useful as it stands. In Neilsen's construction \cite{Nielsen:2005mn1,Nielsen:2006mn2,Nielsen:2007mn3}, one works with unitary matrices acting on qubits, rather than operators acting on wave functions. Hence the components of the velocity analagous to eq.~\reef{v12} can be isolated by simply tracing over the corresponding matrix generators. This procedure does not immediately lend itself to eq.~\reef{v12}, so in order to make progress, we shall re-express our problem in terms of matrices. 

Recall that we reduced the problem to evaluating the complexity of the ground state \eqref{eq:targetPhys} of two coupled harmonic oscillators, starting from a factorized Gaussian reference state \eqref{eq:refPhys}. That is, we begin and end with a Gaussian wave function; furthermore, it is straightforward to show that the scaling and entangling operators preserve the general Gaussian form of the wave function, \ie all of the intermediate wave functions take a form analogous to eq.~\eqref{eq:targetPhys}. Therefore, since we're only working with Gaussian states, we may think of the space of states as the space of (positive) quadratic forms. In other words, the states under consideration are all of the form
\beq
\psi \simeq \exp\!\left[-\frac{1}{2}x_a\,A_{ab}\,x_b\right]~,
\label{gauss}
\eeq
and thus we may think of the relevant space of states as the three-dimensional space of 2$\times$2 positive symmetric matrices $A$, with $A_{ab}=A_{ba}$, det$A>0$, and $A_{11},A_{22}>0$.\footnote{These positivity constraints ensure that both eigenvalues of $A_{ab}$ are positive.} In particular, the reference and target states become, respectively,
\beq
A_\mt{R}=\omega_0\mathbb{1}~,\qquad
A_\mt{T}=\left(
\begin{matrix}
\omega_1&\beta\\
\beta&\omega_2
\end{matrix}
\right)~,
\label{eq:stateMatrix}
\eeq
where $\omega_1,\ \omega_2$ and $\beta$ are given by eq.~\reef{eq:omega12pm}.

We now translate the scaling and entangling gates to this matrix representation. That is, we build a representation of these operators as 2$\times$2 matrices which act on the symmetric matrices $A$. In particular, one finds that the gate matrices act as 
\beq
A' = Q_{ab}\, A\,\, Q_{ab}^T\,,
\label{matrix0}
\eeq
where
\beq
Q_{ab} = \exp\!\left[\eps\,M_{ab}\right]
\qquad{\rm with}\quad
\left[M_{ab}\right]{}_{cd} = \delta_{ac}\,\delta_{bd}\,.
\label{matrix}
\eeq
In this notation, $\left[M_{ab}\right]{}_{cd}$ is a 2$\times$2 matrix, where $c$ and $d$ denote row and column indices, respectively.\footnote{A quick way to construct these matrices is to consider the action of $\cO_{ab}$ on the column vector $\lp x_1,x_2\rp^T$, and then build the matrix $M^T_{ab}$ which yields the same result. One can verify that the commutators of the $M_{ab}$ match those of the $\cO_{ab}$. Note that, while the action of the $Q_{ab}$ in eq.~\reef{operate} leaves the wave functions properly normalized at each step, we lose track of this normalization when working with the $A_{ab}$.  \label{footy}} Explicitly, we shall denote the basis of generators $M_I$ as
\beq
\bal
M_{11}&=\begin{pmatrix}1 & 0\\0 & 0\end{pmatrix}\,,\qquad
M_{12}=\begin{pmatrix}0 & 1\\0 & 0\end{pmatrix}\,,\qquad\\
M_{21}&=\begin{pmatrix}0 & 0\\1 & 0\end{pmatrix}\,,\qquad
M_{22}=\begin{pmatrix}0 & 0\\0 & 1\end{pmatrix}~.
\eal\label{Msimple}
\eeq

With this new matrix formulation of our problem, we readily observe that the action of the gates $Q_{ij}$ -- or more generally, circuits constructed from $Q_{ij}$ -- on the vector $(x_1,x_2)^T$ produces a vector whose elements are linear combinations of $x_1$ and $x_2$. Furthermore, since the gates are invertible, this is precisely the definition of the group of transformations $\GLtwo$.\footnote{Note that one can also see the emergence of this group by observing that the algebra of the original operator generators $\cO_{ab}$ in eq.~\reef{operate} close to form the algebra $\frak{gl}(2,\mathbb{R})$.} Thus our circuits form a representation of $\GLtwo$, \ie the $U(s)$ are trajectories in the space of $\GLtwo$ transformations. 

Now, in this matrix formulation, the path-ordered exponentials in eq.~\eqref{eq:pathPsi} are replaced by
\beq
U(s)=\cev{\mathcal{P}}\,\mathrm{exp}\int_0^s\dd \tilde s\ Y^I\!(\tilde s)\,M_I~,\qquad{\rm with}\ \ 
A_\mt{T}=U(s=1)\,A_\mt{R}\,U^T(s=1)~,\label{pathA}
\eeq
where $M_I$ are the generators given in eq.~\reef{Msimple}. The advantage of this formulation is that eq.~\reef{v12} becomes
\beq
Y^I(s)\,M_I=\pd_sU(s)\,U^{-1}(s)
\quad
\implies\quad
Y^I(s)=\tr\lp\pd_sU(s)\,U^{-1}(s)M^T_I\rp~.\label{eq:metricRightInv}
\eeq
That is, we now have a simple expression which yields the components of the velocity vector $Y^I(s)$. Before we can utilize this expression however, we must explicitly construct a parametrization of the $\GLtwo$ transformations. We proceed with this task in the next subsection, but first let us make a few comments. 

Our task will be to find the shortest geodesic in some right-invariant metric on $\GLtwo$ that connects the initial and final states, $A_\mt{R}$ and $A_\mt{T}$, as in eq.~\reef{pathA}. We emphasize \textit{shortest} geodesic because in fact, we will find that there is a continuous family of geodesics connecting the desired states. This non-uniqueness arises because our space of circuits is four-dimensional (since $\mathrm{dim}\lp\GLtwo\rp=4$) whereas our space of states is only three-dimensional (since the $2\times2$ matrices $A_{ij}$ are symmetric). As a result of this mismatch, we should expect to find a one-parameter family of geodesics $U(s)$ which yield the desired transformation $A_\mt{T}=U(s\!=\!1)\,A_\mt{R}\,U^T(s\!=\!1)$. However, as we have explained, the complexity is defined as the cost of the minimal or optimal circuit that obtains the specified target state. Hence this one-parameter family of solutions is merely the set of all possible circuits within this class. To find the optimal circuit, we simply need to find the geodesic within this family with the shortest length \reef{cost2}.

Since our ultimate aim will be to return to free scalar field theory, we note in passing that the notation introduced in the last two subsections generalizes very easily from two coupled oscillators to $N$ coupled oscillators. We would then build a right-invariant metric on $\mathrm{GL}(N,\mathbb{R})$. Furthermore, note that the dimension of the space of circuits becomes $N^2$, while the dimension of the space of Gaussian states or quadratic forms is only $N(N+1)/2$. Hence the non-uniqueness involved in finding the most efficient circuit $U(s)$ which produces the desired transformation grows quickly. We shall discuss the extension to a lattice of oscillators in section \ref{sec:N}.

\subsection{Geodesics on circuit space} \label{geod}
To proceed with constructing the desired geodesics, we must choose an explicit parametrization of a general element $U\in\GLtwo= \mathbb{R}\times\mathrm{SL}(2,\mathbb{R})$. Let us first consider $\widetilde U\in\mathrm{SL}(2,\mathbb{R})$, which can be written as
\beq
\widetilde U=\begin{pmatrix}x_0-x_3~ & x_2-x_1 \\ x_2+x_1~ & x_0+x_3\end{pmatrix}~,
\qquad\mathrm{with}\qquad
x_0^2+x_1^2-x_2^2-x_3^2=1~.
\ee
We recognize the constraint imposing det$\,\widetilde U=1$ as the embedding of (Lorentzian) AdS$_3$ in $\mathbb{R}^{2,2}$. Indeed, the appearance of AdS$_3$ could have been anticipated since the latter is the universal cover of $SL(2,\mathbb{R})$. Our familiarity with this embedding then motivates the following choice of coordinates:
\be
x_0=\cos\tau\cosh\rho~,\;\;\;
x_1=\sin\tau\cosh\rho~,\;\;\;
x_2=\cos\theta\sinh\rho~,\;\;\;
x_3=\sin\theta\sinh\rho~, \label{slr}
\ee
where $\tau,\ \rho$ and $\theta$ are the usual time, radius, and angle, respectively, of global coordinates on AdS$_3$. We can easily extend this parametrization to $U\in\GLtwo= \mathbb{R}\times\mathrm{SL}(2,\mathbb{R})$ by introducing an additional coordinate to parameterize the determinant of $U$, \ie
\beq
U=\begin{pmatrix}x_0-x_3~ & x_2-x_1 \\ x_2+x_1~ & x_0+x_3\end{pmatrix}~,
\qquad\mathrm{with}\qquad
x_0^2+x_1^2-x_2^2-x_3^2=e^{2y}~.
\ee
Hence we extend eq.~\reef{slr} to
\be
x_0=e^y\cos\tau\cosh\rho~,\;\;\;
x_1=e^y\sin\tau\cosh\rho~,\;\;\;
x_2=e^y\cos\theta\sinh\rho~,\;\;\;
x_3=e^y\sin\theta\sinh\rho~, \label{glr}
\ee
where, as before, $\tau,\,\rho,\,\theta$ are coordinates on the $\mathrm{SL}(2,\mathbb{R})$ subgroup, and $y$ parametrizes the $\mathbb{R}$ fibre. With these coordinates, we can express a general $U\in\GLtwo$ as
\beq
U
=e^{y}\,\begin{pmatrix}\cos\tau\cosh\rho-\sin\theta\sinh\rho~ & -\sin\tau\cosh\rho+\cos\theta\sinh\rho \\ \sin\tau\cosh\rho+\cos\theta\sinh\rho~ & \cos\tau\cosh\rho+\sin\theta\sinh\rho \end{pmatrix}~.\label{Umatrix}
\eeq

We are now equipped to construct the geometry implicit in the cost function \reef{cost2}, where the velocity components are given by eq.~\eqref{eq:metricRightInv}. As mentioned above, we begin by choosing $G_{IJ}=\delta_{IJ}$, which assigns an equal cost or weight to every gate. This choice then defines the following right-invariant metric:
\beq
\bal
\dd s^2&=\delta_{IJ} \, \tr\lp\dd U\,U^{-1}\, M^T_I\rp\,\tr\lp\dd U\,U^{-1}\,M^T_J\rp\\
&=2\dd y^2+2\dd\rho^2+2\cosh(2\rho)\cosh^2\!\rho\,\dd\tau^2+2\cosh(2\rho)\sinh^2\!\rho\,\dd\theta^2-2\sinh^2\!\lp2\rho\rp\,\dd\tau\dd\theta~.
\eal\label{metric1}
\eeq
For later use, it is also convenient to express this in the form
\beq
\dd s^2=2\dd y^2+2\dd\rho^2+2\dd x^2+2\cosh(4\rho)\,\dd z^2-4\cosh(2\rho)
\,\dd x\,\dd z~,
\label{metric2}
\eeq
where we have defined the pseudo-lightcone coordinates
\beq
x\equiv\frac{1}{2}(\theta+\tau)~,\qquad
z\equiv\frac{1}{2}(\theta-\tau)~.\label{lcCoords}
\eeq
Note that our metric \reef{metric1} is Euclidean, as is appropriate for defining a cost function, and so does not contain the (Lorentzian) AdS$_3$ geometry noted above. Indeed, a Lorentzian signature would not be suitable for the problem at hand, since certain directions would then carry negative or zero cost. We discuss the relation between our geometry and that of AdS$_3$ in appendix \ref{sec:appxKilling}.

With the geometry in hand, we now wish to find the geodesics, and thereby the optimal circuit. Inspecting the metric \reef{metric1}, we can see three obvious Killing vectors: $\pd_y,\, \pd_\tau,\, \pd_\theta$. However, the metric is right-invariant by construction, meaning eq.~\reef{metric1} remains unchanged if we right-multiply $U(s)$ by a constant $\GLtwo$ transformation. Therefore there must be one Killing vector for each generator of $\GLtwo$, namely, \textit{four}.\footnote{We thank Lucas Hackl for discussions on this point.} In fact, it turns out that choosing $G_{IJ}=\delta_{IJ}$ results in an extra ``accidental'' symmetry, and so the metric above has a total of five Killing vectors. These Killing vectors $ (\hat k_I)^i\pd_i$ are explicitly constructed in appendix \ref{sec:appxKilling}, and are given in eqs.~\eqref{Killed2} and \eqref{Killed5}.

Of course, the existence of five Killing vectors implies an equal number of conserved momenta, $c_I\equiv (\hat k_I)^i\,g_{ij}\,\dot x^j$, which we will use to solve for the geodesics. Given the Killing vectors in eqs.~\eqref{Killed2} and \eqref{Killed5}, it is straightforward to evaluate the corresponding conserved quantities:
\beq
\bal
c_1&=2\,\dot y~,\\
c_2&=2 \sin (\theta -\tau )\dot\rho+\cos (\theta -\tau ) \left[\lp\sinh(4\rho)-\sinh(2\rho)\rp\dot\theta-\lp\sinh(4\rho)+\sinh(2\rho)\rp\dot\tau\right]~,\\
c_3&=2\cos (\theta -\tau )\dot\rho-\sin (\theta -\tau ) \left[\lp\sinh (4 \rho )-\sinh (2 \rho )\rp\dot\theta-\lp\sinh (4 \rho )+\sinh (2 \rho )\rp \dot\tau\right]~,\\
c_4&=\lp\cosh\!\lp4\rho\rp-\cosh\!\lp2\rho\rp\rp\dot\theta-\lp\cosh\!\lp4\rho\rp+\cosh\!\lp2\rho\rp\rp\dot\tau~,\\
c_5&=(1-\cosh\!\lp2\rho\rp)\,\dot\theta+(1+\cosh\!\lp2\rho\rp)\,\dot\tau~,
\eal\label{eq:conserved}
\eeq
where the dot denotes differentiation with respect to some affine parameter $s$ along the geodesic. We are free to choose this parameter such that the normalization of the tangent vector is constrained to be constant, \ie
\beq
g_{ij}\dot x^i\dot x^j
=2\dot y^2+2\dot\rho^2+2\cosh(2\rho)\lp \sinh^2\!\rho\ \dot\theta^2+\cosh^2\!\rho\ \dot\tau^2\rp-2\sinh^2(2\rho)\,\dot\theta\,\dot\tau
\equiv k^2\,.\label{eq:geodesic}
\eeq
In a GR calculation, we would typically choose the normalization (for a spatial geodesic) to be +1, but this choice would leave the final value of $s$ at the end of the circuit undetermined. However, recall that our notation for the path-ordered exponentials above is such that the circuits run over $0\le s\le1$, \cf \reef{eq:pathPsi}. Hence we shall scale the affine parameter $s$ to lie in this range. The normalization constant $k$ then gives the length of the geodesic, \ie the depth of the corresponding circuit, since from eq.~\reef{cost2} we have
\beq
{\cal D}(U)=\int_0^1\dd s\sqrt{g_{ij}\,\dot x^i\,\dot x^j}\equiv k\,.
\label{costa4}
\eeq
The minimum value of $k$ is then the depth of the optimal circuit, and by extension, the complexity of the target state $\psi_\mt{T}$.

Next, we must establish the boundary conditions for our geodesics. The geodesics (and paths in the circuit geometry in general) are described by $\bx(s)=\{\tau(s),\,\rho(s),\,\theta(s),\,y(s)\}$. Now, our initial condition is that $U=\mathbb{1}$ at $s=0$, and by comparing with the parametrization in eq.~\eqref{Umatrix}, we find that all coordinates except $\theta$ are initially zero, \ie
\beq
\bx(s=0)=\{0,\,0,\,\theta_0,\,0\}~.\label{eq:coordinit}
\eeq
Note that the fact that $\theta=\theta_0$ is undetermined is not surprising since this is an angular coordinate, but the geodesic starts at the origin $\rho=0$. Hence the freedom to specify $\theta_0$ is the freedom that the geodesic leave the origin in any direction. In part, this freedom reflects the fact that we do not expect the boundary conditions to uniquely fix the geodesic, but to instead give rise to a one-parameter family thereof---see the discussion at the end of the previous subsection. 

\begin{figure}[h!]
\centering
\includegraphics[width=0.3\textwidth]{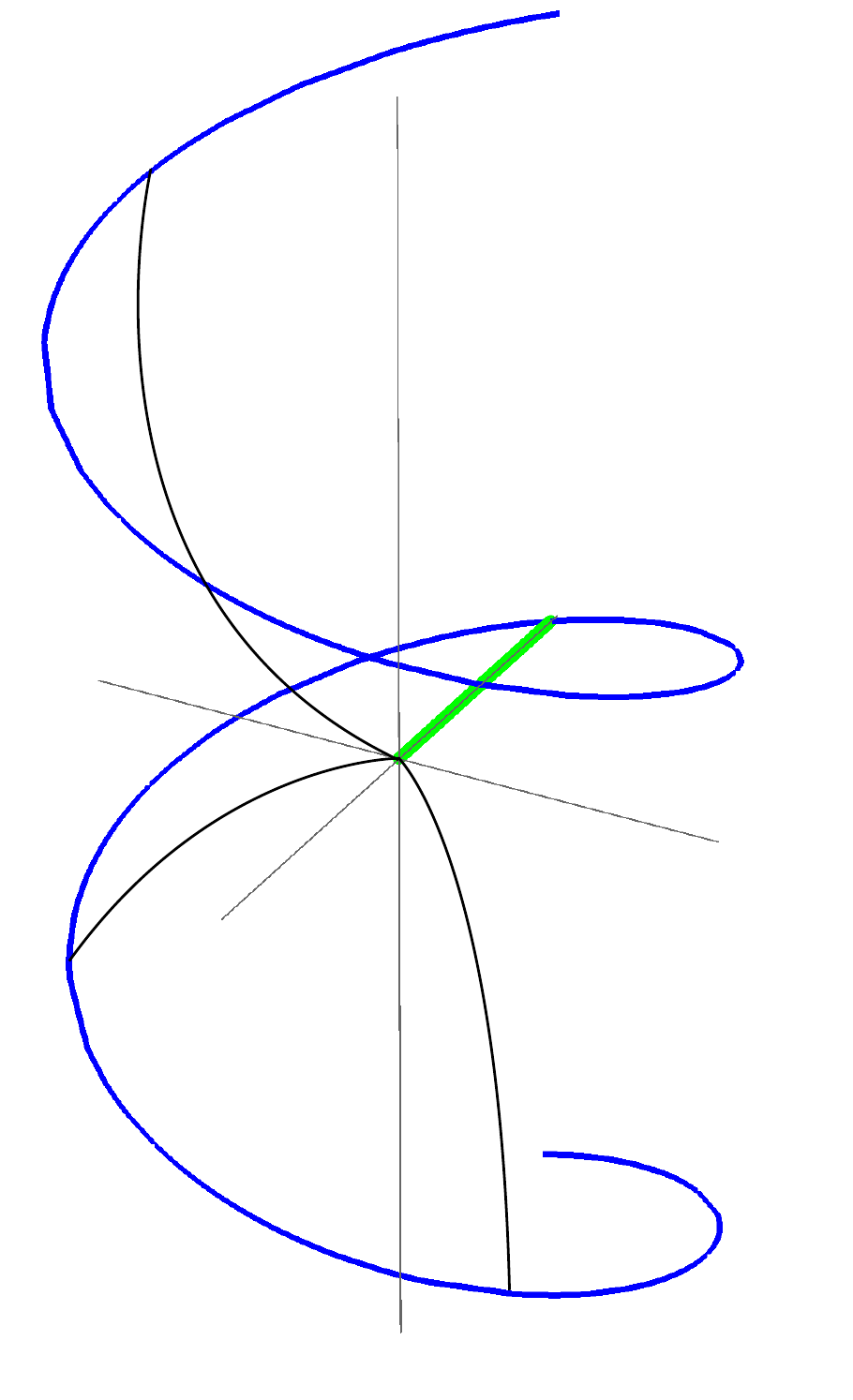}
\caption{Sketch of the one-parameter family of geodesics. The vertical axis is $\tau$, the horizontal plane is described by the radius $\rho$ and the azimuthal angle $\theta$, and the $y$ direction is suppressed. The circuits which produce the transformation from $A_\mt{R}$ to $A_\mt{T}$ are described by geodesics running from the origin to the blue spiral at $\theta+\tau=\theta_1+\tau_1$ and $\rho=\rho_1$ (shown here for the special case $\theta_1+\tau_1=\pi$, which appears in eq.~\reef{fin2} below). The black curves represent (non-minimal) geodesics within the one-parameter family of solutions with different values of $\theta_0$. The minimum geodesic corresponds to the green line in the $\tau=0$ plane with $\Delta\theta=0$ (\ie $\theta_0=\theta_1$), whose length is given by eq.~\eqref{solver4b}. \label{fig:spiral}}
\end{figure}

Now, the end-point of the geodesic is determined by $A_\mt{T}=U(s\!=\!1)\,A_\mt{R}\,U^T(s\!=\!1)$, as in eq.~\reef{pathA}, where the quadratic forms for the reference and target states are given in eq.~\reef{eq:stateMatrix}. Substituting the initial state $A_\mt{R}=\omega_0\,\mathbb{1}$ and the explicit representation of the unitaries \eqref{Umatrix} into this relation, we have
\beq
A_\mt{T}=\omega_0\,UU^T=
\omega_0\, e^{2y_1}\begin{pmatrix}
\cosh(2\rho_1)-\sin(\theta_1+\tau_1)\sinh(2\rho_1) &
\cos(\theta_1+\tau_1)\sinh(2\rho_1) \\
\cos(\theta_1+\tau_1)\sinh(2\rho_1) &
\cosh(2\rho_1)+\sin(\theta_1+\tau_1)\sinh(2\rho_1)
\end{pmatrix}~,
\label{endA}
\eeq
where the subscript 1 denotes the value of the coordinate at $s\!=\!1$, \eg $y_1=y(s\!=\!1)$. Comparing the entries of the matrix on the right-hand side with those of $A_\mt{T}$ in eq.~\reef{eq:stateMatrix}, we arrive at the following
boundary conditions for the end of the geodesic:
\beq
\bal
\omega_1/\omega_0&=e^{2y_1}\left[\cosh(2\rho_1)-\sin(\theta_1+\tau_1)\,\sinh(2\rho_1)\right]~,\\
\omega_2/\omega_0&=e^{2y_1}\left[\cosh(2\rho_1)+\sin(\theta_1+\tau_1)\,\sinh(2\rho_1)\right]~,\\
\beta/\omega_0&=e^{2y_1}\cos(\theta_1+\tau_1)\,\sinh(2\rho_1)~.
\eal\label{eq:finalCoords}
\eeq
Implicitly, these constraints allow us to identify the final coordinates $\bx(s=1)$ for the geodesics corresponding to circuits which produce the desired transformation. Explicitly, we may solve this system to obtain
\beq
e^{2y_1}=\frac{\sqrt{\omega_1\omega_2-\beta^2}}{\omega_0}~,\quad
\cosh(2\rho_1)=\frac{\omega_1+\omega_2}{2\sqrt{\omega_1\omega_2-\beta^2}}~,\quad
\tan(\theta_1+\tau_1)=\frac{\omega_2-\omega_1}{2\beta}~.
\label{fini}
\eeq
However, there is an obvious ambiguity here since $\theta_1$ and $\tau_1$ appear only in the combination $\theta_1+\tau_1$. Since only this linear combination is fixed by eq.~\reef{fini}, we have a one-parameter family of final boundary conditions---the linear combination $\theta_1-\tau_1$ remains unspecified. Na\"ively, this might lead one to suspect a two-parameter family of allowed solutions, since the initial conditions left $\theta_0$ unfixed as well. But this is not the case: rather, the geodesic equations of motion relate the freedom in the boundary conditions at $s=0$ and $s=1$, and the freedom in the initial and final conditions combine to yield the one-parameter family of geodesics anticipated above. This situation is illustrated in figure \ref{fig:spiral}, which shows a one-parameter family of solutions beginning at the origin and ending on the spiral given by $\theta+\tau=\theta_1+\tau_1$ and radius $\rho=\rho_1$. To determine the complexity of the final state $A_\mt{T}$, we must find the minimum length geodesic within this family, and thereby the optimal circuit.

Having specified the boundary conditions, we proceed to solve for the geodesics  by examining the conserved momenta \eqref{eq:conserved}. The first of these gives the simplest constraint: $c_1=2\,\dot y$. Integrating with respect to the affine parameter $s$ then yields: $y(s) = c_1\,s/2+y_0$. In this case the undetermined coefficients are easily fixed by the boundary conditions, $y(s=1)=y_1$ and $y(s=0)=0$, hence:
\beq
c_1=2\,y_1\quad\mathrm{and}\quad y_0=0\qquad\implies \qquad y(s) =y_1\,s\,.
\label{yrun}
\eeq 
Next, we consider $c_4$ and $c_5$. These two constraints may be solved to obtain
\beq
\dot\tau=c_5+\frac{c_4-c_5}{4\cosh^2\!\rho}~,\qquad
\dot\theta=c_5+\frac{c_4+c_5}{4\sinh^2\!\rho}~.
\label{trun}
\eeq
We then observe that $\dot\theta$ diverges at the origin $\rho=0$ unless $c_4=-c_5$, which we must therefore impose in order to be compatible with the initial conditions. Implicitly, we are setting the angular momentum, \ie the conserved momentum associated with the Killing vector $\pd_\theta$, to zero, which is characteristic of geodesics passing through the (radial) origin $\rho=0$. With this condition, the $\theta$ equation can be trivially integrated to yield $\theta=c_5\,s+\theta_0$, where we have already imposed $\theta(s=0)=\theta_0$. Imposing the final boundary condition then yields 
\beq
c_5=\Delta\theta\equiv\theta_1-\theta_0
\qquad\implies\qquad \theta(s)=\Delta\theta\,s+\theta_0~.
\label{gonzo}
\eeq
Furthermore, the above allows us to simplify the $\dot\tau$ equation to
\beq
\dot\tau=\Delta\theta\lp1-\frac{1}{2\cosh^2\!\rho}\rp~.\label{tauDot4}
\eeq
Now, combining our expressions for $\dot\theta$ and $\dot\tau$ with the constraints $c_2$ and $c_3$ in eq.~\reef{eq:conserved}, we find a relatively simple equation for $\dot\rho$:
\beq
\dot\rho^2=\frac{c_2^2+c_3^2}4-\frac{\Delta\theta^2}4\,\tanh^2\!\rho~.\label{rhoDot4}
\eeq

In principle, we should now solve for the general solutions of eqs.~\reef{tauDot4} and \reef{rhoDot4} subject to the boundary conditions in eqs.~\reef{eq:coordinit} and \reef{fini}. While it is possible to carry out this exercise, the final solutions are not particularly illuminating.\footnote{The general solution for eq.~\reef{rhoDot4} is given by
\beq
\sinh\rho=\frac{c}{\sqrt{c^2-\Delta\theta^2}}\,\sinh\lp\frac{s}{2}\sqrt{c^2-\Delta\theta^2}\rp\,,
\label{new4}
\eeq
where $c^2=c_2^2+c_3^2$ is fixed by substituting the boundary condition $\rho=\rho_1$ at $s=1$ into this equation. Furthermore, given this result, it is possible to integrate eq.~\reef{tauDot4} to obtain $\tau(s)$; one finds
\beq
\tau=\Delta\theta\,s-\tan^{-1}\lp\frac{\Delta\theta}{\sqrt{c^2-\Delta\theta^2}}\tanh\lp\frac{s}{2}\sqrt{c^2-\Delta\theta^2}\rp\rp~.
\eeq
Substituting $\tau=\tau_1$ at $s=1$ into this expression fixes $\tau_1$ in terms of $\Delta\theta$ and $c^2$. Combining this result with the boundary condition for $\theta_1+\tau_1$ in eq.~\reef{fini}, we can then determine $\theta_1$. In turn, $\theta_0$ is now fixed since we know $\Delta\theta$ and $\theta_1$. \label{rocket}} Instead, let us point out the particularly simple solution that arises for $\Delta\theta=0$. In this case the expressions for $\dot\tau$ and $\dot\rho$ reduce to
\beq
\bal
\dot\tau=0\qquad&\implies\quad\tau=0\,,\\
\dot\rho=\frac12\sqrt{c_2^2+c_3^2}\quad&\implies\quad \rho=\rho_1\,s\,,
\eal\label{solver4}
\eeq
which combine with $y=y_1\,s$ and $\theta=\theta_0$ from eqs.~\reef{yrun} and \reef{gonzo} to describe a simple ``straight-line'' geodesic. Substituting this solution into eq.~\reef{Umatrix}, we can write the corresponding circuit as
\beq
\bal
U_0(s)&=e^{y_1s}\begin{pmatrix}\cosh\lp\rho_1s\rp- \sin\theta_0\,\sinh\lp \rho_1s\rp~ & \cos\theta_0\,\sinh\lp\rho_1s\rp \\ \cos\theta_0\,\sinh\lp\rho_1s\rp~ & \cosh\lp\rho_1s\rp+\sin\theta_0\,\sinh\lp\rho_1s\rp \end{pmatrix}\\
&=\exp\left[\begin{pmatrix}1 & 0 \\ 0 & 1 \end{pmatrix} y_1\,s+\begin{pmatrix}-\sin\theta_0 & \cos\theta_0 \\ \cos\theta_0 & \sin\theta_0\end{pmatrix} \rho_1\,s\right]\,.
\eal\label{solver4a}
\eeq
Note that an explicit path-ordering is not needed in the second expression since it is simply the exponential of a fixed matrix.\footnote{For this simple case, it is straightforward to identify the exponential form in the second line of eq.~\eqref{solver4a} given the expression appearing in the first. In general however, one would apply eq.~\reef{eq:metricRightInv} to identify the components of $Y^I(s)$ and then substitute these into eq.~\eqref{pathA}.} The circuit depth of $U_0$, \ie the length of the geodesic, is given by eqs.~\reef{eq:geodesic} and \reef{costa4}, which for this simple solution yields
\beq
{\cal D}(U_0)=\sqrt{2(y_1^2+\rho_1^2)}\,.
\label{solver4b}
\eeq
Again, in principle, we should determine all of the other geodesics satisfying the appropriate boundary conditions, and compare their respective circuit depths to ${\cal D}(U_0)$ in order to determine the minimum. However, we shall instead provide a more indirect but less technically challenging proof that this simple straight-line solution is in fact the shortest possible geodesic, and hence that it describes the optimal circuit.

To prove that the straight-line solution above is the geodesic whose length is the (global) minimum, we recall from eq.~\reef{eq:geodesic} that the length of any geodesic is given by the normalization constant $k$. Now into this expression, we substitute our general solutions for $y(s)$ and $\theta(s)$ from eqs.~\reef{yrun} and \reef{gonzo}, respectively, as well as the expression for $\dot\tau$ from eq.~\reef{tauDot4}, whereupon we find
\beq
k^2
=2y_1^2+2\dot\rho^2+\lp1-\frac{1}{2\cosh^2\!\rho}\rp\,\Delta\theta^2\,.\label{eq:geodesic3}
\eeq
This equation holds point-by-point along any geodesic satisfying the appropriate boundary conditions, but what we would like to argue (without explicitly solving for $\rho(s)$) is that $k^2$ is minimized by choosing $\Delta\theta=0$.

To begin, consider motion in the $\rho$-direction along any of our geodesics. The average velocity is given by
\beq
\int_0^1\dd s \,\dot \rho =\rho_1~. \label{fun1}
\eeq
Additionally, we have
\beq
0\le\int_0^1\dd s\, (\dot \rho-\rho_1)^2=\int_0^1\dd s\,\dot\rho^2\ -\rho_1^2 \,,
\label{fun2}
\eeq
and hence we may conclude that $\int_0^1\dd s\,\dot\rho^2\ge\rho_1^2$, and that this inequality is only saturated when $\dot\rho=\rho_1$ along the entire geodesic. Now, examining the coefficient of $\Delta\theta^2$ in eqn.~\reef{eq:geodesic3}, we have
\beq
\frac12\le 1-\frac{1}{2\cosh^2\!\rho}\le 1~,
\label{fun3}
\eeq
where the lower inequality is only saturated at $\rho=0$, and the upper inequality is saturated at $\rho\to\infty$.\footnote{In accordance with its interpretation as a radial coordinate, we do not consider negative values of $\rho$.} Given that all of our geodesics must start at $\rho=0$ and end at $\rho=\rho_1$, upon averaging over any of these geodesics, we find
\beq
\frac12<\int_0^1 \dd s\lp 1-\frac{1}{2\cosh^2\!\rho} \rp< 1\,.
\label{fun5}
\eeq
Finally, let us average eq.~\reef{eq:geodesic3} over any of our geodesics:
\beq
\bal
k^2&=2y_1^2+2\int_0^1\dd s\,\dot\rho^2\ +\ \Delta\theta^2\,\int_0^1 \dd s\lp 1-\frac{1}{2\cosh^2\!\rho}\rp\\
&\ge2y_1^2+2\rho_1^2 +\frac{\Delta\theta^2}2
\ge 2(y_1^2+\rho_1^2)\,.
\eal\label{eq:geodesic4}
\eeq
Comparing this to eq.~\reef{solver4b}, we have established the inequality $k\ge{\cal D}(U_0)$. Furthermore, our argument has established that this inequality can only be saturated with $\dot\rho(s)=\rho_1$ (\ie $\rho=\rho_1 s$) and $\Delta\theta=0$ (\ie $\theta(s)=\theta_0$, which implies $\tau(s)=0$ via eq.~\reef{tauDot4}). We have therefore proved that the simple straight-line geodesic indeed constitutes the global minimum for our cost function \reef{costa4}, and hence that eq.~\reef{solver4b} is in fact the complexity of the Gaussian wave function in this framework:
\beq
\CC(A_\mt{T})=\sqrt{2\lp y_1^2+\rho_1^2\rp}~.
\eeq

As an exercise, we can compare the above result for ${\cal D}(U_0)$ in eq.~\reef{solver4b}, which was evaluated using eq.~\reef{costa4}, with the result found by evaluating eq.~\reef{cost2}. In this case, we must identify the components $Y^I(s)$, which is easily done by examining the exponential expression in eq.~\reef{solver4a}:\footnote{Recall that our $\GLtwo$ generators are given in eq.~\reef{Msimple}.}
\beqa
Y^{11}&=&y_1-\rho_1\,\sin\theta_1\,,\quad
Y^{22}=y_1+\rho_1\,\sin\theta_1\,,
\nonumber\\
&&\qquad Y^{12}=Y^{21}=\rho_1\,\cos\theta_1\,.
\label{components}
\eeqa
Since these components are all constant, the integral over $s$ in eq.~\reef{cost2} is trivial, and the circuit depth (with $G_{IJ}=\delta_{IJ}$) reduces to
\beqa
{\cal D}(U_0)&=&\sqrt{(Y^{11})^2 +(Y^{12})^2+(Y^{21})^2+(Y^{22})^2}
\label{eq:Ccoords}\\
&=&\sqrt{
\lp y_1-\rho_1\sin\theta_1\rp^2+2\lp\rho_1\cos\theta_1\rp^2
+\lp y_1+\rho_1\sin\theta_1\rp^2}
=\sqrt{2\lp y_1^2+\rho_1^2\rp}~,
\nonumber
\eeqa
in agreement with eq.~\eqref{solver4b}.

\subsection{Normal-mode subspace}\label{sec:normalmodes}

To properly interpret the complexity, we must re-express our result \reef{solver4b} in terms of the physical parameters of the two coupled oscillators \reef{qm1}, as well as the frequency $\omega_0$ in the reference state \reef{eq:omega12pm}. However, one finds that the complexity is most elegantly described in terms of the normal-mode frequencies $\tom_+$ and $\tom_-$ given in eqs.~\reef{qm2} and \reef{qm3}. Using eq.~\reef{eq:omega12pm}, the final boundary conditions \reef{fini} simplify to
\beq
y_1=\frac14\,\log\frac{\tom_+\tom_-}{\omega_0^2}~,\qquad
\rho_1=\frac14\,\log\frac{\tom_-}{\tom_+}~,\qquad
\theta_1+\tau_1=\pi~.
\label{fin2}
\eeq
Substituting these expressions for $y_1$ and $\rho_1$ into eq.~\reef{solver4b} then yields the complexity of the ground state,
\beq
\CC(A_\mt{T})   ={\cal D}(U_0)=
\frac{1}{2}\sqrt{\log^2\!\lp\frac{\tilde\omega_+}{\omega_0}\rp+\log^2\!\lp\frac{\tilde\omega_-}{\omega_0}\rp}~.\label{eq:CpmPre}
\eeq
At this point, let us also note that the boundary condition $\theta_1+\tau_1=\pi$ (along with $\Delta\theta=0$ and $\tau(s)=0$) implies that the initial angle is $\theta_0=\pi$. This straight-line geodesic is illustrated by the green line in figure \ref{fig:spiral}. The corresponding circuit \reef{solver4a} simplifies to
\beq
U_0(s)=e^{y_1s}\begin{pmatrix}\cosh\lp \rho_1s\rp~ & -\sinh\lp\rho_1s\rp \\ -\sinh\lp\rho_1s\rp~ & \cosh\lp\rho_1s\rp \end{pmatrix}
=\exp\left[\begin{pmatrix}y_1 & -\rho_1 \\ -\rho_1 & \ y_1 \end{pmatrix} s
\right]\,,
\label{solver5}
\eeq
with $y_1$ and $\rho_1$ given by eq.~\reef{fin2}.

The simple and elegant form \reef{eq:CpmPre} of the complexity in terms of the normal-mode frequencies suggests that we should investigate the optimal circuit \reef{solver4a} in terms of the normal modes. The relationship between the physical positions of the masses and the normal-mode coordinates was given in eq.~\reef{qm3}, but we can understand this change of coordinates in terms of a simple rotation. In particular, we can perform the coordinate transformation via the orthogonal rotation matrix $R$,\footnote{Our transformation matrix certainly satisfies $R\,R^T=R^T\,R=\mathbb{1}$. However, with the conventions adopted above, we note that det$R=-1$ and as a result, we actually have that as a numerical matrix $R$ is symmetric, as shown with the eq.~\reef{eq:rot}. However, we still distinguish $R$ and $R^T$ in the following because $R$ provides a mapping from the physical positions to the normal coordinates, while $R^{-1}=R^T$ provides the inverse mapping. In other words, the columns of $R$ are labeled 1,2 while the rows are labeled +,-- and vice versa for $R^T$. \label{footy99a}} 
\beq
R=\frac{1}{\sqrt{2}}\begin{pmatrix} 1\, & 1 \\ 1\, & -1 \end{pmatrix}
\;\;\;\implies\;\;\;
\begin{bmatrix}\tilde x_+\\ \tilde x_-\end{bmatrix}=R\begin{bmatrix}x_1\\ x_2\end{bmatrix}~.\label{eq:rot}
\eeq
Introducing the short-hand notation $x=\lp  x_1, x_2\rp^T$ and $\tilde x=\lp \tilde x_+,\tilde x_-\rp^T$, the transformation \eqref{eq:rot} may be concisely written $\tilde x =R\, x$, and the inverse transformation becomes $x =R^T \tilde x$. Of course, we can also use this transformation to re-express the target Gaussian wave function in terms of the normal-mode coordinates,
\beq
\psi_\mt{T}\sim\mathrm{exp}\left[-\frac12\,x^TA_\mt{T} \,x\right]=\mathrm{exp}\left[-\frac12\, \tilde x^T R\, A_\mt{T}R^T \tilde x\right]\quad\implies
\quad \tilde A_\mt{T}=R\,A_\mt{T}\,R^T~,
\eeq
where $\tilde A_\mt{T}$ denotes the quadratic form describing the ground state in the normal-mode space. Explicitly performing this rotation, one finds
\beq
\tilde A_\mt{T}=\begin{pmatrix}\tilde\omega_+ & 0 \\ 0 & \tilde\omega_-\end{pmatrix}~.
\label{Atilde}
\eeq
That is, the target state becomes a factorized Gaussian in the normal-mode basis, \cf eq.~\reef{qm2}. Of course, this decoupling was the essential point of introducing the normal-mode coordinates in the first place. Furthermore, if we apply this transformation to the reference state in eq.~\reef{eq:stateMatrix}, we see that it retains its simple form, \ie
\beq
\tilde A_\mt{R} = R \,A_\mt{R} \,R^T =\omega_0\, \mathbb{1}\,.
\label{Atilde2}
\eeq
That is, the reference state remains a factorized Gaussian when written in terms of the normal modes. 

Now, given the action of the gates and circuits on the quadratic forms, \cf eq.~\reef{pathA}, we can transform our minimal circuit \reef{solver5} to act in the normal-mode space:
\beq
\tilde U_0(s) \equiv R\,U_0(s) R^T\qquad
{\rm where}\ \ \ \tilde A_\mt{T}=\tilde U_0(s=1)
\,\tilde A_\mt{R}\,\tilde U^T_0(s=1)\,.\label{rotateU}
\eeq
This transformation effects a remarkable simplification of the circuit \reef{solver5} to
\beq
\bal
\tilde U_0(s)&=\exp\left[\begin{pmatrix}y_1-\rho_1 & 0 \\ 0 & \ y_1+\rho_1 \end{pmatrix} s\right]\\
&=\exp\left[\begin{pmatrix}\frac12\,\log\frac{\tom_+}{\omega_0} & 0 \\ 0 & \ \frac12\,\log\frac{\tom_-}{\omega_0}  \end{pmatrix} s\right]=\begin{bmatrix}\lp\frac{\tom_+}{\omega_0}\rp^{\!s/2} & 0 \\ 0 & \lp\frac{\tom_-}{\omega_0}\rp^{\!s/2} \end{bmatrix} \,,
\eal\label{solver5a}
\eeq
where in the second line we have used eq.~\reef{fin2}.

The important lesson learned here is as follows: from the perspective of the normal modes, both the target state and the reference state are factorized Gaussians, as shown in eqs.~\reef{Atilde} and \reef{Atilde2}. The optimal circuit $\tilde U_0(s)$ then simply acts in a diagonal fashion to ``amplify'' each of the diagonal entries in the corresponding quadratic forms, taking $\omega_0$ to $\tom_\pm$ in a simple linear manner. It is rather intuitive that this should be the optimal way to prepare $\tilde A_\mt{T}$ from $\tilde A_\mt{R}$, since if any off-diagonal entries (\ie entanglement) were introduced along the circuit, they would simply have to be removed by the time the trajectory reaches its end-point. This feature of the optimal circuit will greatly simplify our considerations of a lattice of coupled oscillators in the next section.

Before turning to this generalization however, we wish to emphasize that the original circuit \reef{solver4a} is performing the same operation of amplifying the normal modes---this is simply a matter of re-expressing $U_0$ in an alternative basis of generators. To properly clarify this, we need to introduce some additional notation. In the above, we adopted a tilde to denote various quantities in the normal-modes basis.\footnote{At this point, we wish to alert the reader to a subtle distinction that arises in our notation here: as established in footnote \ref{footy78}, we have introduced tilde's to distinguish quantities related to the normal modes from similar quantities in the position basis. Beginning with eq.~\reef{Atilde}, a state, circuit, or generator carrying a tilde acts in the normal-mode space, \ie on wave functions written in terms of normal modes. However, this should be distinguished from the instances described here, where we place the tilde's on the indices. These tilded indices indicate that a normal-mode ``basis'' may still appear on objects acting in the oscillator position space. For example, above eq.~\reef{bark}, $M_{\tilde I}$ indicates certain linear combinations of the standard generators \reef{Msimple}, which still act on wave functions written in terms of $x_1,x_2$, but in a way that scales or entangles the normal modes.} We also introduced the index notation $I=\{11,22,12,21\}$ to label the components of the velocity $Y^I(s)$ and the generators $M_I$. Here we would like to combine these two conventions to introduce a new index label $\tilde I=\{++,+-,-+,--\}$ to denote the same objects with components acting in the normal-mode basis. Thus the natural basis of generators $\tilde M_{\tilde I}$ with which to construct the circuits acting on the states described in the normal-mode basis are 
\beq
\tilde M_{++}=\begin{pmatrix}1 & 0\\0 & 0\end{pmatrix}\,,\ \quad
\tilde M_{+-}=\begin{pmatrix}0 & 1\\0 & 0\end{pmatrix}\,,\ \quad
\tilde M_{-+}=\begin{pmatrix}0 & 0\\1 & 0\end{pmatrix}\,,
\ \quad
\tilde M_{--}=\begin{pmatrix}0 & 0\\0 & 1\end{pmatrix}~.
\label{Msimple2}
\eeq
As numerical matrices, these $\tilde M_{\tilde I}$ are of course identical to the $M_I$ given in eq.~\reef{Msimple}, but the two sets of generators act in different spaces. Via the transformation \reef{eq:rot}, we can also transform these generators to act on the states in the original position basis, \ie $M_{\tilde I}=R^T\,\tilde M_{\tilde I}\,R$:
\beqa
M_{++}=&\frac12\begin{pmatrix}1 & 1\\1 &1 \end{pmatrix}\quad\   &=\,\frac12\left(M_{11}+M_{22}+M_{12}+M_{21} \right)\,,\nonumber\\
M_{+-}=&\frac12\begin{pmatrix}1 & -1\\1 &-1 \end{pmatrix}\ \ &=\,\frac12\left(M_{11}-M_{22}-M_{12}+M_{21} \right)\,,
\label{bark}\\
M_{-+}=&\frac12\begin{pmatrix}1 & 1\\-1 &-1 \end{pmatrix} &=\,\frac12\left(M_{11}-M_{22}+M_{12}-M_{21}\right)\,,
\nonumber\\
M_{--}=&\frac12\begin{pmatrix}1 & -1\\-1 &\ 1 \end{pmatrix} 
&=\,\frac12\left(M_{11}+M_{22}-M_{12}-M_{21}\right)\,.
\nonumber
\eeqa
The action of these generators can be read off from the indices, \eg $M_{++}$ scales the $x_+$ coordinate or amplifies the corresponding normal mode. Of course, we could also transform the original generators $M_I$ in eq.~\reef{Msimple} with $\tilde M_I=R\, M_I\, R^T$ to construct the corresponding normal-mode basis. For example, $\tilde M_{11}$ would still scale the $x_1$ coordinate but would act on states in the normal-mode basis, \ie it acts on Gaussian wave functions written in terms of $\tilde x_\pm$.

With this new notation in hand, we would like to express our optimal circuit $U_0$ in terms of the generators $M_{\tilde I}$. It is easily shown, either by examining eq.~\reef{solver5} directly or by transforming the expression in eq.~\reef{solver5a} with $U_0(s)=R^T\,\tilde U_0(s)\,R$, that the optimal circuit can be expressed as
\beq
U_0(s)=\exp\left[\lp M_{++}\, (y_1-\rho_1)+M_{--}\, (y_1+\rho_1)\rp \,s\right]
~,
\label{swer3}
\eeq
where $M_{\pm\pm}$ are the linear combinations of the original generators given in eq.~\reef{bark}. In this form, we again recognize that the optimal circuit is simply amplifying the two normal modes, without introducing (and then having to remove) any entanglement between $x_\pm$.

We can also observe that this simple circuit only involves two commuting generators, $M_{++}$ and $M_{--}$. Since the generators commute, it is straightforward to show that the geometry of corresponding normal-mode subspace is flat. That is, if we consider general circuits of the form
\beq
U(y,\rho)=\exp\left[M_{++}\, (y-\rho)+M_{--}\, (y+\rho)\right]
~,
\label{swer4}
\eeq
then the corresponding metric becomes\footnote{This conclusion is slightly premature, since we have not shown that the metric \reef{metric1} is invariant under the change of basis from the original generators \reef{Msimple} to those in eq.~\reef{bark}, but we shall prove this below in eq.~\reef{metric1a}. Note that we have also used that the new basis of generators still satisfies $\tr\lp M_{\tilde I}\,M_{\tilde J}^T\rp=\delta_{\tilde I\tilde J}$.}
\beq
\bal
\dd s_\mt{n-m}^2&=\delta_{\tilde I\tilde J} \, \tr\lp\dd U\,U^{-1}\, M^T_{\tilde I}\rp\,\tr\lp\dd U\,U^{-1}\,M^T_{\tilde J}\rp\\
&=\dd (y-\rho)^2+\dd(y+\rho)^2=2\dd y^2+2\dd\rho^2~.
\eal\label{metric5x}
\eeq
Hence we recognize the normal-mode subspace as precisely the $\lp\theta,\tau\rp=\lp\pi,0\rp$ plane in our extended geometry \reef{metric2}.\footnote{Implictly, we may allow $\rho$ to run over positive and negative values in eq.~\reef{swer4}. Hence this subspace also includes $\lp\theta,\tau\rp=\lp0,0\rp$.} This perspective also makes clear why the optimal geodesic remains in the normal-mode subspace. Examining the full metric \reef{metric2}, it is clear that motion in the $\theta$ and $\tau$ directions only extends the length of the trajectory. Thus since the start and end points both lie in this plane, there is no advantage to be gained by moving out of the normal-mode subspace. This argument also relies on the fact that $g_{yy}$ and $g_{\rho\rho}$ in the full metric \reef{metric2} are constants, independent of $\theta$ and $\tau$, which precludes the existence of ``short-cuts'' to be found by moving off the normal-mode subspace (we return to this point in section \ref{sec:penalty}). This is another important feature that extends to the case of a lattice of coupled oscillators in the next section.

To close this section, we wish to introduce some additional technology which will prove useful in those that follow. Thus far, we have two particularly useful sets of generators for our gates and circuits, namely, $M_I$ and $M_{\tilde I}$ given in eqs.~\reef{Msimple} and \reef{Msimple2}, respectively. While these generators all act on states and circuits in the physical basis, $M_I$ acts to scale or entangle the physical positions $x_{1,2}$, while $M_{\tilde I}$ scales or entangles the normal-mode coordinates $x_\pm$. The transformation between the two bases is given in eq.~\reef{bark}, but we would like to build an explicit transformation matrix $\hat R$:
\beq
M_{\tilde I} = \widehat R_{\tilde I J}\,M_J\qquad{\rm where}\qquad
\widehat R_{\tilde I J} =\frac12\begin{pmatrix}
1 &\ 1&\ 1&\ 1\\
1 &-1&\ 1&-1\\
1 &\ 1&-1&-1\\
1 &-1&-1&\ 1
 \end{pmatrix}=R_{ka}\otimes R_{\ell b}\,.
\label{bark2}
\eeq
Note that in the final equality, $R$ is the rotation matrix in eq.~\reef{eq:rot}, and we are identifying the indices as follows: $\tilde I=(k\ell)$ with $k,\ell\in\{+,-\}$, and $J=(ab)$ with $a,b\in\{1,2\}$.\footnote{ Recall that as defined in eq.~\reef{eq:rot}, $R$ is the matrix which transforms the `1,2' indices of the oscillator position basis to the `+,--' indices of the normal-mode basis---see footnote \ref{footy99a}.} This identification is really the origin of the interesting tensor product structure $\widehat R=R\otimes R$. The expression in eq.~\reef{bark2} indicates that the first (second) $R$ is rotating the first (second) component of the pairs which comprise the $\tilde I$ and $J$ indices on the two generators. Given this expression, we immediately see that  $\widehat R$ is also an orthogonal rotation matrix. Hence we can easily invert the transformation between the basis generators via $M_I=(\widehat R^T)_{I\tilde J}M_{\tilde J}=\widehat R_{\tilde J I} M_{\tilde J}$. Similarly, this transformation acts on the velocity components as $Y^I=Y^{\tilde J}\widehat R_{\tilde J I}$. These transformations will prove useful in examining the complexity with cost functions written in different bases. For example, in the present context, we can see that the cost function remains unchanged if we express it directly in the normal-mode basis. We can also transform the metric \reef{metric1} as follows:
\beq
\bal
\dd s^2&=\delta_{IJ} \, \tr\lp \dd U\,U^{-1}\, M^T_I\rp\,\tr\lp \dd U\,U^{-1}\,M^T_J\rp\\
&=\widehat R_{\tilde I I} \,\widehat R_{\tilde J J}\,\delta_{IJ} \, \tr\lp \dd U\,U^{-1}\, M^T_{\tilde I}\rp\,\tr\lp \dd U\,U^{-1}\,M^T_{\tilde J}\rp\\
 &=\delta_{\tilde I \tilde J} \, \tr\lp \dd U\,U^{-1}\, M^T_{\tilde I}\rp\,\tr\lp \dd U\,U^{-1}\,M^T_{\tilde J}\rp~,
\eal\label{metric1a}
\eeq
where we have used the fact that $\widehat R$ is an orthogonal matrix. In going from the second to third line, we have used the identity $\widehat R_{\tilde I I} \,\delta_{IJ\vphantom{\tilde I}}\,(\widehat R^T)_{J\tilde J }=\delta_{\tilde I \tilde J} $. Note that the invariance of the metric under this change of basis was already used in evaluating the metric on the normal-mode subspace in eq.~\reef{metric5x}. We extend this discussion of changing between the position and normal-mode bases to the case of a linear lattice of $N$ oscillators in appendix \ref{nmsub}.

\section{A lattice of oscillators}\label{sec:N}

In this section, we wish to return to the original problem of a free scalar field regulated by a lattice, \cf \reef{eq:Hlattice}. That is, we will consider evaluating the complexity of the ground state of a lattice of coupled oscillators \reef{hqm}. Drawing on our experience with the two coupled oscillators, this becomes a straightforward calculation. In particular, as we saw above, both the ground state and the reference state are described by factorized Gaussians in the normal-mode space. And in this space, the optimal circuit simply amplifies each of the diagonal entries in the corresponding quadratic forms in a linear manner. To simplify the technicalities in the following discussion, we will explicitly consider the case of a one-dimensional lattice, and discuss more general dimensions in the next subsection. 

Hence, we begin with $N$ oscillators on a one-dimensional circular lattice,
\be
H=\frac{1}{2}\sum_{a=0}^{N-1}\left[  p_a^2+\omega^2\, x_a^2+\Omega^2\lp x_a-x_{a+1}\rp^2\right]~,
\label{qm88}
\ee
with periodic boundary conditions $x_{a+N}=x_a$.\footnote{Note that for convenience, we have labeled the first oscillator with $a=0$, rather than $a=1$, \ie the sum in eq.~\reef{qm88} runs over $a\in\{0,1,\cdots,N-1\}$.} As in the two oscillator problem, we have set the masses $M_a=1$ for simplicity but we should think of the frequencies as being related to the field theory parameters by $\omega=m$ and $\Omega=1/\delta$, as in eq.~\reef{hqm}. The Hamiltonian \reef{qm88} then corresponds to the lattice version of a (one-dimensional) free scalar field on a circle of length $L=N\,\delta$. Of course, to solve the above system, one simply rewrites the Hamiltonian in terms of the normal modes,
\be
H=\frac{1}{2}\sum_{k=0}^{N-1}\left[ \, |\tilde p_k|^2+\tom_k^2\  |\tilde x_k|^2\,\right]~,
\label{qm288}
\ee
where the transformation to the normal-mode basis is achieved by a (discrete) Fourier transform,
\beq
\tilde x_k\equiv \frac{1}{\sqrt{N}}\sum_{a=0}^{N-1}\mathrm{exp}\lp-\frac{2\pi i\,k}{N}\,a\rp x_a~.\label{eq:Fourier}
\eeq
where $k\in\{0,\ldots,N\!-\!1\}$, and we note that $\tilde x_k^\dagger=\tilde x_{N-k}$.\footnote{We can see this result as a combination of two simpler identities: $\tilde x_k^\dagger=\tilde x_{-k}$, which follows from the complex conjugation of eq.~\reef{eq:Fourier}, and $\tilde x_k=\tilde x_{k+N}$, which follows from the periodicity of the lattice. Note that our convention for the range of $k$ was chosen to match the range of the position labels $a$, rather than shifting the range of $k$ to run over positive and negative values, \ie $k\in\{-\lceil N/2\rceil+1,-\lceil N/2\rceil+2,  ,\cdots, \lfloor N/2\rfloor\}$, which is a more typical convention. Furthermore, for future reference, note that we can define $\vec u_k\equiv [u_k]_a=\mathrm{exp}\lp -2\pi i\,k\,a/N\rp$ as the orthogonal basis of an $N$-dimensional vector space,  satisfying the normalization condition
\beq
\vec u^\dagger{}_{\! k}\cdot\vec u_k'=\sum_{a=1}^N[u^\dagger{}_{\! k}]_a [u_{k'}]_a=\sum_{a=0}^{N-1}\exp\!\lp-\frac{2\pi i(k-k')}{N}\,a\rp=N\,\delta_{k,k'}~.\label{eq:FourierNorm}
\eeq
Hence we use the usual definition for the normal-mode momenta
\beq
\tilde p_k\equiv \frac{1}{\sqrt{N}}\sum_{a=0}^{N-1}\mathrm{exp}\lp\frac{2\pi i\,k}{N}\,\,a\rp p_a~.
\eeq
Note the change in the sign in the exponential in comparison to eq.~\reef{eq:Fourier}. This definition then produces
the standard commutation relations: $[\tilde x_k,\tilde p_{k'}]=i\delta_{kk'}$ and $[\tilde x_k,\tilde x_{k'}]=0=[\tilde p_k,\tilde p_{k'}]$. \label{crap}} 
The normal-mode frequencies $\tom_k$ are defined in terms of the physical frequencies $\omega$ and $\Omega$ in the Hamiltonian \reef{qm88} as follows:
\beq
\tom_k^2=\omega^2+4\Omega^2\,\sin^2\!\frac{\pi k}{N}\,,\label{eq:eigenfreq}
\eeq
(see appendix \ref{sec:omegaDeriv}). As desired, eq.~\reef{qm288} reduces the problem to $N$ decoupled harmonic oscillators, which enables us to easily write the ground-state wave function as
\beq
\psi_0(\tilde x_0,\tilde x_1, \tilde x_2,\cdots)
=\prod_{k=0}^{N-1}\,\left(\frac{\tom_k }{\pi}\right)^{1/4}\ \mathrm{exp}\!\left[-\frac{1}{2}\,\tom_k\,|\tilde x_k|^2\right]~.\label{targetk}
\eeq
As before, this ground state will be the target state in our complexity computations. 

While eq.~\reef{targetk} will suffice to describe the ground state, in principle, one would also like to express the wave function in terms of the original variables $x_a$ in the position basis. This transformation is facilitated using notation introduced in section \ref{sec:normalmodes}. In particular, following eq.~\reef{eq:rot}, we write the Fourier transformation \reef{eq:Fourier} between the position and normal-mode bases as $\tilde x = R_{\ssc N}\, x$, with
\beq
R_{\ssc N}\equiv\frac{1}{\sqrt{N}}\begin{pmatrix}
1 & 1 & 1 & \ldots & 1 \\
1 & \mu & \mu^2 & \ldots & \mu^{N-1} \\
1 & \mu^2 & \mu^4 & \ldots & \mu^{2(N-1)} \\
\vdots & \vdots & \vdots & \ddots & \vdots \\
1 & \mu^{N-1} & \mu^{2(N-1)} & \ldots & \mu^{(N-1)^2} \\
\end{pmatrix}
~,
\label{logan}
\eeq
where $\mu\equiv\exp\lp-2\pi i/N\rp$.\footnote{As discussed in footnote \ref{footy99a} for the matrix $R$ in eq.~\reef{eq:rot}, we distinguish $\RN$ from $\RN^T$ even though the numerical matrix in eq.~\reef{logan} is symmetric. Note that if we write out the transformation to show the indices, we have $\tilde x_k = [R_{\ssc N}]_{ka}\, x_a$. That is, the row index of $\RN$ has values in the momenta $k$ while the column index has values in the lattice position $a$.  In passing, we also observe that eq.~\reef{logan} reduces to eq.~\eqref{eq:rot} for the special case $N=2$, for which we have $\mu=\exp\lp-i\pi\rp=-1$.} Since $\RN$ is a unitary matrix, \ie $\RN\!{}^\dagger \RN=\mathbb{1}$, the inverse transformation is given by $x =\RN\!{}^\dagger\,\tilde x$. 

Now let us adopt the notation of section \ref{sec:geometry} (and in particular, of eq.~\reef{gauss}) to write the target state \reef{targetk} as
\beq
\psi_\mt{T}(\tilde x_k)=\prod_{k=0}^{N-1}\lp\frac{\tom_k}{\pi}\rp^{\frac14}\ \mathrm{exp}\left[-\frac12\,\tilde x^\dagger  \tilde A_\mt{T}\, \tilde x\right]\qquad
{\rm with}\qquad \tilde A_\mt{T}=\mathrm{diag}\lp\tom_0,\ldots,\tom_{N-1}\rp~.
\label{raffle}
\eeq
Using the rotation \reef{logan}, we can write this target state in terms of the physical coordinates,\footnote{The relation $\tom_k=\tom_{N-k}$ ensures that $A_\mt{T}$ is real. \label{footygg}}
\beq
\psi_\mt{T}(x_a)=\prod_{k=0}^{N-1}\lp\frac{\omega_k}{\pi}\rp^{\frac14}\ \exp\left[-\frac12\, x^T A_\mt{T}\,  x \right]\qquad
{\rm with}\qquad A_\mt{T}=\RN\!{}^\dagger \tilde A_\mt{T}\,\RN~.
\label{targetX}
\eeq

We are now prepared to extend our complexity calculations to this lattice of coupled oscillators. We have already identified the target state as the ground state \reef{targetk}. In analogy with eq.~\reef{eq:refPhys}, the reference state will be a factorized Gaussian state,
\beq
\psi_\mt{R}(x_a)=\lp\frac{\omega_0}{\pi}\rp^{N/4}\,
\exp\left[-\frac12\, x^T  A_\mt{R} \, x \right]\qquad
{\rm with}\qquad A_\mt{R}=\omega_0\,\mathbb{1}~.
\label{refX}
\eeq
where the individual oscillators are completely unentangled.\footnote{Recall our tilde notation to distinguish the normal-mode space from the physical space. In particular, the reference frequency $\omega_0$ is independent of the normal-mode frequency with $k=0$, \ie $\omega_0\neq\tom_0$!}  An important feature of our reference state is that it is invariant under translations on the lattice, \ie the Gaussian of each oscillator has the same width $\omega_0$. As a result, it remains a factorized Gaussian when expressed in terms of the normal-mode coordinates:
\beq
\psi_\mt{R}(\tilde x_k)=\lp\frac{\omega_0}{\pi}\rp^{N/4}\,
\exp\left[-\frac12\, \tilde x^\dagger   \tilde A_\mt{R} \, \tilde x \right]\qquad
{\rm with}\qquad \tilde A_\mt{R}=\RN\, A_\mt{R}\,\RN\!{}^\dagger =\omega_0\,\mathbb{1}~.
\label{refk}
\eeq

Lastly, we need to consider the elementary gates with which we will build the circuit $U$ that implements the desired transformation $\psi_\mt{T}=U\,\psi_\mt{R}$. With the notation introduced in eq.~\reef{eq:gates}, the set of gates (particularly the entangling and scaling gates) is easily enlarged for the present problem by simply extending the range of the indices: $a,b\in\{1,2\}\ \longrightarrow\ a,b\in\{0,1,2,\cdots,N-1\}$. These discrete gates are then easily extended to the path-ordered exponentials introduced in eqs.~\reef{eq:pathPsi} and \reef{path2}, \ie $U(s)=\cev{\mathcal{P}}\,\exp\left[\int_0^s\dd \tilde s\,Y^I\!(\tilde s)\,\op_I\right]$, where the index $I$ runs over the $N^2$ values corresponding to pairs $(ab)$, and the operators $\op_I$ take the same form as in eq.~\reef{operate}. In discussing the target and reference states with the notation of eq.~\reef{gauss}, we also anticipated mapping these exponentials to the matrix formulation introduced in eqs.~\reef{matrix0}, \reef{matrix} and \reef{pathA} for Gaussian states. In fact, the generators have precisely the form given in eq.~\reef{matrix}, where again the indices run over the range $a,b,c,d\in\{0,1,2,\cdots,N-1\}$. That is, we now have $N^2$ generators which are $N\times N$ matrices. This extends the $\GLtwo$ group found in section \ref{sec:geometry} to the group $\mathrm{GL}(N,\mathbb{R})$ in the present problem.

Following the analysis in section \ref{sec:geometry}, we use the analogous $F_2$ cost function, {\it i.e.},\footnote{In the position basis, there is no need for the complex conjugations appearing in eqs.~\reef{cost5} or \reef{metric5} since all of the relevant quantities are real. However, we are including them here in anticipation that later on, we will transform these formulae to the normal-mode space. These  transformations are accomplished with $\RN$ in eq.~\reef{logan}, which is a complex unitary matrix.  Hence using, \eg $M^\dagger$ rather than $M^T$ allows us to use precisely the same expressions without change.  Of course, as defined in eq.~\reef{eq:Fourier}, the normal modes are generally complex, but as we commented above, they also satisfy the ``reality condition'' $\tilde x_k^\dagger=\tilde x_{N-k}$---which ensures that we have not doubled the number of degrees of freedom.}
\beq
\mathcal{D}(U)=\int_0^1\dd s\sqrt{\delta_{IJ}\,Y^I(s)\lp Y^J(s)\rp^*}\,, \qquad{\rm where}\qquad Y^I(s)=\tr\lp\pd_sU(s)\,U^{-1}(s)M^\dagger_I\rp\,.
\label{cost5}
\eeq
Hence the optimal circuit will correspond to a geodesic in the $\mathrm{GL}(N,\mathbb{R})$ geometry given by a right-invariant metric, analogous to eq.~\reef{metric1}. To simplify the discussion of the metric here (and in the next section), we introduce the following notation:
\beq 
\dd s^2=\delta_{IJ} \  \dd Y^I\!\lp \dd Y^J\rp^* \qquad
{\rm with}\qquad \dd Y^I=\tr\lp \dd U\,U^{-1}\, M^\dagger_I\rp\,.
\label{metric5}
\eeq
However, extending the detailed calculations above to the full $N^2$-dimensional geometry would be very involved. In particular, the next step would require finding the analog of eq.~\reef{Umatrix}, \ie a convenient parametrization of a general group element $U\in \mathrm{GL}(N,\mathbb{R})$, which would naturally involve $N^2$ coordinates. Thus at this point, we rely on the lessons learned from the case of two coupled oscillators in the previous section. 

In particular, there we found that since both the ground state and the reference state are described by factorized Gaussians in the normal-mode basis, the optimal circuit simply acts to amplify each of the diagonal entries in the corresponding quadratic forms in a simple linear manner. We have already noted by way of eqs.~\reef{raffle} and \reef{refk} that the former statement about factorized Gaussians also applies in our lattice problem. Hence it is natural that the most efficient circuit simply amplifies the Gaussian width for each of the normal-mode coordinates, \ie $\omega_0\to\tom_k$. In particular, the circuit does not introduce any entanglement between the normal modes at any stage, since this entanglement would have to be removed before arriving at the final target state \reef{raffle}. Via eq.~\reef{rotateU}, let us write the optimal circuit acting in the normal-mode basis; we have
\beq
U_0(s) = \RN\!{}^\dagger\,\tilde U_0(s)\, \RN\qquad
{\rm where}\qquad \tilde A_\mt{T}=\tilde U_0(s=1)
\,\tilde A_\mt{R}\,\tilde U^\dagger_0(s=1)\,,\label{nmU}
\eeq
and thus the straight-line circuit $\tilde U_0(s)$ becomes
\beq
\tilde U_0(s)=\exp\left[\tilde M_0\,s\right]
\qquad{\rm with}\ \ 
\tilde M_0={\rm diag}\!\left({\textstyle \frac{1}{2}\log\frac{\tom_0}{\omega_0},\,
\frac{1}{2}\log\frac{\tom_1}{\omega_0},
\cdots, \frac{1}{2}\log\frac{\tom_{\ssc N-1}}{\omega_0}} \right)\,.
\label{raffle2}
\eeq
This circuit certainly accomplishes the desired transformation with $\tilde U_0(s=1) = \exp\left[\tilde M_0\right]$, but the intuition from the previous analysis of two coupled oscillators suggests that it is also the optimal circuit.

We can add to this intuitive picture as follows: in the discussion around eqs.~\reef{swer4} and \reef{metric5}, we identified the normal-mode subspace as consisting of those circuits $U$ which only involve the scaling generators for the normal modes. Consequently, it is straightforward to show that the geometry of the normal-mode subspace is flat since these generators all commute with one another. In the present case, the normal-mode subspace becomes a $N$-dimensional subspace of $U\in \mathrm{GL}(N,\mathbb{R})$ with the form $U=\RN\,\tilde U \RN\!{}^\dagger$, where\footnote{In general, the coordinates $\tilde y_k$ are complex but satisfy the normal-mode ``reality condition'' $\tilde y_k^\dagger=\tilde y_{N-k}$.}
\beq
\tilde U_\mt{n-m}=\exp\left[\tilde M_\mt{n-m}\right]
\qquad{\rm with}\qquad
\tilde M_\mt{n-m}={\rm diag}\!\left(\tilde y_0,\,\tilde y_1,\,\cdots,\, \tilde y_{\ssc N-1} \right)\,.
\label{raffle4}
\eeq
Substituting this expression into eq.~\reef{metric5}, one finds the following flat Cartesian metric induced on this subspace:
\beq
\dd s^2_\mt{n-m}=|\dd \tilde y_0|^2+|\dd \tilde y_1|^2+\cdots+|\dd \tilde y_{\ssc N-1}|^2\,.
\label{metric6}
\eeq
Therefore any geodesic within the normal-mode subspace will simply take the form of a straight line. It is then straightforward to show that if we confine the circuit to this normal-mode subspace \reef{metric6}, the optimal circuit is described by the simple circuit in eq.~\reef{nmU}, which we write as
\beq
U_0(s) = \RN\!{}^\dagger\,\tilde U_0(s)\, \RN 
= \exp\!\left[\RN\!{}^\dagger \, \tilde M_0\, \RN\,s\right]\,,
\label{straight9}
\eeq
where $\tilde U_0(s)$ and $\tilde M_0$ are defined in eq.~\reef{raffle2}. 

There are actually some subtleties in the preceding argument which make the conclusion somewhat premature. The first is that eq.~\reef{metric5} which defines the metric is written in the position basis, whereas eq.~\reef{metric6} was implicitly calculated for an expression \reef{raffle4} written in the normal-mode basis. That is, in eq.~\reef{raffle4}, we worked with $\tilde M_\mt{n-m}=\tilde Y^{\tilde I} \tilde M_{\tilde I}$ with a particular choice of $\tilde Y^{\tilde I}$.\footnote{Again, the tilde on the index $\tilde I$ indicates that it runs over pairs of momentum labels $(k\ell)$, while the tilde on $M$ indicates that these generators act on Gaussian wave functions written with the normal-mode coordinates $\tilde x_k$.} However, we show in appendix \ref{nmsub} that this was nonetheless a valid approach since the metric takes precisely the same form when written in terms of the normal-mode space. This requires extending the discussion around eq.~\reef{bark2} describing the change of bases for the case of two coupled oscillators to the analogous transformation for our linear lattice of $N$ oscillators. 

Secondly, to properly establish that the optimal circuit follows a straight line in the normal-mode subspace, as in eq.~\reef{straight9}, we must show that no shorter path can be found by making an excursion outside this subspace. To begin, we note that implicitly we assumed in eq.~\reef{raffle4} that all of the other coordinates in the $\GLN$ geometry could be set to zero. Recall that in the $\GLtwo$ metric \reef{metric1}, the metric on the normal-mode subspace was completely independent of the other coordinates, \ie we had $\dd s^2_\mt{n-m}=2\dd y^2 +2\dd \rho^2$ irrespective of the values of $\theta$ and $\tau$. In particular, recall that the optimal circuit was a straight line in this subspace with $\theta=\pi$ and $\tau=0$.

We would like to establish a similar result for the present $N^2$-dimensional geometry. For simplicity, we will work in the normal-mode space. We proceed by expressing general circuits $\tilde U$ using the Iwasawa (or KAN) decomposition of $\GLN$; see for example \cite{bump}. This states that any $\tilde U\in \GLN$ can be uniquely written as the product of three matrices, $\tilde U = K\, A\, N$, where $K$ is an orthogonal matrix, $A$ is a diagonal matrix with positive entries,\footnote{We denote this diagonal matrix with the traditional $A$, but it should not be confused with the quadratic forms specifying the Gaussian states, \cf eq.~\reef{gauss}. Similarly, $K$ here should not be confused with the gates producing a momentum shift in eq.~\reef{eq:gates}, nor should $N$ be confused with the total number of oscillators. We trust that these distinctions will be clear from context.} and $N$ is an upper triangular matrix with every diagonal element equal to 1. Clearly, we are interested in the $A$ component as this describes the normal-mode subspace, as in eq.~\reef{raffle4}.

As a warm up exercise, let us consider translating $\tilde U_\mt{n-m}$ by some fixed angles and shifts. In particular, we write $\tilde U=K_0\,\tilde U_\mt{n-m}\,N_0$ where only the $\tilde y_k$ in $\tilde U_\mt{n-m}$ vary (\cf eq.~\reef{raffle4}) and ask what is the metric on the corresponding subspace. Since $N_0$ acts on the right, and the metric is right-invariant by construction, it has no effect on the geometry. Our experience in changing bases in appendix \ref{nmsub} allows us the eliminate the $K_0$ rotation as well: following eq.~\reef{tran1}, we write the differentials $\dd\tilde Y^{\tilde I}=\tr\big(\dd\tilde U \,\tilde U\,\tilde M^\dagger_{\tilde I} \big)$ as
\beq
\dd\tilde Y^{\tilde I}=
\tr\lp \dd\tilde U_\mt{n-m} \,\tilde U_\mt{n-m}^{-1}\,\big[K_0^T \tilde M_{\tilde I}\,K_0\big]^\dagger \rp
\,.\label{tran4}
\eeq
Now in the last factor, $K_0$ acts by a similarity transformation on the generators which effectively produces a change of basis. However, using the special form of the generators \reef{lot1}, it is straightforward to show -- following a series of steps analogous to those given in eqs.~\reef{lot2} or \reef{lot2a} -- that
\beq
K_0^T \tilde M_{\tilde I}\,K_0 = \big[\widehat K_0\big]_{\tilde I\tilde J}\,\tilde M_{\tilde J}\qquad
{\rm where}\qquad \widehat K_0=K_0\otimes K_0\,.
\label{tran5}
\eeq
It follows that $\widehat K_0$ is an orthogonal matrix since $K_0$ is orthogonal, and hence this rotation of the generator basis leaves the metric unchanged. Therefore we find that the induced metric on this subspace is
\beq
\dd s^2_\mt{n-m}=\delta_{\tilde I\tilde J}\,\dd\tilde Y^{\tilde I}_\mt{n-m}
\big( \dd\tilde Y^{\tilde J}_\mt{n-m}\big)^*\qquad{\rm where}\qquad
\dd\tilde Y^{\tilde I}_\mt{n-m}=\tr\lp \dd\tilde U_\mt{n-m} \,\tilde U_\mt{n-m}^{-1}\, \tilde M_{\tilde I}^\dagger \rp\,,
\label{tran6}
\eeq
which again yields the simple answer given in eq.~\reef{metric6}.

This result establishes that there are indeed no short-cuts to be found by running the circuit through the angle and shift directions. That is, the circuit must run from $\tilde y_k=0$ to $\tilde y_k=\frac12\log(\tilde \omega_k/\omega_0)$ as in eq.~\reef{raffle2}. Eq.~\reef{tran6} further establishes that there will be a fixed distance or cost associated with this displacement, irrespective of the orientation of the normal-mode subspace in the full geometry, that is, irrespective of the angles and shifts chosen in $K_0$ and $N_0$. Since the full geometry is Euclidean, moving in these ``orientation directions'' will only add to the distance. Thus the best strategy is to fix the shifts and angles at the beginning of the circuit (to zero, as required by $U(s\!=\!0)=\mathbb{1}$) and then move only in the normal-mode directions.

This argument is still not quite sufficient to establish that the simple straight-line circuit is a geodesic in the full $N^2$-dimensional geometry. In particular, non-vanishing off-diagonal terms in the metric which mix $\tilde y_k$ with the other coordinates would force the geodesic to move away from the normal-mode subspace in the additional angle and shift directions. But since evaluating the full metric would require a rather lengthy and involved calculation, we instead consider small deviations of the circuits around the subspace specified by $\tilde U=K_0\,\tilde U_\mt{n-m}\,N_0$, \ie we extend our initial ansatz to allow small excursions in the $K$ and $N$ directions,
\beq
\tilde U=K_0\,\exp\big[ \tilde M^\mt{rot}_{\tilde I}\theta_{\tilde I}\big]\,\tilde U_\mt{n-m}\exp\big[M^\mt{shift}_{\tilde I}\eta_{\tilde I}\big]\,N_0\,,
\label{extend0}
\eeq
where $\theta_{\tilde I},\,\eta_{\tilde I}\ll1$. Here, the (small) change in $K$ only involves the
(antisymmetric) rotation generators
\beq
\big[M^\mt{rot}_{k\ell}\big]_{pq}=\lp\delta_{kp}\delta_{\ell q}- \delta_{\ell p}\delta_{kq}\rp \qquad{\rm with}\qquad k<\ell
\,,\label{rotgen}
\eeq
while the (small) change in $N$ only involves the shift generators 
\beq
\big[M^\mt{shift}_{k\ell}\big]_{pq}=\delta_{kp}\delta_{\ell q}\qquad{\rm with}\qquad k<\ell
\,.\label{shiftgen}
\eeq
The rotation generators are, of course, a linear combination of the original generators given in eq.~\reef{lot1}, and hence are not orthogonal to the shift generators in the sense that
\beq
\tr\lp M^\mt{rot}_{k\ell}\ [M^\mt{shift}_{pq}]^\dagger\rp =\delta_{kp}\,\delta_{\ell q}\,.
\label{notortho}
\eeq
Of course, all of these generators are orthogonal to the diagonal generators appearing in $\tilde U_\mt{n-m}$, which will become the key point momentarily.

With our extended circuits \reef{extend0}, we now evaluate the differentials $\dd\tilde Y^{\tilde I}=\tr\big( \dd\tilde U \,\tilde U\,\tilde M^\dagger_{\tilde I} \big)$ on the normal-mode subspace, \ie at $\theta_{\tilde I}=0=\eta_{\tilde I}$,
\beq
\dd\tilde Y^{\tilde I}=\big[\widehat K_0\big]_{\tilde I\tilde J}\ \Big[\,
\dd\tilde Y^{\tilde J}_\mt{n-m}
+\tr\!\lp \tilde  M^\mt{rot}_{\tilde K}\,d\theta_{\tilde K}\,\tilde M_{\tilde J}^\dagger \rp
+\ \tr\!\lp \tilde U_\mt{n-m}\,M^\mt{shift}_{\tilde I} \,\tilde U_\mt{n-m}^{-1}\,d\eta_{\tilde I}\,\tilde M_{\tilde J}^\dagger \rp  \Big]
\,,\label{tran7}
\eeq
where $\widehat K_0$ is the orthogonal matrix given in eq.~\reef{tran5} and $\dd\tilde Y^{\tilde I}_\mt{n-m}$ are the differentials along the normal-mode directions identified in eq.~\reef{tran6}. As before, the rotation of the differentials by $\widehat K_0$ can be ignored since this transformation leaves $\delta_{IJ}$ in the metric unchanged. Next we observe that the only non-vanishing components of $\dd\tilde Y^{\tilde J}_\mt{n-m}$ are along the diagonal directions, \ie $\tilde J=(kk)$. It is then easy to show that the other two differentials are orthogonal to these. Given the explicit form of the rotation generators in eq.~\reef{rotgen}, it is clear that the second term only contributes in the off-diagonal directions, \ie $\tilde J=(k\ell)$ with $k\ne\ell$. Similarly, one can show that the same is true of the third term via eqs.~\reef{raffle4} and \reef{shiftgen},
\beq
\tilde U_\mt{n-m}\,M^\mt{shift}_{k\ell}\,\tilde U_\mt{n-m}^{-1}=e^{\tilde y_k-\tilde y_\ell}
\,M^\mt{shift}_{k\ell}\qquad
{\rm with}\qquad k<\ell\,.
\label{shiftgen2}
\eeq
The key point then is that $\dd\tilde Y^{\tilde J}_\mt{n-m}$ are orthogonal to the other two differentials in eq.~\reef{tran7}. In fact, it is straightforward to show that the full metric on the normal-mode subspace (\ie $\theta_{\tilde I}=0=\eta_{\tilde I}$) becomes
\beq
\dd s^2_\mt{n-m}=|\dd \tilde y_0|^2+|\dd \tilde y_1|^2+\cdots+|\dd \tilde y_{\ssc N-1}|^2
+\sum_{k<\ell}\left[(d\theta_{k\ell})^2 + |d\theta_{k\ell}+e^{\tilde y_k-\tilde y_\ell}\,d\eta_{k\ell}|^2
\right]\,.
\label{metric6a}
\eeq
Hence there are no off-diagonal terms in the metric, which would drive the geodesic away from the normal-mode subspace.\footnote{In general, we are asking that there are no source terms in the linearized equations for the $\theta_{\tilde I}$ and $\eta_{\tilde I}$. In turn, this means that we are asking that there are no linear terms in the cost function \reef{cost5} when expanding about the straight-line trajectories \reef{raffle2}. Here we have explicitly shown that no such linear terms arise as off-diagonal terms in the metric, involving the differentials of $\theta_{\tilde I}$ and $\eta_{\tilde I}$. In principle, we should also verify that the metric components $g_{\tilde y_k\tilde y_k}$ are not varied at linear order in the perturbations. However, our previous analysis shows that the metric on the normal mode subspace is completely independent of the coordinates parametrizing the $K$ and $N$ transformations, which ensures that this class of potential terms linear in $\theta_{\tilde I}$ and $\eta_{\tilde I}$ vanishes. Therefore the above discussion is sufficient to ensure that the geodesic equations in the full $\mathrm{GL}(N,\mathbb{R})$ geometry have no source terms which would push the straight-line trajectories away from the normal-mode subspace.} We may therefore conclude that the optimal circuit indeed takes the form of the simple straight-line circuit in eq.~\reef{raffle2}. 

Thus, for the cost function \eqref{cost5}, the complexity for our lattice of oscillators is obtained by simply summing up the circuit elements in the normal-mode basis.  Using eqs.~\eqref{raffle2} and \eqref{raffle4}, we find
\beq
\CC=\frac{1}{2}\sqrt{\sum_{k=0}^{N-1}\left(
\log\frac{\tilde\omega_k}{\omega_0}\right)^2}\,,\label{complexityN}
\eeq
where the normal-mode frequencies are given in eq.~\reef{eq:eigenfreq}. Recall that in our lattice regularization \reef{eq:Hlattice}, we had $\omega=m$ and $\Omega=1/\delta$, and so we can express the complexity \reef{complexityN} in terms of the field theory parameters via
\beq
\tom_k^2=m^2+\frac{4}{\delta^2}\,\sin^2\!\frac{\pi k}{N}\,.\label{freak1}
\eeq
Furthermore, we can replace $N=L/\delta$ where $L$ is the total length of the one-dimensional lattice of oscillators. Of course, $\omega_0$ remains the (as yet unspecified) frequency which specifies the Gaussian reference state \reef{refX}.

The entire discussion in this section is easily extended (albeit with a somewhat tedious extension of the notation) to the evaluation of the complexity of a $(d\!-\!1)$-dimensional spatial lattice of $N^{d-1}$ oscillators, and the final result is
\beq
\CC=\frac{1}{2}\sqrt{\sum_{\lbrace k_i\rbrace=0}^{N-1}\left(
\log\frac{\tilde\omega_{\vec k}}{\omega_0}\right)^2}\,,\label{complexityNd}
\eeq
where $k_i$ are the components of the momentum vector $\vec k=(k_1,k_2,\cdots,k_{d-1})$, and the normal-mode frequencies are given by
\beq
\tom_{\vec k}^2=m^2+\frac{4}{\delta^2}\,\sum_{i=1}^{d-1}\sin^2\!\frac{\pi k_i}{N}\,.\label{freakd}
\eeq
The linear size of each spatial direction here is $L=N\delta$, and so the total (spatial) volume of the system is $V=L^{d-1}=N^{d-1}\delta^{d-1}$. Hence the total number of oscillators can be expressed as
\beq
N^{d-1}=\frac{V}{\delta^{d-1}}\,,
\label{vold}
\eeq
which will prove useful below.

\subsection{Comparison with holography}\label{sec:continuum}

Eq.~\reef{complexityNd} gives our result for the complexity of the ground state of a free scalar field in $d$ spacetime dimensions. We would now like to compare this result with the analogous results arising from the proposals for holographic complexity discussed in the introduction. Of course, we must note that we are trying to compare complexities for disparate QFTs, \ie a free theory with a single degree of freedom in the present case versus a strongly coupled theory with a large number of degrees of freedom in holography. Hence there is no \textit{a priori} reason to expect that the results should agree in the two cases. Nevertheless, we will find that with certain choices, our QFT calculations share a number of qualitative features with holographic complexity. We can interpret these similarities as providing guidance towards understanding the cost function that underlies the holographic complexity conjectures.

Examining eq.~\reef{complexityNd}, we see that the expression under the square root essentially involves an integration over the spatial momenta. Our experience with QFT thus suggests that the result will be dominated by the UV modes, \ie by modes with  $\tilde\omega_{\vec k}\sim 1/\delta$. Hence as an approximation which allows us to identify the leading contribution to the complexity, we may replace all of the $\tilde\omega_{\vec k}$ with $1/\delta$ in eq.~\reef{complexityNd} to obtain\footnote{See appendix \ref{sec:approx} for more accurate estimates of the large $N$ behaviour of the complexity \reef{complexityNd}.}
\beq
\CC\approx\frac{N^{\frac{d-1}2}}{2}\ \log\!\lp\frac{1}{\delta\,\omega_0}\rp\sim\lp\frac{V}{\delta^{d-1}}\rp^{1/2}\,,
\label{croot}
\eeq
where we have used eq.~\reef{vold} to re-express the leading power of $N$ in terms of $V/\delta^{d-1}$. 

The leading UV divergence in holographic complexity for both the CA and CV proposals was studied in some detail in \cite{Carmi:2016wjl}. Hence we can compare our QFT result \reef{croot} with the analogous results for holographic complexity; denoting the latter collectively as $\CC_\mathrm{holo}$, these were found to take the form 
\beq
\CC_\mathrm{holo}\sim\frac{V}{\delta^{d-1}}~.\label{Cholo}
\eeq
Thus we see that the leading terms in the QFT and holographic complexities differ by the power of $1/2$ appearing in eq.~\reef{croot}. However, the origin of this square root is clear: it is simply the overall square root appearing in eq.~\reef{complexityNd}, which in turn arises from our use of the $F_2$ cost function in eq.~\reef{cost5}. Now, there is nothing wrong with the result in eq.~\reef{croot} per se, but it does suggest that if our QFT complexity is to emulate the leading behaviour found in holographic complexity, then we should make an alternative choice for the cost function.\footnote{An alternative approach \cite{Brown:2017jil} would be to simply assign each gate the cost $N^{\frac{d-1}2}$. However, it seems this may be problematic if, \eg we wish to compare complexities for different UV cut-offs. \label{footy99}} For example, the cost function defined by the $F_1$ measure in eq.~\reef{eq:Fmetrics}, which involves the first power of a single sum over the modes, would produce the desired behaviour. More generally, a natural family of cost functions which would reproduce the divergence in eq.~\reef{Cholo} is
\beq
\widetilde{\mathcal{D}}_\kappa=\int_0^1\dd s\sum
\,\left|Y^{\tilde I}(s)\right|^\kappa~,\label{Dalpha}
\eeq
where the natural choice would be that $\kappa$ is a positive integer, but any positive real value (with $\kappa\ge1$) will suffice for most of the following discussion. Note that we have defined the cost function here with the sum running over the normal-mode basis---we return to this point below in section \ref{base8}.

Here, we should note that only $\kappa=1$ (equivalently, the $F_1$ cost function), satisfies the condition of positive homogeneity as described in the introduction; that is, for general $\kappa>1$, doubling the amplitude of $Y^{\tilde I}$ does not double the cost. This issue can be described as saying that only the case $\kappa=1$ yields a reparametrization-invariant cost function, \ie that replacing $s\to \hat s(s)$ leaves the cost unchanged. However, we may proceed with the physics intuition that we can think of $\widetilde{\mathcal{D}}_\kappa$ as different kinds of actions describing the motion of a particle in the space of circuits.

Now it is relatively straightforward to show that in fact the straight-line circuit in eq.~\reef{raffle2} minimizes all of these cost functions. This circuit only acts with scaling gates on the various normal modes. It is clear that if the circuit were to make an excursion away from the normal-mode subspace, \ie if the path also moved in the entangling directions, this would only turn on new components of the velocity $Y^{\tilde I}$ and thereby increase the cost of the circuit. To establish that the straight-line path is favoured by the general $\kappa$ cost function, let us consider a more general trajectory in the normal-mode subspace, $\tilde U_1(s)=\exp\left[\tilde M_1(s)\right]$, where
\beq
\tilde M_1={\rm diag}\!\left({\textstyle \frac{f_0(s)}{2}\log\frac{\tom_0}{\omega_0} ,\,
\frac{f_1(s)}{2}\log\frac{\tom_1}{\omega_0},
\cdots, \frac{f_{\ssc N-1}(s)}{2}\log\frac{\tom_{\ssc N-1}}{\omega_0}} \right)\,,
\label{raffle2a}
\eeq
and each of the $f_k(s)$ is an arbitrary function satisfying
\beq
f_k(s=0)=0\qquad{\rm and}\qquad f_k(s=1)=1\,.
\label{bnd1}
\eeq
With this ansatz, the cost function \reef{Dalpha} evaluates to
\beq
\widetilde{\mathcal{D}}_\kappa(\tilde U_1)=\frac{1}{2^\kappa}\sum_k \left|\log{\textstyle \frac{\tom_k}{\omega_0}}\right|^\kappa\times
\int_0^1\dd s \left|\partial_s f_k(s)\right|^\kappa\, \,.
\label{Dalpha1}
\eeq
However, with reasoning along the lines of that in eqs.~\reef{fun1} and \reef{fun2}, one can argue that $\int_0^1\dd s \left|\partial_s f_k(s)\right|^\kappa\ge1$ (for $\kappa\ge1$), and that the inequality is saturated if and only if $\partial_s f_k(s)=1$, \ie $f_k(s)=s$.\footnote{An exception to this result arises for $\kappa=1$. In this case, the bound is saturated by any functions $f_k(s)$ satisfying $\partial_s f_k(s)\ge0$ everywhere. We return to this point in the discussion section \ref{discuss}.} That is, the general $\kappa$ cost function \reef{Dalpha} is minimized by the straight-line circuit $\tilde U_0(s) = \exp\!\left[\tilde M_0\,s\right]$. Furthermore, working with the UV approximation $\tilde\omega_{\vec k}= 1/\delta$, the leading contribution to the complexity then becomes
\beq
\CC\approx\frac{V}{\delta^{d-1}}\ 
\left| \log\!\lp\frac{1}{\omega_0\,\delta}\rp\right|^\kappa\,.
\label{calpha}
\eeq
An interesting feature of this result is that in limit $\delta\to0$, this contribution appears to diverge faster than the power law $1/\delta^{d-1}$ in the first factor.

This last observation, however, depends on the choice of $\omega_0$ which defines our reference state \reef{refX}, which we have hitherto left unspecified. Considering this choice, there seem to be a number of reasonable options. First, $\omega_0$ could be associated with some ultraviolet frequency at the lattice scale. For example, $\omega_0=e^{-\sigma}/\delta$ where $e^{-\sigma}$ provides a numerical scale that ensures $\omega_0> \tilde\omega_{\vec k}$ for all $\vec k$. In this case, the leading contribution in eq.~\reef{calpha} reduces to
\beq
\CC\approx\sigma^\kappa\,\frac{V}{\delta^{d-1}}\,.
\label{calpha2}
\eeq
With this choice, the extra logarithmic factors in the $\delta\to0$ divergence have been eliminated and we are only left with the $1/\delta^{d-1}$ factor. However, this choice also entails the interesting feature that the (subleading) infrared contributions to the complexity will involve the UV cut-off scale. That is, the full sum over momenta in eq.~\reef{Dalpha} includes summing over the infrared modes, \ie modes with $\tilde\omega_{\vec k}\sim m$. These infrared contributions will take the form
\beq
\CC_\mt{IR}\approx-  \log^\kappa(m\delta)\,.
\label{calpha3}
\eeq

An alternative choice would be to associate $\omega_0$ with some infrared scale, \ie $\omega_0\ll1/\delta$. One might choose this scale to be a physical scale in the problem, such as the mass $m$ or the volume $V$, but this would tie the reference state to the properties of the QFT.\footnote{Furthermore, if we choose $\omega_0\sim V^{-1/(d-1)}$, the complexity becomes superextensive.} This appears problematic if we wish to compare the complexities of states in different theories, \eg with different masses---see below. Hence it seems that we are instead led to choose some arbitrary IR scale to define the reference frequency, which then becomes a part of our definition of the complexity of QFT states. In this sense, the appearance of $\omega_0$ here is not very different from the appearance of the arbitrary numerical factor $\sigma$ in the complexity with the previous UV choice. Of course, if $\omega_0$ is a fixed IR frequency, the additional logarithmic factor in the complexity \reef{calpha} survives and contributes to the leading divergence in the limit $\delta\to 0$.

With the above in mind, we find surprising similarities when we compare our result with the CA proposal for holographic complexity. The leading divergence appearing in the latter \reef{ccaction} takes the form \cite{Carmi:2016wjl}
\beq
\CC_\mt{A}\sim\frac{V}{\delta^{d-1}}\ \log\!\left(\frac{L_\mt{AdS}}{\alpha\,\delta}\right)\,,
\label{leader}
\eeq
where $\delta$ is the short-distance cut-off scale in the boundary CFT, $L_\mt{AdS}$ is the AdS curvature scale of the bulk spacetime, and $\alpha$ is an arbitrary (dimensionless) coefficient which fixes the normalization of the null normals on the boundary of the WDW patch. Since $\CC_\mt{A}$ is a quantity which is to be defined in the boundary CFT, it should not depend on the bulk AdS scale. However, we can eliminate this factor with the freedom in choosing $\alpha$, \ie we set $\alpha = \omega_0 L_\mt{AdS}$ where $\omega_0$ is some arbitrary frequency. In this case,  eq.~\reef{leader} reduces to
\beq
\CC_\mt{A}\sim\frac{V}{\delta^{d-1}}\ \log\!\left(\frac{1}{\omega_0\,\delta}\right)\,. 
\label{leader2}
\eeq
While this choice eliminates the AdS scale, the holographic complexity still depends on the choice of  $\omega_0$, just as in our QFT result \reef{calpha}. Furthermore, all of the issues discussed above with respect to this choice also appear in the case of the CA conjecture. In particular, we emphasize that with the choice $\omega_0=e^{-\sigma}/\delta$, the UV cut-off appears in infrared contributions to the holographic complexity arising from joint terms deep in the bulk \cite{Carmi:2016wjl,CCMMS}. Whereas in \cite{Carmi:2016wjl} this ambiguity and the associated issues seemed problematic for holographic complexity, here can view them as a natural feature of complexity for QFTs.

We can go further in comparing our QFT results with holography. In particular, in order for the leading divergence in eq.~\reef{calpha} to match the holographic result \reef{leader2} more closely, we should choose $\kappa=1$ in eq.~\reef{Dalpha}, \ie the $F_1$ cost function. Of course, this reasoning only applies when the reference frequency is chosen in the IR and the logarithmic factor modifies the form of the leading divergence. In the case where $\omega_0$ is set by the cut-off scale, the leading divergence is a simple power law and the exponent $\kappa$ only modifies the overall numerical pre-factor, which we have not specified here. However, $\kappa$ also appears in the IR contribution in eq.~\reef{calpha3}. Again the analogous contributions in holography would be linear in $\log(\delta)$ because of the form of the corresponding boundary terms in the gravitational action \cite{Lehner:2016vdi}. Hence this reasoning again favours the choice $\kappa=1$. That said, given the aforementioned disparity between the field theories that we are comparing, it is not clear how much weight to give this observation. 

One can also look at the form of subleading corrections to the leading divergence. As discussed in appendix \ref{sec:approx},  the first subleading correction comes from the mass and has the form ${V\,m^2}/{\delta^{d-3}}$. This form could be anticipated by simple dimensional analysis, and analogous results can be found in holographic complexity as well \cite{pratik}. Specifically, if the boundary CFT is perturbed by a relevant operator of dimension $\Delta$, the corresponding coupling $\lambda$ will have dimension $d-\Delta$, and the first subleading correction to the holographic complexity then takes the form $V\,\lambda^2/\delta^{2\Delta-(d+1)}$. For the CV conjecture, these calculations follow in close parallel with the analogous calculations of corrections to holographic entanglement entropy induced by relevant operators \cite{Hung:2011ta}. However, we expect that analogous results will appear for the CA conjecture. Such corrections to holographic complexity were considered in \cite{Carmi:2016wjl} as arising from placing the boundary theory in a curved spacetime or from evaluating the state of a curved time slice. It may be interesting to understand how to extend our present QFT calculations of complexity to incorporate such situations. 

\newpage

\section{Penalty factors}\label{sec:penalty}

In eq.~\reef{cost2}, the cost function was written with a general metric $G_{IJ}$, which allows us the freedom to include penalty factors to weight certain directions or classes of gates more heavily than others. This is particularly relevant for the lattice of oscillators representing the regulated scalar field. In the previous section, our circuits implicitly included entangling gates $Q_{ab}$, which coupled points on the lattice which were arbitrarily far apart. However, if we want complexity to be a physical attribute of a QFT, then we would expect it to reflect the notion of locality. That is, gates which couple far-separated points should be more expensive -- \ie incur a higher cost in the geometric distance function -- than those which couple nearest neighbours. 

To gain some experience with this idea, we return to the problem of two coupled oscillators and we introduce a penalty factor weighting the entangling gates, which act on two oscillators (or sites), more heavily than the scaling gates, which act on a single oscillator (or site). Specifically, we may penalize the ``off-diagonal'' directions by choosing
\beq
G_{IJ}=\mathrm{diag}(1,\pen^2,\pen^2,1)~,
\label{pen99}
\eeq
with $\pen>1$.  As a result, our original metric \reef{metric1} is replaced by the following more complicated metric:
\beqa
\dd s^2&=&\,G_{IJ} \ \tr\lp \dd U(s)\,U^{-1}(s)M^T_I\rp\,\tr\lp \dd U(s)\,U^{-1}(s)M^T_J\rp
\nonumber\\
&=&\,2\,\dd y^2+2\left[\pen^2-(\pen^2-1)\sin^2(\theta+\tau)\right]\dd\rho^2+(\pen^2-1)\sin\lp2(\theta+\tau)\rp\sinh(2\rho)\,\dd\rho\,\left(\dd\tau-\dd\theta\right)
\nonumber\\
&&+2\left[\pen^2\cosh(2\rho)-(\pen^2-1)\cos^2(\theta+\tau)
\sinh^2\rho\right]\cosh^2\!\rho\,\dd\tau^2
\label{penalty1}\\
&&+2\left[\pen^2\cosh(2\rho)-(\pen^2-1)\cos^2(\theta+\tau)
\cosh^2\rho\right]\sinh^2\!\rho\,\dd\theta^2\nonumber\\
&&-\left[2\pen^2-(\pen^2-1)\cos^2(\theta+\tau)\right]\sinh^2(2\rho)\,\dd\tau\,\dd\theta~.
\nonumber
\eeqa
Of course, this geometry reduces to eq.~\eqref{metric1} upon setting $\pen=1$. The metric has a slightly simpler expression in terms of the pseudo-lightcone coordinates \eqref{lcCoords}, where $\theta=x+z$, $\tau=x-z$:
\beqa
\dd s^2&=&2\dd y^2+2\left[\pen^2-\lp \pen^2-1\rp\sin^2\!\lp2x\rp\right]\dd\rho^2-2 \left(\pen^2-1\right)\sin (4 x)
 \sinh (2 \rho ) \dd\rho\dd z
 \label{penalty2}\\
 &&+\,2 \pen^2\dd x^2+2\left[\pen^2\cosh(4\rho)-(\pen^2-1)\cos^2(2x)\sinh^2\!\lp2\rho\rp\right]\dd z^2-4 \pen^2\cosh (2 \rho )\,\dd x\dd z~,
\nonumber
\eeqa
which reduces to eq.~\reef{metric2} when $\pen=1$. Although these coordinates somewhat obscure our physical intuition for the geometry, they are computationally much simpler. Therefore we will work with the metric in the form \reef{penalty2} for most of the following. 

As in the unpenalized case, this metric enjoys the four Killing vectors $(\hat k_I)^i$ given in eq.~\eqref{Killed2}.\footnote{Again, these arise from the right-invariance of the expression in the first line of eq.~\reef{penalty1}. The fifth ``accidental'' Killing vector $\partial_x$ in eq.~\reef{Killed5} no longer gives rise to a symmetry for the penalized metric, as is clear from eq.~\reef{penalty2}.} For the metric \eqref{penalty2}, the associated conserved quantities $\hat c_I=(\hat k_I)^ig_{ij}\dot x^i$ are

\beq
\bal
\hat c_1\equiv&\ 2\dot y~,\\
\hat c_2\equiv&  -\dot x\left[2 \pen^2 \sinh (2 \rho )  \cos (2 z)\right]\\
&+\dot z \Big[\cos (2 z) \sinh (4 \rho )  \left(2\pen^2-\left(\pen^2-1\right) \cos^2 (2 x)\right) -\left(\pen^2-1\right)\sin (2 z) \sinh (2 \rho ) \sin (4 x) \Big]\\
&+ \dot\rho \Big[2\sin (2 z) \left(\pen^2-\left(\pen^2-1\right) \sin^2 (2 x)\right)-\left(\pen^2-1\right) \cos (2 z)\cosh (2 \rho ) \sin (4 x) \Big]~,\\
\hat c_3\equiv&\ \dot x\left[2 \pen^2 \sinh (2 \rho )  \sin (2 z)\right]\\
&- \dot z \Big[ \sin (2 z) \sinh (4 \rho )  \left(2\pen^2-\left(\pen^2-1\right) \cos^2 (2 x)\right)+\left(\pen^2-1\right) \cos (2 z) \sinh (2 \rho ) \sin (4 x) \Big]\\
&+ \dot\rho \Big[2\cos (2 z) \left(\pen^2-\left(\pen^2-1\right) \sin^2 (2 x)\right)+\left(\pen^2-1\right) \sin (2 z) \cosh (2 \rho ) \sin (4 x) \Big]~,\\
\hat c_4\equiv&\  -2 \pen^2 \cosh (2 \rho ) \,\dot x
+2 \left(\pen^2 \cosh (4 \rho )-\left(\pen^2-1\right) \cos^2 (2 x)\sinh ^2(2 \rho ) \right)\,\dot z\\
&\ - \left(\pen^2-1\right)  \sinh (2 \rho ) \sin (4 x)\,\dot\rho~.\\
\eal\label{revel0}
\eeq
One can check that these quantities indeed reduce to those given in eq.~\eqref{eq:conserved} when $\pen=1$. Solving the first equation for $y$ is trivial, and we simply recover eq.~\eqref{yrun}, \ie $y=y_1\,s$. The next three equations may be solved for $\dot\rho$, $\dot x$, and $\dot z$:
\beq
\bal
\dot\rho=&
\frac{1}{4 \pen^2}\bigg\{\left(\pen^2-1\right) \cosh (2 \rho ) \sin (4 x) (\hat c_2 \cos (2 z)-\hat c_3 \sin (2 z))\\
&+2\left(\pen^2-\lp \pen^2-1\rp\cos^2 (2 x)\right) (\hat c_2 \sin (2 z)+\hat c_3 \cos (2 z))-\left(\pen^2-1\right) \hat c_4 \sinh (2 \rho ) \sin (4 x)\bigg\}\\
\dot x=&\frac{1}{4 \pen^2}\bigg\{\frac{1}{\sinh (2 \rho )}\bigg[2\lp\cosh(4\rho)+\lp \pen^2-1\rp\cos^2\!(2x)\cosh^2\!\lp2\rho\rp\rp\lp\hat c_2\cos(2z)-\hat c_3\sin(2 z)\rp\\
&+\lp \pen^2-1\rp\sin(4x)\cosh\lp2\rho\rp\lp\hat c_2\sin(2z)+\hat c_3\cos(2z)\rp\bigg]
-2\left(1+\left(\pen^2-1\right) \cos^2 (2 x)\right)\cosh\!\lp2\rho\rp\hat c_4\bigg\}\\
\dot z=&\frac{1}{4 \pen^2}\bigg\{\frac{1}{\sinh(2\rho)}\bigg[2\left(1+\left(\pen^2-1\right) \cos^2 (2 x)\right)\cosh\lp2\rho\rp (\hat c_2 \cos (2 z)-\hat c_3 \sin (2 z))\\
&+\left(\pen^2-1\right) \sin (4 x) (\hat c_2 \sin (2 z)+\hat c_3 \cos (2 z))\bigg]
-2\left(1+\left(\pen^2-1\right) \cos^2 (2 x)\right)\hat c_4\bigg\}
\eal
\label{revel1}
\eeq

Now recall that in the unpenalized case, the expression for $\dot\theta$ in eq.~\reef{trun} diverged at the origin $\rho(s=0)=0$ unless the conserved quantities were properly tuned. This divergence is simply the usual angular momentum barrier at the origin, and the tuning amounts to setting the angular momentum to zero. The same issue arises here, as reflected in the fact that both $\dot x$ and $\dot z$ have the same pole structure as $\rho\to0$, \ie in this limit, $\dot \theta=\dot x+\dot z$ diverges but $\dot \tau=\dot x-\dot z$ does not. Taking the limit $\rho\rightarrow0$, and setting $x_0=z_0$ (since $\tau(s=0)=0$) in eq.~\reef{revel1}, one finds that this divergence is avoided by choosing
\beq
\hat c_2=\frac{\hat c_3}{\pen^2}\tan\lp2z_0\rp~.
\label{lov2}
\eeq
Substituting eq.~\reef{lov2} back into the expressions for derivatives thus renders them well-behaved at the origin, as required by the initial condition $\rho(s=0)=0$, but since their forms are not appreciably simpler we shall not write them out here.

Now with the metric \reef{penalty2}, the normalization of the tangent vector $k^2=g_{ij}\,\dot x^i\dot x^j$ becomes
\beq
\bal
k^2=&\ 2 y_1^2+2\left[\pen^2-\lp \pen^2-1\rp\sin^2\!\lp2x\rp\right]\dot\rho^2-2 \left(\pen^2-1\right)\sin (4 x)
 \sinh (2 \rho ) \dot\rho\,\dot z\\
&+\,2 \pen^2\dot x^2 +2\left[\pen^2\cosh(4\rho)-(\pen^2-1)\cos^2(2x)\sinh^2\!\lp2\rho\rp\right]\dot z^2-4 \pen^2\cosh (2 \rho )\,\dot x\,\dot z~.
\eal\label{eq:kPenFull}
\eeq
In principle, one can substitute the expressions in eq.~\reef{revel1} for $\dot\rho$, $\dot x$, and $\dot z$  into this expression to obtain an explicit formula for the geodesic length, with no derivatives. Unfortunately, the resulting expression appears quite intractable, and a general solution remains beyond our reach. 

However, we are ultimately only interested in the optimal trajectory. Given our experience with the unpenalized metric, one might reasonably conjecture that the global minimum is again obtained with the simple straight-line circuit \reef{solver5}, and as we now verify, this trajectory remains a geodesic in the penalized geometry \reef{penalty2}. Recall that the first constraint in eq.~\reef{revel0} yielded the desired behaviour for $y$, \ie $y(s)=y_1\,s$, as in eq.~\reef{yrun}. The straight-line solution also had $\tau$ and $\theta$ fixed with $\tau(s)=0$ and $\theta(s)=\pi$. This then implies that $x$ and $z$ are fixed with $x(s)=\pi/2$ and $z(s)=\pi/2$. Combining the latter with eq.~\reef{lov2} then yields $\hat c_2=0$. Substituting these values of $x$ and $z$ into the last two expressions in eq.~\reef{revel1}, we obtain
\beq
\dot x=-\frac{\hat c_4}2\,\cosh(2\rho)\,\qquad\mathrm{and}\qquad
\dot z=-\frac{\hat c_4}2\,.
\label{rats5}
\eeq
Hence consistency with the condition $\dot x=\dot z=0$ demands that we set
$\hat c_4=0$. Finally, the $\dot \rho$ equation yields
\beq
\dot\rho=-\frac{\hat c_3}{2\pen^2}\quad\implies\quad
\hat c_3=-2\pen^2\,\rho_1\,,
\label{rats4}
\eeq
and we arrive at the desired solution: $\rho(s)=\rho_1\,s$. 

Having shown that this simple trajectory remains a geodesic in the penalized geometry \reef{penalty2}, we substitute this geodesic into eq.~\eqref{eq:kPenFull} to obtain
\beq
k^2=2\lp y_1^2+\pen^2\rho_1^2\rp\equiv k_0^2~,\label{eq:straight}
\eeq
where we have introduced the label $k_0$ to denote the geodesic length of the straight-line circuit to avoid confusion with other lengths considered below. Note that eq.~\eqref{eq:straight} is the natural generalization of eq.~\eqref{solver4b} to the case with $\pen>1$. 

However, it turns out that this is \emph{not} the minimum geodesic for the penalized metric: shorter trajectories can be found. In particular, examining the geometry \reef{penalty2} more closely, we see that 
\beq
g_{\rho\rho}=2\left[\pen^2-\lp \pen^2-1\rp\sin^2\!\lp2x\rp\right]
\label{rats6}
\eeq
depends on the $x$ coordinate. This contrasts with the unpenalized metric \reef{metric1} (or the lattice metric \reef{metric6a}) where we found that the geometry of the normal-mode subspace (\ie the metric for the $y$ and $\rho$ directions) was independent of the other coordinates. The latter property was essential to showing that the straight-line circuit was indeed the optimal trajectory. 

Examining eq.~\reef{rats6}, it is clear that there should be short-cuts for the motion along $\rho$ if we move away from the normal-mode subspace, \ie away from $x=\pi/2$.\footnote{As an amusing observation, let us add that it is also clear that there are no such short-cuts if $\pen<1$. That is, if we weight the scaling gates more heavily than the entangling gates, then the straight-line circuit \reef{solver5} will remain the optimal circuit.} For example, we might consider the following simple path consisting of two segments:
\beqa
a)\ \ 0\le\bar s\le 1\ :&&y=0\,,\ \rho=\rho_1\,\bar s\,,\ x=\pi/4\,,\ z=\pi/4\,;
\label{segments}\\
b)\ \ 1\le\bar s\le 2\ :&&y=y_1(\bar s-1)\,,\ \rho=\rho_1\,,\ x={\textstyle \frac{\pi}4}\,\bar s,\ z=\pi/4\,,
\nonumber
\eeqa
where $\bar s$ provides some arbitrary parametrization of the path. This segmented path is not a geodesic, but does connect the initial point at the origin to the desired end-point at $y=y_1$, $\rho=\rho_1$ and $x=\pi/2$. The first segment moves only in the $\rho$ direction at the optimal value of $x$, and then the second segment moves uniformly in both the $x$ and $y$ directions to arrive at the required end-point. The total length of this path is
\beq
k_s=\sqrt{2}\rho_1+\sqrt{2\lp\frac{\pi}{4}\rp^2\pen^2+2y_1^2}~,\label{eq:ks}
\eeq
where we use the subscript $s$ to denote ``segmented'', in contrast with $k_0$ above. Of course, the relation between $k_0$ and $k_s$ now depends on the details of the various parameters $y_1$, $\rho_1$, and $\pen$. It is natural  that all three of these coefficients are large, in which case one generally finds $k_0>k_s$. However, to simplify the analysis and illustrate this result, let us consider the regime where the penalty factor is the largest constant, \ie $\pen\gg\rho_1,y_1$ and $\rho_1,y_1\gg1$. Then we may approximate the two lengths with
\beqa
k_s&\simeq&\frac{\pi}{2\sqrt{2}}\,\pen+\sqrt{2}\rho_1+ \frac{2\sqrt{2}}{\pi}\,\frac{y_1^2}{\pen^2} +\cdots\,,
\label{compar2}\\
k_0&\simeq&\sqrt{2}\pen\rho_1+ \frac{y_1^2}{\sqrt{2}\pen\rho_1} +\cdots\,,\qquad\implies\quad k_0/ k_s\simeq \frac{4}\pi\,\rho_1\gg1\,.
\nonumber
\eeqa
Thus in the penalized geometry \reef{penalty2}, the length of the segmented path is much shorter than the straight-line geodesic in this regime. Again, the segmented path is not a geodesic and so cannot describe the optimal path, but we shall find that it gives a remarkably good approximation to the optimal geodesic. 

We would like to find the optimal geodesic but as noted below eq.~\reef{eq:kPenFull}, obtaining the general solution seems out of reach. However we can make progress with a simplifying assumption: if we examine the penalized metric \reef{penalty2}, we see that as the radius $\rho$ increases, the most rapidly growing component of the metric is $g_{zz}\sim \pen^2 e^{4\rho}$ (for generic $x$). This suggests that motion in the $z$ direction will quickly be suppressed as the geodesics move out from the origin. Therefore we simplify our problem by considering trajectories confined to a constant-$z$ submanifold, for which the relevant metric is given by
\beq
\dd s^2=2\dd y^2+2\left[\pen^2-\lp \pen^2-1\rp\sin^2\!\lp2x\rp\right]\dd\rho^2+2 \pen^2\dd x^2~.\label{eq:metricSimple}
\eeq
We will return to justify our assumption of no (or little) motion in the $z$ direction below. Obviously, eq.~\eqref{eq:metricSimple} is a much simpler geometry, and the analysis of the geodesics becomes much more tractable. We leave the details of solving for the resulting geodesic to appendix \ref{sec:subspace} and only refer to certain key results in the comparison below. Our expectation is that the new geodesic is the optimal trajectory, at least for large $\pen$, but we must add that we have not provided an irrefutable proof of this result. Furthermore, using the results of appendix \ref{sec:subspace}, we also explicitly show below that the segmented path \reef{segments} provides a good approximation of this optimal geodesic in the regime where the penalty factor is large.

Our analysis in appendix \ref{sec:subspace} suggests we use the following quantity to more conveniently compare the lengths of the various paths:
\beq
\bar k^2\equiv \frac{k^2}2-y_1^2\ .
\label{newer}
\eeq
For the straight-line geodesic, we have simply
\beq
 \bar k_0=\pen\,\rho_1\ ,
\label{kbar0}
\eeq
while for the segmented path, we have
\beqa
\bar k_s&=&\frac{\pi \pen}{4}\left[1+\frac{8\rho_1}{\pi\pen}\lp\frac{2\rho_1}{\pi\pen}+\sqrt{1+\lp\frac{4y_1}{\pi\pen}\rp^2}\rp\right]^{\!\frac12}
\nonumber\\
&\approx&\frac{\pi}{4}\pen+\rho_1+\frac{8}{\pi^2}\,\frac{y_1^2}{\pen^2}\,\rho_1+\ldots~,
\label{kbars}
\eeqa
where as above, the expansion in the second line assumes $\pen\gg\rho_1,y_1$. Now for the optimal geodesic, we have from eq.~\reef{eq:kbarCoords}
\beqa
\bar k&\approx&\frac{\pen}{2}\tan^{-1}\!\sqrt{\pen^2-1}+\rho_1
\nonumber\\
&\approx&\frac{\pi}{4}\pen+\rho_1-\frac{1}{2}-\frac{1}{12\pen^2}+\op\!\lp\frac1{\pen^{4}}\rp~.
\label{kbaropt}
\eeqa
Comparing these results, we see that in this large $\pen$ regime, $\bar k$ for the optimal geodesic is much smaller than $\bar k_0$ for the straight-line geodesic, and extremely close to $\bar k$ for the segmented path. Furthermore, we note that both $\bar k_0$ and $\bar k$ are completely independent of $y_1$, while it only appears in $\bar k_s$ at order  $y_1^2/\pen^2$.

\begin{figure}[h!]
\centering
\includegraphics[width=0.45\textwidth]{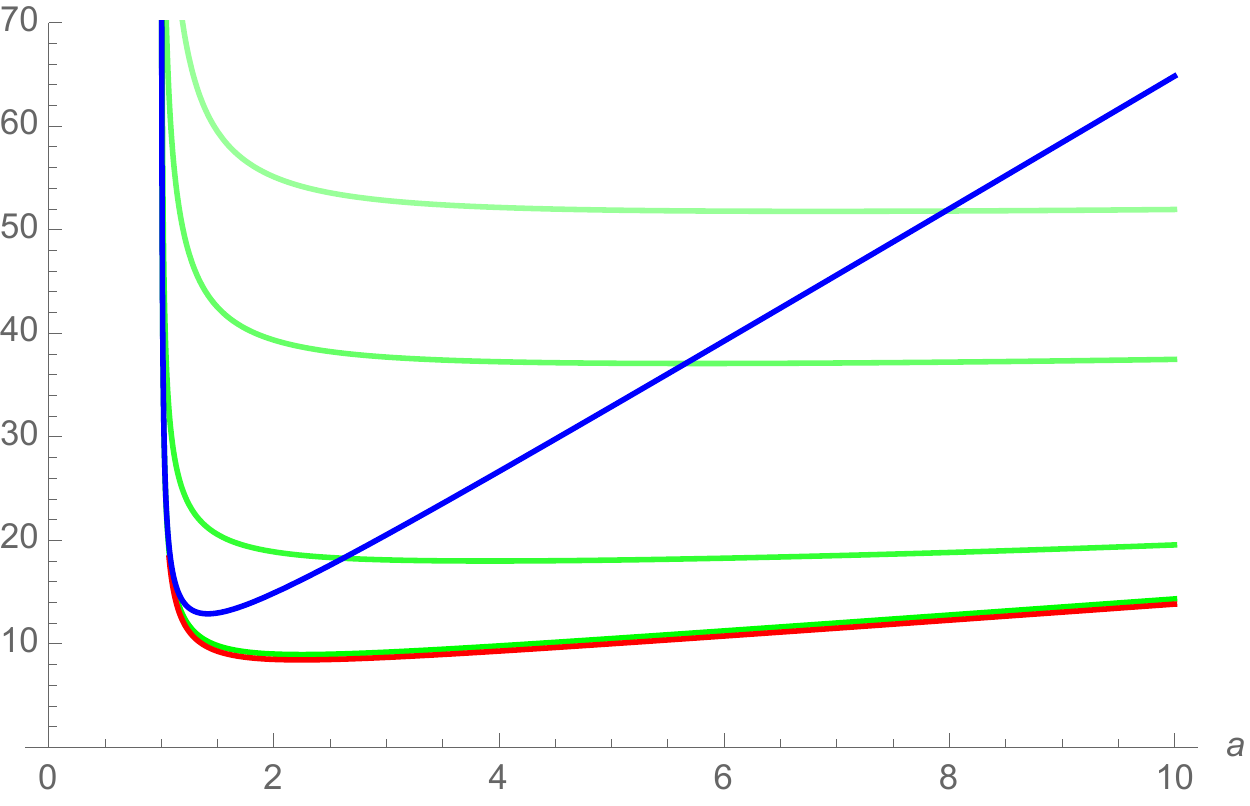}
\includegraphics[width=0.45\textwidth]{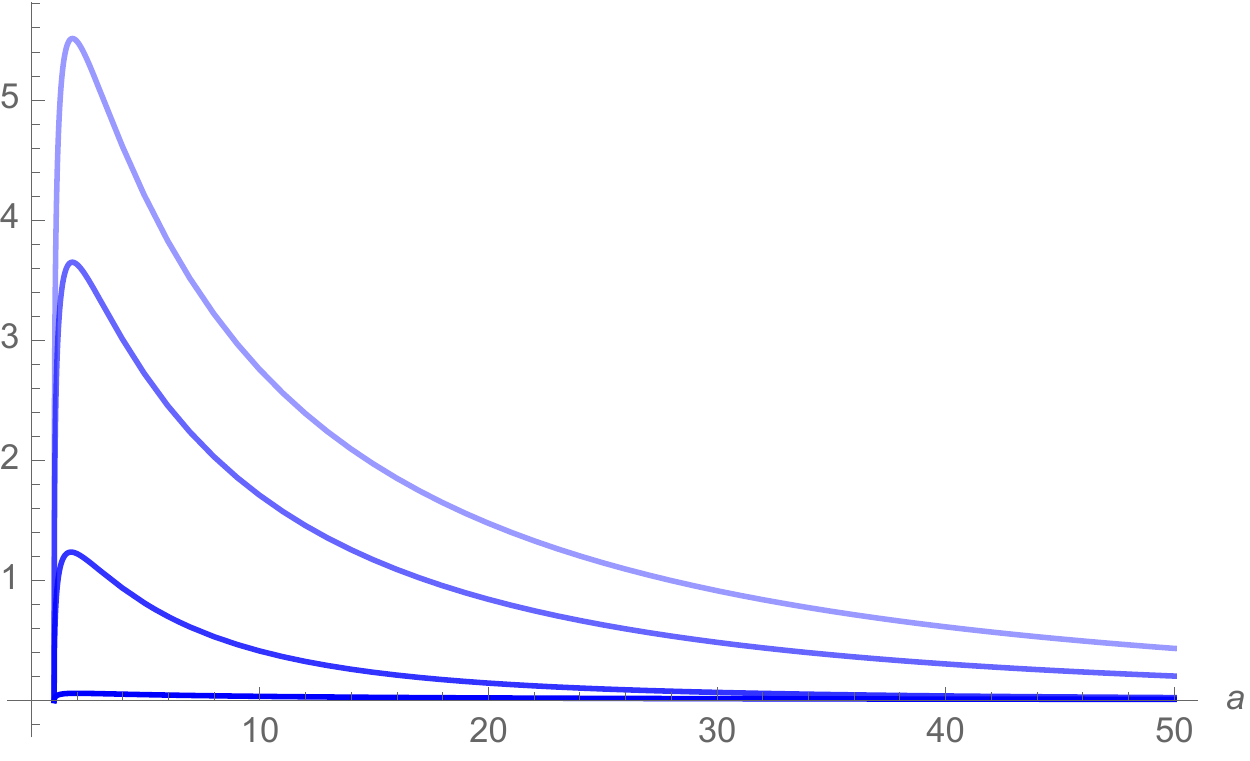}
\caption{\textbf{(Left:)} Plot of $\hat k$ (red), $\hat k_0$ (blue), and $\hat k_s$ (green, $y_1\in\{0,20,100,200\}$) as given in eqs.~\eqref{eq:vs} and \eqref{eq:kStraight} as functions of the penalty factor $\pen$, with $\eps=10^{-10}$. Clearly, $\bar k$ represents a shorter geodesic than the straight-line circuit with $\bar k_0$. This again indicates the existence of short-cuts outside of the normal-mode subspace in the penalized geometry. For $\bar k_s$, the opacity reflects the value of $y_1$, with $y_1=0$ the lowest/darkest curve running parallel to $\hat k$ and $y_1=200$ the highest/faintest. \textbf{(Right:)} Plot of $\lp\bar k_s-\bar k\rp/\bar k$ as a function of $\pen$ for $\eps=10^{-10}$, where the shading runs through the same range of $y_1$ as in the left plot (from $y_1=0$ at the bottom to $y_1=200$ at the top). Though it is not clear at this scale, the lower-most curve, $y_1=0$, follows the same basic shape as the others, peaking at $\pen\sim0.1$ and then slowly approaching to zero as $a\rightarrow\infty$. The curves never become negative, indicating that $\bar k_s>\bar k$ for all $\pen>1$.\label{fig:kCompare}}
\end{figure}

In figure~\ref{fig:kCompare}, we plot $\bar k$, $\bar k_0$, and $\bar k_s$ as functions of $\pen$ for fixed values of the variable $\eps$ (see eq.~\reef{eq:kEps}), as well as for various values of $y_1$ in the case of $\bar k_s$. Recall the definition \reef{newer} which shows that these quantities are giving us direct information about the length of the corresponding paths. Hence one clearly sees in the figure that the new optimal geodesic is shorter than the straight-line circuit for all values of $\pen$.
In the right panel, we also see that the length of the segmented path quickly approaches the length of the optimal geodesic for large values of the penalty factor, in agreement with eqs.~\reef{kbars} and \reef{kbaropt}. In fact, if we take the difference of these two equations in the large $\pen$ limit, we find
\beq
\bar k_s-\bar k\simeq\frac{1}{2}~.\label{eq:overhead}
\eeq
\vskip -2ex

\begin{figure}[h!]
\centering
\includegraphics[width=0.45\textwidth]{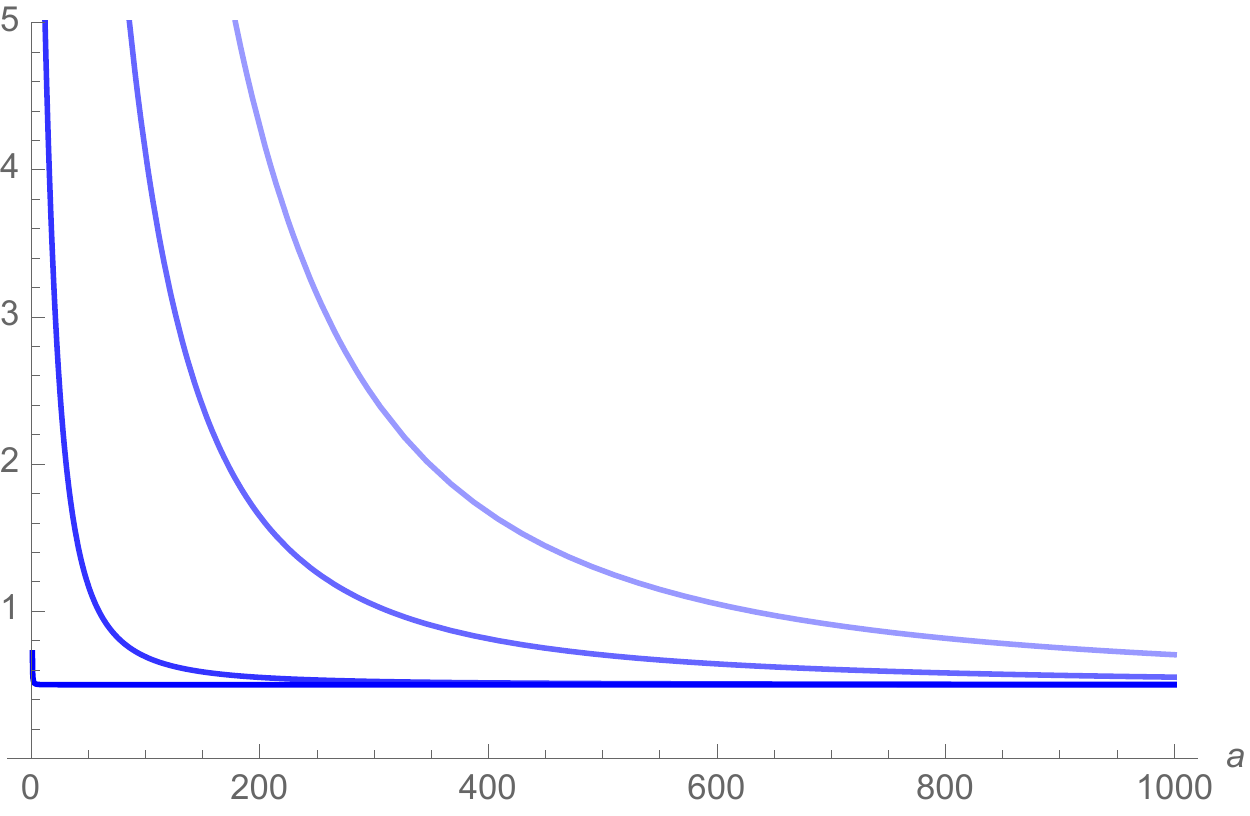}
\caption{Plot of $\bar k_s-\bar k$ as a function of $\pen$ for $\eps=10^{-10}$, where the shading runs through the same range of $y_1$ as in figure \ref{fig:kCompare} (from $y_1=0$ at the bottom to $y_1=200$ at the top). In all cases, the difference eventually approaches $1/2$, consistent with eq.~\eqref{eq:overhead}. For example, when $\pen=1000$, we have $\bar k_s\simeq791.848$ and $792.052$ for $y_1=0$ and $200$, respectively, while $\hat k\simeq 791.348$.\label{fig:kCompare2}}
\end{figure}

\noindent Given that in this limit, both $\bar k$ and $\bar k_s$ are diverging, it is impressive to find the simple $\op(1)$ difference shown above. Figure \ref{fig:kCompare2} examines this difference in more detail numerically.

\newpage

\begin{figure}[h!]
\centering
\includegraphics[width=0.48\textwidth]{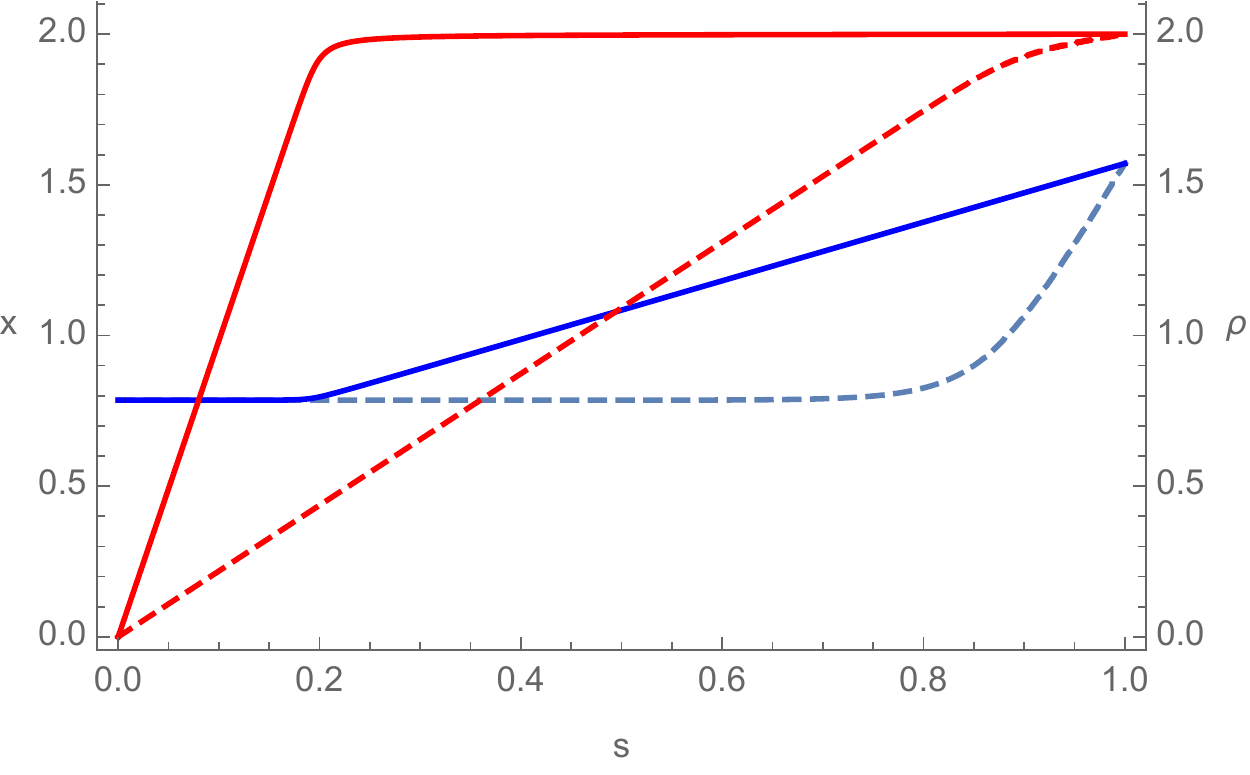}
\;
\includegraphics[width=0.48\textwidth]{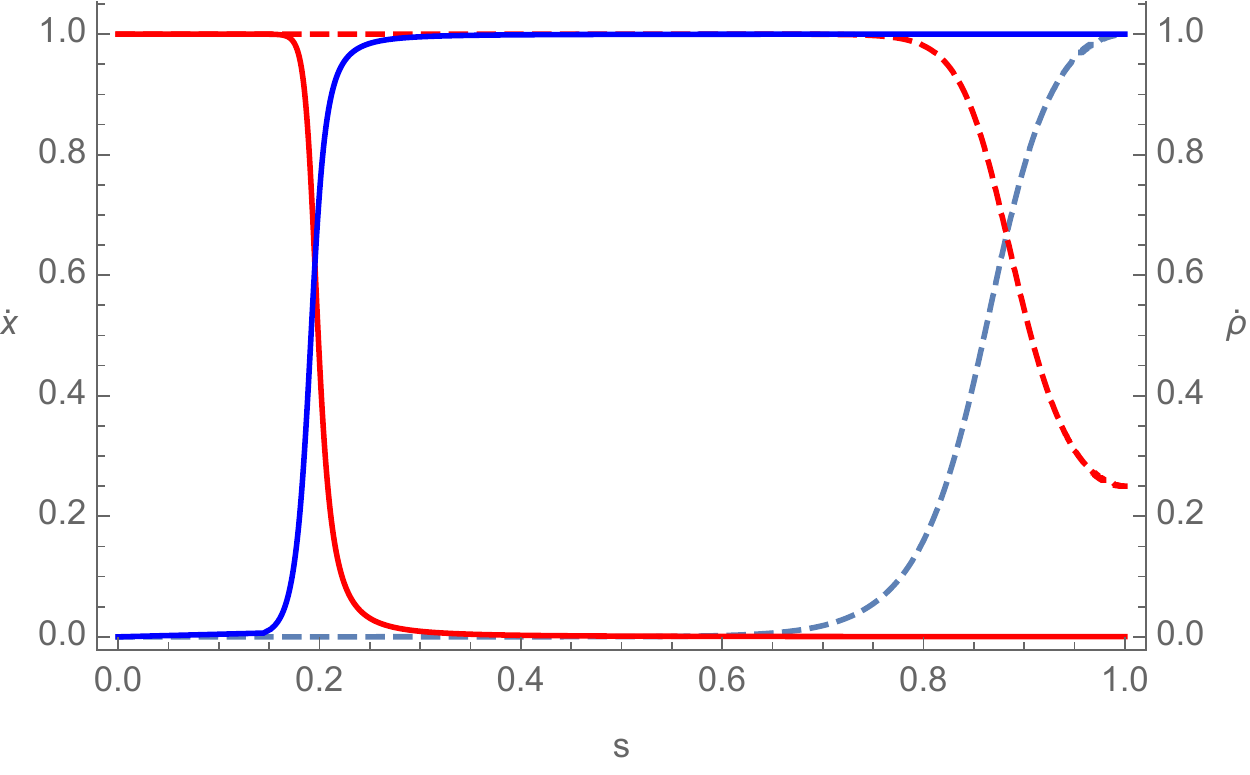}
\caption{ (\textbf{Left}) Plot of $x(s)$ (blue) and $\rho(s)$ (red) for the parameter values set by \eqref{eq:kEps} and \eqref{eq:params}, with $\eps=10^{-16}$. (\textbf{Right}) Plot of $\dot x(s)$ (blue) and $\dot\rho(s)$ (red) for the same. In both plots, the dashed curves correspond to $\pen=2$, while for the solid curves we have set $\pen=50$. One sees that the amount of ``time'' for which the circuit remains on the constant $x=\pi/4$ segment is inversely proportional to the strength of the penalty factor. For illustrative purposes, we have normalized $\rho|_{s=1}$ to 2 in the left plot, and normalized both $\dot x|_{s=1}$ and $\dot\rho|_{s=1}$ to 1 in the right.\label{fig:funcS}}
\end{figure}

Of course, the above results support the conjecture that the new geodesic represents the optimal geodesic and hence yields the shortest possible distance between the origin and the end-point. Since the segmented path \reef{segments} is not itself a geodesic, it must have a longer length. However, the impressive agreement in eq.~\reef{eq:overhead} seems to indicate that this path is coming very close to the optimal geodesic. We can confirm this very clearly by examining $x(s)$ and $\rho(s)$ numerically. As shown in figure \ref{fig:funcS}, the geodesic essentially has two phases: the first in which $\rho$ increases uniformly with fixed $x=\pi/4$ and the second in which $\dot\rho\to0$ and $x$ increases uniformly from $\pi/4$ to $\pi/2$. As shown in the figure, these two distinct phases are separated by an abrupt but smooth transition, which becomes particularly obvious for larger $\pen$. We might note that the growth in $y$ is uniform throughout the entire span $0\le s\le 1$. However, for large $\pen$, the transition occurs for small $s$ (see further comments below) and so $y$ is growing primarily in the second phase where $x$ increases. Hence we can see very explicitly that the behaviour of the optimal geodesic is indeed very similar to that of the segmented path \reef{segments} for large $\pen$. 

\begin{figure}[h!]
\centering
\includegraphics[width=0.49\textwidth]{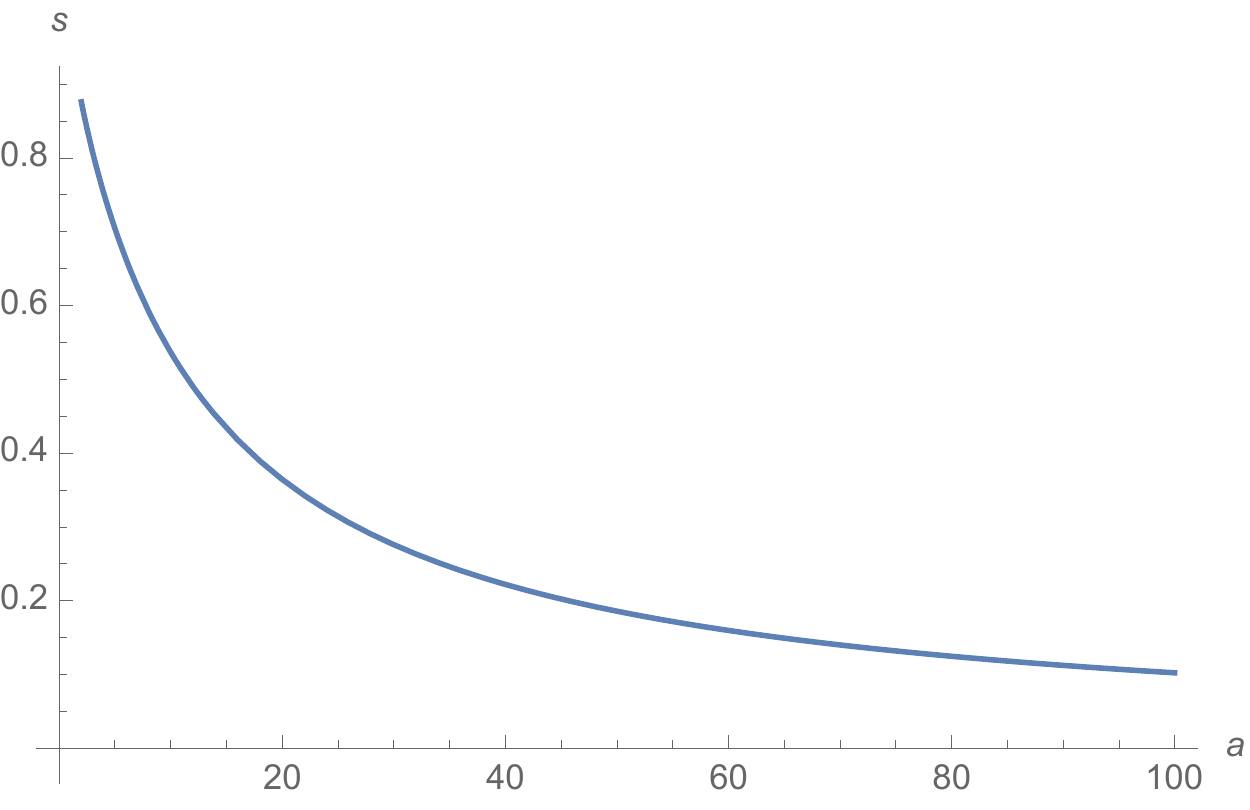}
\caption{Plot of $s_\mt{trans}(\pen)$ in eq.~\eqref{eq:sa}, for $\eps=10^{-15}$. \label{fig:switchover}}
\end{figure}

Let us examine the behaviour of the transition point in more detail. For computational purposes, we define this as the value of $s=s_\mt{trans}$ at which the two (normalized) curves for $\dot x$ and $\dot\rho$ cross in figure \ref{fig:funcS}, \ie $\dot x(s)/\dot x|_\mt{max} =\dot\rho(s)/\dot\rho|_\mt{max}$ at $s=s_\mt{trans}$. In the limit $\eps\ll1$, this point can be well-approximated by\footnote{This expression is obtained by equating the normalized quantities $\dot x/\dot x|_\mathrm{max}$ and $\dot \rho/\dot\rho|_\mathrm{max}$ (from eqs.~\eqref{revel3} and \eqref{revel2}, respectively) to solve for the critical point $x_\mathrm{crit}$ at which the curves in the right plot in figure \ref{fig:funcS} cross. Upon substituting this into \eqref{eq:sX} for $s(x)$, and using $\eqref{eq:vs}$ for $\bar k$ and \eqref{eq:params} for $\bar c_2$, one obtains an expression that depends only on $\pen$ and $\eps$, \ie $s\lp\pen,\eps\rp$, in which we then take $\eps\ll1$.}
\beq
s_\mt{trans}\lp\pen\rp\approx1-\frac{\Pi\!\lp1-\pen^2,\,h(\pen),\,1-\eps\rp}{\Pi\!\lp1-\pen^2,\,1-\eps\rp}~,\label{eq:sa}
\eeq
where
\beq
h(\pen)\equiv\csc^{-1}\left[\frac{\lp\pen^2-1\rp\lp\pen+\sqrt{5\pen^2-4}\rp}{2\pen^3-\pen-\sqrt{5\pen^2-4}}\right]^{\!\frac12}~.
\eeq
We plot $s_\mt{trans}(\pen)$ in figure \ref{fig:switchover}. In conjunction with eq.~\eqref{eq:suppressed}, one sees that a large penalty factor strongly suppresses the duration of the first phase, and thus the circuit spends most of its ``time'' -- in terms of some fixed total affine parameter -- on the second phase. Additionally, one sees that the switchover point appears to go to zero as $\pen \rightarrow \infty$. We can verify this by first approximating $h(\pen)$ for large $\pen$ as
\beq
h(\pen)\approx\csc^{-1}\left[\frac{2}{\sqrt{5}-1}\right]^{\!\frac12}~,
\eeq
and then expanding eq.~\eqref{eq:sa} in the limit $\pen\rightarrow\infty$, $\eps\rightarrow0$:
\beq
s_\mt{trans}(\pen)\approx\frac{1}{\pen}\left[\frac{1}{\pi}\log\lp\frac{1}{\eps}\rp+\op\!\lp\eps^0\rp\right]+\op\!\lp\frac{1}{\pen^2}\rp~,
\eeq
where $\op\!\lp\eps^0\rp\approx1.11502$. Comparing this expression with eq.~\eqref{hous2x} for large $\pen$, we see that the leading-order term in $s_\mt{trans}(\pen)$ above may be written as
\beq
s(\pen)\approx\frac{4\,\rho_1}{\pi\,\pen}~.
\eeq
This result has an intuitive explanation: $\rho_1$ sets the radial distance that must be covered in the first phase (which is large in the $\eps\to0$ limit while the ``angular'' change in $x$ remains fixed), while as explained above, a large penalty factor $\pen$ compels the circuit to complete this motion as quickly as possible.

To close this discussion of the penalized geometry, let us reiterate that we have argued that the geodesic confined to the constant-$z$ subspace \reef{eq:metricSimple} is the optimal geodesic, and hence that its length gives the complexity of the state. That is, when we penalize the entangling gates with eq.~\reef{pen99}, the complexity of the ground state becomes
\beq
\CC=\sqrt{2}\left[\frac\pi4\,\pen+\rho_1-\frac12+\frac{2\,y_1^2}{\pi\,\pen} +\op{\lp\frac1{\pen^2}\rp}\right]\,.
\label{loft}
\eeq
in the regime $\pen\gg\rho_1,y_1$.

\subsection{New optimal circuit}

\begin{figure}[h!]
\centering
\includegraphics[width=0.44\textwidth]{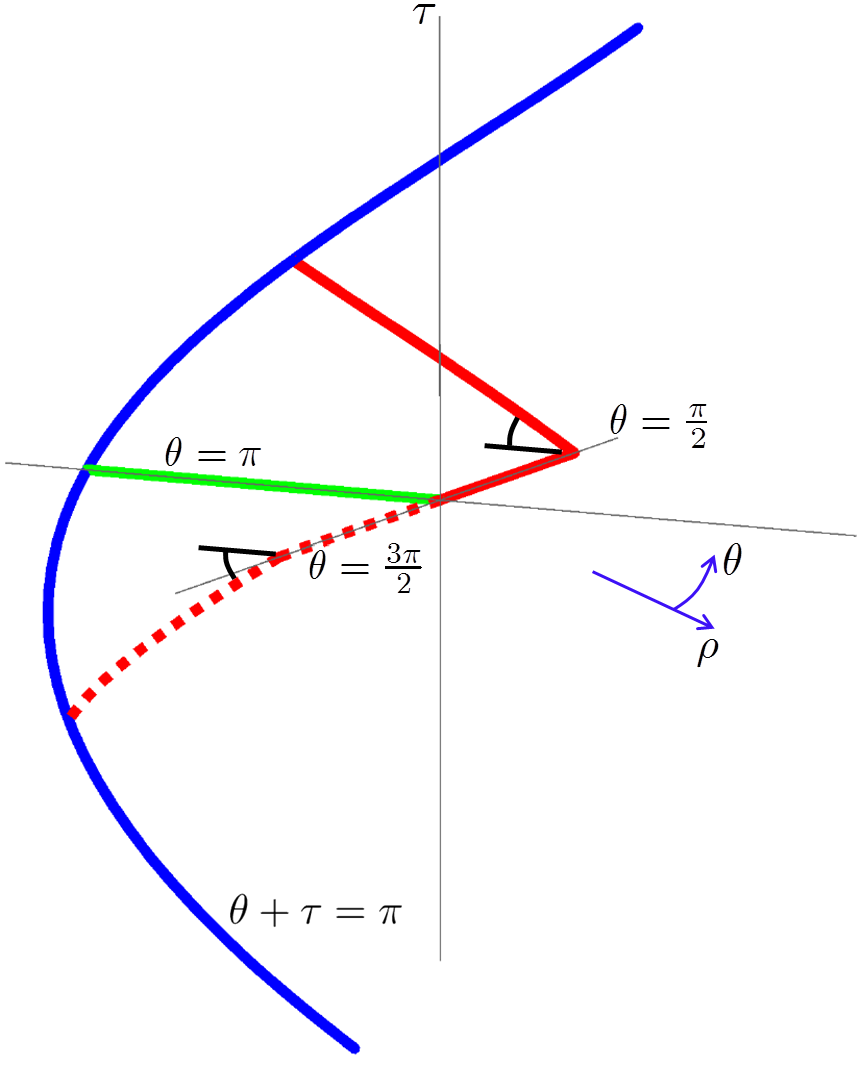}
\caption{Sketch of optimal circuits. The vertical axis is $\tau$ and the horizontal plane is described by the radius $\rho$ and the azimuthal angle  $\theta$, while the $y$ direction is suppressed. The unpenalized minimum (green) goes straight out along $(\theta,\tau)=(\pi,0)$. The optimal circuit in the penalized geometry is well-approximated by the segmented circuit $U_s(\bar s)$ in red: the first segment, $U_a(\bar s)$, goes straight out along $(\theta,\tau)=(\pi/2,0)$ until reaching $\rho_1$, whereupon the second segment, $U_b(\bar s)$, curves  upwards to $(\theta,\tau)=(3\pi/4,\pi/4)$. One sees that, relative to the unpenalized minimum, the segmented circuit arrives at a different but equally valid point along the one-parameter family of allowed end-points given by $\theta+\tau=\pi$ (blue), but with a shorter length. The dashed path has an identical length, and is simply the segmented circuit rotated by 180$^o$ about the $(\theta,\tau)=(\pi,0)$ axis.\label{fig:spiralPi}}
\end{figure}

Given the optimal geodesic for the penalized geometry, we would now like to examine the properties of the corresponding circuit. For simplicity, we rely on the fact that for $\pen\gg1$ the optimal geodesic is well approximated by the segmented path described in eq.~\reef{segments} and explicitly build the circuit for the latter path. First however, let us rewrite eq.~\reef{segments} in terms of $\theta=x+z$ and $\tau=x-z$:
\beqa
a)\ \ 0\le\bar s\le 1\ :&&y=0\,,\ \rho=\rho_1\,\bar s\,,\ \theta=\pi/2\,,\ \tau=0\,;
\label{segmentsX}\\
b)\ \ 1\le\bar s\le 2\ :&&y=y_1(\bar s-1)\,,\ \rho=\rho_1\,,\ \theta={\textstyle \frac{\pi}4}(\bar s+1),\ \tau={\textstyle \frac{\pi}4}(\bar s-1)\,.
\nonumber
\eeqa
Recall that $\bar s$ is some arbitrary parameter along the path. Furthermore, given this form, it is interesting to plot this segmented path in the ($\rho,\,\theta,\,\tau$) space in order to visually compare it to the straight-line geodesic---see figure \ref{fig:spiralPi}. This is essentially a comparison of the optimal geodesic in the original geometry \reef{metric1} to that in the new penalized geometry \reef{penalty1}. One feature that the figure emphasizes is that these two geodesics end at different points along the allowed (blue) spiral at $\theta+\tau=\pi$, $\rho=\rho_1$.

Using the expression for a general element of $\GLtwo$ in eq.~\eqref{Umatrix}, we write the segmented circuit $U_s(\bar s)$ as
\beq
U_s(\bar s)=\begin{cases}
U_a(\bar s)=\begin{pmatrix} e^{-\rho_1 \bar s}~ & 0 \\ 0 & e^{\rho_1\bar s}\end{pmatrix} & {\rm for}\ 0\leq \bar s\leq1\,,\\
U_b(\bar s)=e^{y_1(\bar s-1)}\begin{pmatrix} \cos\lp\frac{\pi}{4}(\bar s-1)\rp e^{-\rho_1}~ & -\sin\lp\frac{\pi}{4}(\bar s-1)\rp e^{\rho_1} \\ \sin\lp\frac{\pi}{4}(\bar s-1)\rp e^{-\rho_1}~ & \cos\lp\frac{\pi}{4}(\bar s-1)\rp e^{\rho_1} \end{pmatrix} & {\rm for}\ 1\leq \bar s\leq2\,,
\end{cases}\label{eq:segU}
\eeq
Note that $U_a(\bar s=1)=U_b(\bar s=1)$, as required by continuity along the path. Furthermore, observe that the circuit along the second segment can be re-expressed as
\beq
U_b(\bar s)=e^{y_1(\bar s-1)}\bar R(\bar s)\,U_a(\bar s=1)~, \label{eq:seg2}
\eeq
where we have defined the rotation matrix
\beq
\bar R(\bar s)\equiv\begin{pmatrix} \cos\lp\frac{\pi}{4}(\bar s-1)\rp ~ & -\sin\lp\frac{\pi}{4}(\bar s-1)\rp  \\ \sin\lp\frac{\pi}{4}(\bar s-1)\rp ~ &\ \cos\lp\frac{\pi}{4}(\bar s-1)\rp  \end{pmatrix}~.
\eeq
The interpretation of eq.~\eqref{eq:seg2} is that upon completing the first segment with $U_a(\bar s=1)$, the circuit performs a rotation (as well as multiplying by the exponential involving $y_1$) along the second segment until we reach the desired target state. The additional evolution along this second segment is therefore captured entirely by $e^{y_1(\bar s-1)}\bar R(\bar s)$. In passing, we also note that at the end-point, \ie $\bar s=2$, the rotation matrix reduces to
\beq
\bar R(\bar s=2)=\frac{1}{\sqrt{2}}\begin{pmatrix} 1 & -1 \\ 1 &\ \  1 \end{pmatrix}~,
\label{dead}
\eeq
which is closely related but distinct from the rotation matrix $R$ defined previously in eq.~\eqref{eq:rot}. 

At its end-point, this new circuit becomes
\beq
U_s(\bar s=2)=\frac{e^{y_1}}{\sqrt{2}}\begin{pmatrix} e^{-\rho_1}~ & -e^{\rho_1} \\  e^{-\rho_1}~ & \ \  e^{\rho_1} \end{pmatrix}\,,
\label{ends1}
\eeq
which we might compare to the end-point of the straight-line circuit \reef{solver5},
\beq
U_0(s=1)=e^{y_1}\begin{pmatrix}\cosh \rho_1~ & -\sinh\rho_1 \\ -\sinh\rho_1~ & \ \cosh\rho_1\end{pmatrix}\,.
\label{ends2}
\eeq
These are clearly different, in accordance with our comment about the geodesics ending at different points in  figure \ref{fig:spiralPi}. However, it is straightforward to show that the transformations implemented in the segmented and straight-line circuits are related by the rotation in eq.~\reef{dead}, \ie
\beq
U_s(\bar s=2)=U_0(s=1)\ \bar R(\bar s=2)~.\label{eq:U2phys}
\eeq
Both of these transformations act on the reference state to produce the target state (see eq.~\reef{eq:stateMatrix}) as $A_\mt{T} =U\,A_\mt{R}\,U^T$. Since the reference state is proportional to the identity, the additional rotation in eq.~\reef{eq:U2phys} leaves this state invariant, \ie  $A_\mt{R}=\bar R(\bar s=2) \,A_\mt{R} \,\bar R^T(\bar s=2)$, and so both $U_s(\bar s=2)$ and $U_0(s=1)$ will produce the same target state, as required.

It is useful to re-express the new circuit \reef{eq:segU} in the normal-mode space using eq.~\reef{rotateU}, \ie $\tilde U(s)=R\,U(s)\,R^T$, which yields
\beq
\tilde U_s(\bar s)=\begin{cases}
\tilde U_a(\bar s)=\begin{pmatrix} \cosh\lp\rho_1 \bar s\rp& -\sinh\lp\rho_1 \bar s\rp \\ -\sinh\lp\rho_1 \bar s\rp & \cosh\lp\rho_1 \bar s\rp \end{pmatrix} & {\rm for}\ 0\leq \bar s\leq1\,,\\
\tilde U_b(\bar s)=e^{y_1(\bar s-1)}\begin{pmatrix} \cos\lp\frac{\pi}{4}(\bar s-1)\rp ~ & \sin\lp\frac{\pi}{4}(\bar s-1)\rp  \\ -\sin\lp\frac{\pi}{4}(\bar s-1)\rp ~ &\ \cos\lp\frac{\pi}{4}(\bar s-1)\rp  \end{pmatrix}\,\tilde U_a(\bar s=1) & {\rm for}\ 1\leq \bar s\leq2\,,
\end{cases}\label{segUnm}
\eeq
where we have expressed $\tilde U_b(\bar s)$ in a form analogous to eq.~\reef{eq:seg2}. The key observation to note here is that the optimal circuit involves off-diagonal components when expressed in the normal-mode space. That is, in terms of the normal modes, the new circuit is utilizing entangling gates, \ie gates which entangle (or disentangle) the normal-mode coordinates. Since we know that the reference state \reef{Atilde} and the target state \reef{Atilde2} are unentangled in this space, it must be that along the first segment $U_a(\bar s)$, the circuit is introducing entanglement in the state, but then this entanglement is removed along the second segment $U_b(\bar s)$ on the second segment. We can see this explicitly by examining the state $\tilde A$ along the trajectory. In particular, along the first segment (\ie for $0\leq \bar s\leq1$), we find
\beq
\tilde A(\bar s)=
\omega_0\,\tilde U_a(\bar s)\,\tilde U^T_a(\bar s)=\omega_0 \begin{pmatrix}\ \cosh\lp2\rho_1 \bar s\rp& -\sinh\lp2\rho_1 \bar s\rp \\ -\sinh\lp2\rho_1 \bar s\rp & \ \cosh\lp2\rho_1 \bar s\rp \end{pmatrix} \,.
\label{state2a}
\eeq
Here we see the entanglement (\ie the off-diagonal terms) begins at zero and steadily grows to a maximum at  $\bar s=1$ at the end of the first segment. Subsequently, along the second segment (\ie for $1\leq \bar s\leq2$), we find
\beqa
\tilde A(\bar s)&=&\omega_0\,\tilde U_b(\bar s)\,\tilde U^T_b(\bar s)
\label{state2b}\\
&=&\omega_0 e^{2y_1(\bar s-1)}\begin{pmatrix} \cosh2\rho_1 -\sinh2\rho_1 \,\sin\lp\frac{\pi}{2}(\bar s-1)\rp & -\sinh2\rho_1 \,\cos\lp\frac{\pi}{2}(\bar s-1)\rp  \\ 
-\sinh2\rho_1 \,\cos\lp\frac{\pi}{2}(\bar s-1)\rp &\cosh2\rho_1 +\sinh2\rho_1 \,\sin\lp\frac{\pi}{2}(\bar s-1)\rp  \end{pmatrix} \,.
\nonumber
\eeqa
Here, the entanglement shrinks steadily back to zero as $\bar s$ runs over this second interval. Recall that $y_1$ and $\rho_1$ are given in terms of the normal-mode frequencies in eq.~\reef{fin2}.

To reiterate our key observation, with eq.~\reef{pen99} penalty factors are introduced to increase the cost of the entangling gates in the position space. As a result (for large $\pen$), the optimal geodesic is deformed to be close to the segmented path described in eqs.~\reef{segments} and \reef{segmentsX}. However, in the normal-mode space, the new geodesic is driven off of the normal-mode subspace. That is, even though the initial and final states are unentangled when written in terms of the normal modes, the optimal circuit still introduces entanglement (among the normal modes) at intermediate steps along the trajectory. One gains some insight into this behaviour by transforming the penalized metric \reef{pen99} to the normal-mode basis using the orthogonal matrix $\wR$ defined in eq.~\reef{bark2}. The new metric then becomes
\beq
G_{\tilde I\tilde J} = \wR_{\tilde I I} \,G_{I J}\,\wR^T_{J\tilde J}=\frac12 
\begin{pmatrix}
1+\pen^2 &0&0&1-\pen^2\\
0&1+\pen^2 &1-\pen^2&0\\
0&1-\pen^2 &1+\pen^2&0\\
1-\pen^2 &0&0&1+\pen^2
\end{pmatrix}
\,.
\label{pen99a}
\eeq
Here we see that in the normal-mode basis, there is no extra cost attributed to the entangling gates relative to the scaling gates. There are also a number of curious negative entries in the off-diagonal components, but this does not fundamentally distinguish the scaling and entangling gates.

\section{Discussion}\label{discuss}

In this paper, we took the first steps towards defining circuit complexity in quantum field theory.  The key idea, due to Nielsen \cite{Nielsen:2005mn1}, was to endow the space of circuits with an appropriate geometry which allows one to translate the task of finding the optimal circuit into the task of finding the minimum geodesic (with appropriate boundary conditions). We implemented this approach for a simple free scalar field theory. The first step however was to introduce a UV regulator by placing the theory on a lattice, which reduced  the scalar field theory to a family of coupled harmonic oscillators. In this context, we were able to construct a interesting set of elementary gates \reef{eq:gates}, in particular, scaling and entangling gates. We also chose our reference state to be a factorized Gaussian state \reef{refX}, whose simplicity lies in the fact that there is no entanglement between different points on the lattice. For the purposes of this preliminary study, we chose the target state to be the ground state \reef{targetX} of the system, which is also a Gaussian state. 

To gain some intuition for the problem, we began by studying the simple case of a pair of harmonic oscillators. The fact that both the reference and target states were Gaussian allowed for the simplification that we could translate from an operator language to a matrix language. It was then straightforward to show that with the $F_2$ cost function, the desired geometry was given by a right-invariant metric \reef{metric1} on $\GLtwo$. The optimal geodesic was then a simple straight line, which only moved through a flat two-dimensional subspace of the full, more complicated geometry. Translating this geodesic to the optimal circuit, the latter had a particularly simple interpretation in the normal-mode basis, where it only consisted of scaling gates amplifying the individual normal modes. This was a reflection of the fact that the ground state also takes the form of a factorized Gaussian when written in terms of the normal modes. These results for the two coupled oscillators were then extended to the full problem of a lattice of coupled oscillators with relative ease. In particular, we were able to show that the optimal circuit was given by the analogous straight-line geodesic moving in the normal-mode subspace, without constructing the full right-invariant metric on the $N^2$-dimensional geometry of $\GLN$.

\subsection*{Comparison with holography:}
 
As discussed in the introduction, a primary motivation for this paper came from recent efforts to understand ``holographic complexity,'' and so it was interesting in section \ref{sec:continuum} to compare our results to those obtained from the holographic proposals. Here we must reiterate the caveat that this comparison involves two very different QFTs, namely a free theory with a single degree of freedom in our scalar field model versus a strongly coupled theory with a large number of degrees of freedom in holography. Hence there is no \textit{a priori} reason to expect that the results should agree in the two cases. Nevertheless, we found that if the cost function is chosen appropriately, the scalar field complexity exhibits remarkable similarities with holographic complexity. Our tentative interpretation of this concordance is that it provides insight into the implicit cost functions that underly the holographic complexity conjectures.

In particular, the leading divergences \reef{Cholo} in both the CV and CA proposals are extensive, \ie they are proportional to the volume of the time slice on which the boundary state is evaluated, as shown in \cite{Carmi:2016wjl}. While the $F_2$ cost function gave a result proportional to $V^{1/2}$ in the scalar field theory, it is straightforward to construct a family of cost functions in eq.~\reef{Dalpha}, all of which yield an extensive complexity for the scalar field theory.\footnote{Again, we remind the reader that one can continue to work with the $F_2$ measure if the cost of the individual gates is set proportional to $\lp V/\delta^{d-1}\rp\!{}^{1/2}$---see footnote \ref{footy99}.} 

With the new cost functions \reef{Dalpha}, the leading contribution also contained a logarithmic factor, which was ambiguous in that it depended on the choice of the frequency $\omega_0$ specifying the reference state \reef{refX}. However, this precisely matched an ambiguity in the holographic complexity \cite{Carmi:2016wjl} found for the CA construction \reef{ccaction}. In the latter case, the logarithmic factor came from joint terms \cite{Lehner:2016vdi} in the gravitational action, and the ambiguity arose from the freedom to choose the normalization of the null normals on the boundary of the WDW patch.  Whereas this ambiguity had originally been seen as problematic for the CA conjecture, our scalar field calculation indicates that it is a perfectly natural feature associated with the freedom in the choice of the reference state that we can anticipate in any definition of complexity for a QFT. 

It might then seem mysterious that no such ambiguity arises for the CV conjecture \reef{volver}. However, as explained in section \ref{sec:continuum}, the additional logarithmic factor in the leading term \reef{calpha} becomes a simple numerical coefficient if we choose $\omega_0= e^{-\sigma}/\delta$, and so such a choice may indeed be an integral part of the microscopic rules implicit in the CV construction. Unfortunately, this does not explain the absence of infrared terms of the form given in eq.~\reef{calpha3} which might be expected with this choice. While it would be premature to conclude that the CV conjecture is incorrect, we might note that there is an alternative proposal in the literature suggesting that the volume of a maximal time slice in the bulk should be dual to the information metric rather than the complexity in the boundary theory \cite{MIyaji:2015mia}.

As further noted in section \ref{sec:continuum}, our scalar field complexity emulates the CA proposal \reef{ccaction} most closely if we choose the $F_1$ cost function, \ie $\kappa=1$ in eq.~\reef{Dalpha}. Given the aforementioned disparity in the two field theories in question, it is unclear how much weight to give this observation. However, we might add that the $F_1$ measure is a natural choice since it adheres most closely to the original definition of complexity, which involved simply counting the number of gates in the optimal circuit. Furthermore, let us add that the $F_1$ cost function will also feature again in the discussion of cMERA networks below.

Of course, it would be interesting if a more precise connection can be found between the holographic and QFT calculations with regards to the ambiguity in the reference state discussed above. At present, it is actually not clear how the reference state enters in the holographic calculations at all, but perhaps one can draw upon the proposal for a state-surface conjecture in \cite{nothing}. Undoubtedly, making this connection concrete would bring us closer to an explicit translation for the complexity between the bulk and boundary. 

\subsection*{Ambiguities and other miscellaneous complaints:}

In the introduction, our ``definition'' of complexity was rather imprecise, as it left open the choice of the reference state $\ket{\psi_\mt{R}}$, the choice of the set of elementary gates which would be used to construct $U$, and the choice of the tolerance (and measure) in eq.~\reef{scotch}. Clearly, even though it is easy to set out interesting questions for complexity (\eg what is the complexity of a particular state in a particular QFT?), the precise value of the complexity will depend on the details of all of these choices.\footnote{Of course, the tolerance does not play a role in the geometric framework adopted here because the gates are no longer discrete---see discussion around eq.~\reef{success}.} The ambiguity in our reference state, \ie the choice $\omega_0$, was already seen to modify the complexity in an interesting way in the preceding discussion. 

Here one might recall early discussions of entanglement entropy from a hep-th perspective: the explicit dependence of the leading contributions on the UV cut-off was certainly seen as problematic (or at least, it was by one of the present authors). However, with some experience, we learned to find universal information in the entanglement entropy and to apply it as a useful diagnostic of QFTs in various ways, \eg \cite{Myers:2010xs,Myers:2010tj}. In fact given our experience with entanglement entropy, the non-universality of the leading contributions to the complexity (because of the power-law dependence on $\delta$, as in eqs.~\reef{croot} and \reef{calpha}), was assumed to be self-evident in the present discussion and not even commented upon. Analogously, we would advocate that complexity is again a new quantity with what initially seems to be unusual and perhaps undesirable features, but that we must develop our experience to learn how complexity can inform us about interesting physics and universal properties of QFTs and holography. Hence rather than regarding the ambiguities discussed above as a problem per se, they should be collectively considered as a new feature which we must learn to accommodate in working with complexity.

In the context of the present calculations, clearly evaluating the complexity for a single state, \eg the ground state, will not be particularly informative. Instead we might compare the complexities of different states, and while extending our calculations to the complexity of excited states would allow such a comparison, it is beyond the scope of the present paper. However, we can certainly compare the complexities of the ground states of different scalar theories, in particular theories with different masses. Here our experience with holographic complexity \cite{pratik} suggests that there may be interesting information that could be extracted from the finite or logarithmic contributions. For example, for even spacetime dimensions (even $d$), there will be an interesting contribution that is intrinsically independent of the cut-off, although it may depend on the reference frequency $\omega_0$. This explicitly appears in eq.~\reef{eq:Cd1}, where we are examining the case $d=2$ and $\kappa=1$. In this instance, we may isolate this constant in\footnote{Of course, we are inspired to formulate this quantity by the constructions using entanglement entropy to examine RG flows of three-dimensional QFTs \cite{run2,run1}.}
\beq
\left(L\,\partial_L-1\right)\CC=-a_0\,.
\label{difff}
\eeq
where $L$ is the linear (spatial) size of the system. This result is independent of both the short-distance cut-off scale and the reference frequency. Hence it would be interesting to better understand the meaning of the coefficient of $a_0$, and whether it carries some universal information about the underlying theories. Of course, it is straightforward to extend this simple example to higher dimensions. It would also be interesting to compare these results to similar calculations for holographic complexity where the boundary CFT is deformed by a relevant operator---see discussion at the end of section \ref{sec:continuum}.

\vskip 1.5ex

Let us also remark here that the reference state \reef{refX} is an unusual state from the textbook perspective of QFT. Recall that  this state was chosen since it has no entanglement between different points (on the lattice). Such a factorized Gaussian is precisely the kind of reference state that appears in the cMERA construction \cite{Haegeman:2011uy}---see the discussion below. However, such an unentangled state is a very unusual state  in standard QFT, which for example would typically have a divergent energy density. Of course, the vacuum energy density of the ground state is also divergent, and so to make this statement meaningful, we may evaluate the difference in the energies of $\ket{\psi_\mt{R}}$ and the ground state $\ket{\psi_\mt{T}}$:
\beq
\bal
\bra{\psi_\mt{R}}\,H\,\ket{\psi_\mt{R}}-\bra{\psi_\mt{T}}\,H\,\ket{\psi_\mt{T}}&= \frac14\,N^{d-1} \omega_0+\frac{1}{4\omega_0}\,\sum \omega_{\vec k}^2-\frac12\,\sum \omega_{\vec k}\\
&\approx \frac{V}{\delta^d} \,\left[\frac{\omega_0\delta}{4}+\frac{1}{\omega_0\delta}-1\right]\,.
\eal
\eeq
We therefore see that that generically, if $\omega_0\delta\ll 1$ or $\omega\delta\gg1$ (which were advocated to be the natural choices in section \ref{sec:continuum}), the ``renormalized'' energy density of $\ket{\psi_\mt{R}}$ diverges as $\delta\to0$.\footnote{Note that with some fine-tuning of $\omega_0$, we could arrange the difference of energy densities to be finite.} Hence this would not be a state that would be considered to be part of the standard Hilbert space that one builds with particle excitations on top of the vacuum. However, one should simply regard this as another unusual feature of complexity. As we have seen both here with our QFT calculations as well as in holographic complexity, the complexity can only be sensibly defined with a finite value of the regulator, in which case the reference state is certainly a sensible state within the associated Hilbert space. 

\vskip 1.5ex

While introducing a UV regulator was an essential step in sensibly defining the complexity in the scalar field theory, let us add that this does not regulate the size of the Hilbert space in the present case.\footnote{We thank Edward Witten for making this observation and raising the following question.} With the lattice regulator, the scalar field theory is reduced to $N^{d-1}$ normal-mode oscillators, but the Hilbert space of each of these oscillators is infinite! It is an interesting question whether or not an additional regulator should be introduced to render the total number of states finite as well. Otherwise it would seem that even within the UV regulated theory, there will be states of infinite complexity.

\subsection*{Penalty factors and locality:}

In section \ref{sec:penalty}, we experimented with the introduction of penalty factors in the case of two coupled oscillators. In particular, we gave a higher cost to the entangling gates than the scaling gates with \reef{pen99}. This certainly resulted in a different optimal circuit, but ultimately the circuit could not avoid incorporating the entangling gates, and so the complexity increased to $\op{(\pen)}$, as shown in eq.~\reef{loft}. Perhaps the most interesting lesson to be learned from these calculations is that the introduction of penalty factors (in the position-space cost function) tends to drive the optimal circuit away from the normal-mode subspace, the restriction to which played a central role in the previous analysis of section \ref{sec:N}.

However, in our simple experiment in section \ref{sec:penalty}, the optimal circuit was still required to introduce entanglement using the entangling gates, and therefore our calculations did not really address the motivation discussed at the beginning of that section. Namely, we expected that penalty factors could be used to introduce a notion of locality in the complexity of the scalar field theory. In particular, our calculations in section \ref{sec:N} included entangling gates $Q_{ab}$, which coupled points on the lattice that were arbitrarily far apart, all with equal cost. It seems natural that using gates which couple far-separated points should incur a higher cost than using those which couple nearest neighbors. 

To gain some insight into this problem,  let us return for a moment to the discrete gates in eq.~\eqref{eq:gates} and consider a one-dimensional lattice of $N$ coupled oscillators. We will show that our set of entangling gates is over-complete in the sense that any of these gates can be constructed from nearest-neighbour entangling gates.  For example, it is a straightforward calculation (using the Baker-Campbell-Hausdorff formula) to show that
\beq
Q_{13}=\lp Q_{12}^{-1}Q_{23}^{-1}Q_{12}Q_{23}\rp^{1/\veps}~.\label{eq:Q13}
\eeq
In other words, the next-to-nearest neighbour entangling gate $Q_{13}$ is equivalent to $4/\veps$ nearest-neighbour entangling gates. A simple generalization of the above result is
\beq
Q_{14}=\lp Q_{13}^{-1}\,Q_{34}^{-1}\,Q_{13}\,Q_{34}\rp^{1/\veps}~,
\eeq
which implies that the next-to-next-to-nearest neighbour entangling gates are equivalent to $8/\veps^2+2/\veps$ nearest-neighbour gates ($8/\veps^2$ from the use of the $Q_{13}$'s and another $2/\veps$ from the $Q_{34}$'s). These calculations can be easily generalized to show that  nonlocal gates $Q_{a,a+1+n}$ or $Q_{a+1+n,a}$, which entangle oscillators that are separated by $n$ intermediate sites, can be constructed by the use of
\beq
c(n)=\frac{2}{\veps}\lp1+c(n-1)\rp
\label{tsoc}
\eeq
nearest-neighbor entangling gates, where $c(0)\equiv 1$. Thus to leading order, the ``cost'' of these nonlocal gates in terms of nearest-neighbour gates grows like $c(n)\sim1/\veps^n$.

Following Neilsen's approach \cite{Nielsen:2005mn1,Nielsen:2006mn2,Nielsen:2007mn3}, we would not eliminate these nonlocal gates from the elementary gate set, but would instead modify the geometry by introducing (heavy) penalty factors to discourage the geodesics from moving along the corresponding directions. The structure of eq.~\reef{tsoc} suggests increasing the penalty factors as a power law to match the growth of the nonlocality, \ie the directions corresponding to $I=(a,a+1+n)$ and $(a+1+n,a)$ would be assigned a penalty factor $\pen^{2n}$. Note that this does not penalize the nearest-neighbour gates at all, in contrast to our experiment in section \ref{sec:penalty}. Of course, for a periodic chain of oscillators, the maximum penalty factor would be $\pen^{N-2}$ and $\pen^{N-3}$ for even and odd $N$, respectively.

We can gain further insight by translating eq.~\reef{eq:Q13} into a macroscopic circuit described by a path-ordered exponential \reef{pathA}. In particular, consider the following path:
\beqa
Y^{23}(s)&=& \alpha\,\left[ 1-\Theta(s-1) -\Theta(s-2) +\Theta(s-3)\right]\,,
\nonumber\\
Y^{12}(s)&=& \alpha\,\left[ \Theta(s-1)-\Theta(s-2) +\Theta(s-3) -\Theta(s-4) \right]\,,
\label{pathA9}
\eeqa
where $0\le s\le4$, $\Theta(x)$ is the Heaviside theta-function, and the $Y^{I}$ are implicitly zero for all other values of $I$. That is, we turn on the $M_{23}$ generator with amplitude $\alpha$ for the interval $0\le s\le 1$;  $M_{12}$ is then turned on with amplitude $\alpha$ for $1\le s\le 2$;
next, $M_{23}$ is turned on with amplitude $-\alpha$ for $2\le s\le 3$; and  finally $M_{12}$ is turned on with amplitude $-\alpha$ for $3\le s\le 4$. Note that the precise parametrization of this path is not important. The circuit 
\beq
U_1(s)=\cev{\mathcal{P}}\,\mathrm{exp}\int_0^x\dd \tilde s\,Y^I\!(\tilde s)\,M_I
\label{ramp}
\eeq
then yields $U_1(s=4)=\exp\left[\alpha^2 M_{13}\right]$, following the same calculation that yields eq.~\reef{eq:Q13}. Hence we could accomplish the same transformation with
\beq
U_2(s)=\cev{\mathcal{P}}\,\mathrm{exp}\int_0^x\dd \tilde s\,Y^{13}(\tilde s)\,M_{13}
\qquad{\rm where}\qquad Y^{13}(s)=\alpha^2/4 \ \ {\rm for}\ \ 
0\le s\le 4\,.
\label{ramp2}
\eeq
Now let us compare the costs of these two circuits using the $F_q$ measure \reef{eq:Fmetrics}, where the nearest neighbour gates are assigned cost 1 while the next-to-nearest neighbour gates are assigned cost $\pen$. The cost functions are then easily evaluated to be
\beq
{\cal D}(U_1)=\int_0^4\dd s \sqrt{\delta_{IJ}Y^I(s)Y^J(s)}=4\alpha~,\qquad
{\cal D}(U_2) = 4\left|Y^{13}\right|=\pen\,\alpha^2~.
\eeq
Hence with an appropriate penalty factor, we can suppress the use of the nonlocal gates in favour of the nearest neighbour gates.  While it would be interesting to examine the effect of the above scheme of penalty factors in more detail, we leave this for future work.

\subsection*{cMERA:}

\begin{figure}[h!]
\centering
\includegraphics[width=0.89\textwidth]{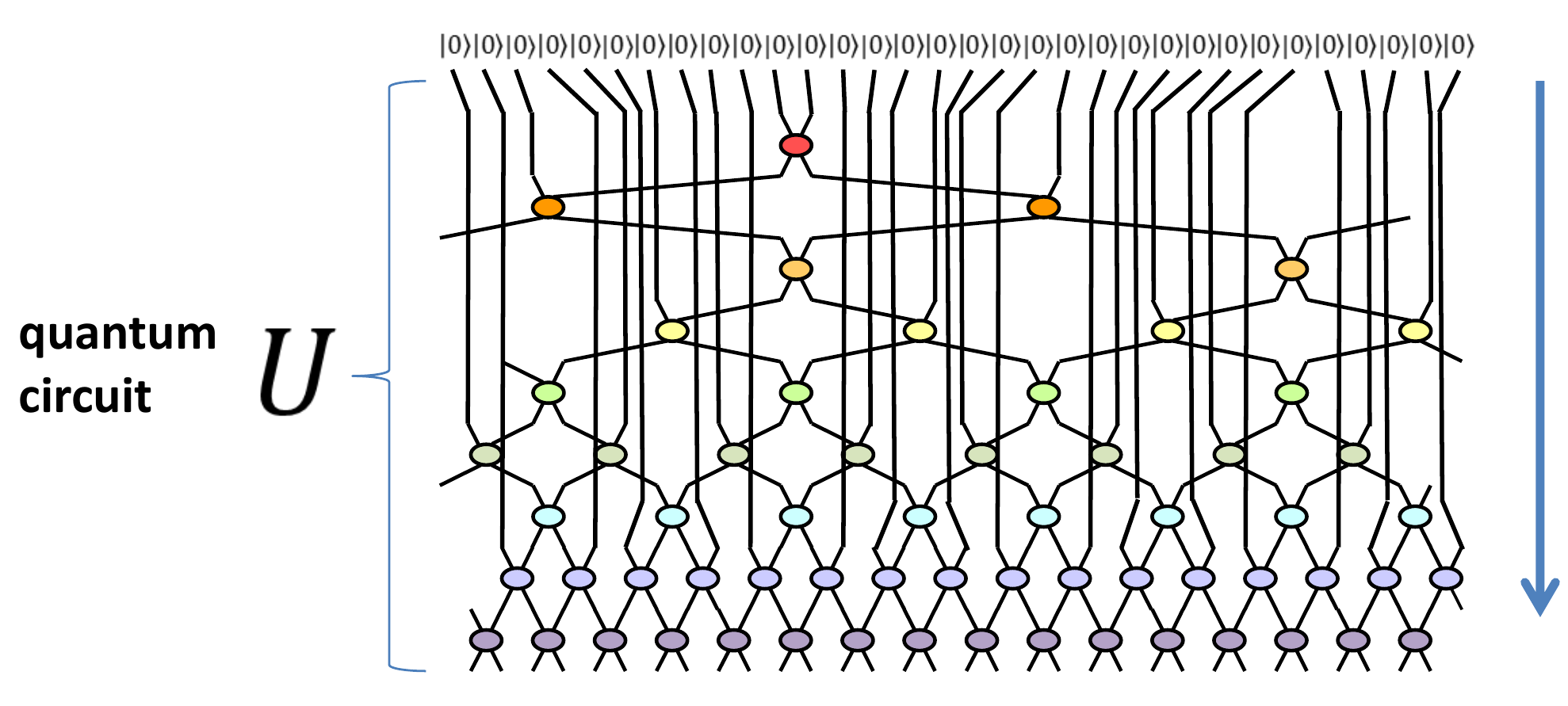}
\caption{Illustration of MERA as a quantum circuit. Starting from the tensor product state $\left|0\right>\left|0\right>\ldots\left|0\right>$ (top), the sequential application of entanglers and isometries (colored dots) efficiently generates the ground state (bottom). These operators can be thought of as unitary gates (although not simple elementary gates), and hence the tensor network can be thought of as the quantum circuit  that connects the reference (product) and target (ground) states $\psi_0$ and $\psi_1$. Figure courtesy of Guifr\'e Vidal \cite{VidalKITP2015}. \label{fig:MERA}}
\end{figure}

The AdS/MERA correspondence was the first proposal for a novel connection between holography and tensor networks \cite{Swingle:2009bg,Swingle:2012wq}. This proposal suggests that the MERA (Multiscale Entanglement Renormalization Ansatz)  tensor network \cite{Vidal:2007hda,vidal2008class, vidal2009entanglement} provides a discrete representation of a time slice of (three-dimensional) AdS space. As illustrated in figure \ref{fig:MERA}, the MERA network consists of unitary operators which, starting from the simple product state $\left|0\right>\otimes\ldots\otimes\left|0\right>$, generate the ground state in $d=2$ critical systems.  In other words, the MERA network can be thought of as a quantum circuit. The AdS/MERA correspondence was certainly a source of motivation/inspiration for the early discussions of holographic complexity, in particular, of the CV conjecture \cite{Stanford:2014jda,Susskind:2014rva}. Furthermore, in these discussions, it was implicitly considered the optimal circuit for the preparation of the CFT ground state.

There has been some progress towards developing a continuum version of MERA, however, these constructions are limited to describing very simple QFTs \cite{Haegeman:2011uy}---see also \cite{Nozaki:2012zj,Franco-Rubio:2017tkt,Hu:2017rsp}. In particular, one example is the cMERA description of the ground state of a free scalar field. That is, there is a cMERA circuit which more-or-less performs precisely the transformation for which the circuits studied herein were constructed. Hence, our original expectation was that our analysis would find that the optimal circuit was something like a cMERA network. However, we instead found the straight-line circuit described in section \ref{sec:N}. The key difference between the two circuits is that the cMERA circuit is organized to systematically introduce entanglement scale-by-scale, \ie to order the amplification of the normal modes according to their wavelength \cite{Hu:2017rsp}. However, by almost all of the measures considered in section \ref{sec:N}, including the $F_2$ cost function and the $\kappa$ cost functions in eq.~\reef{Dalpha}, the straight-line circuit is the optimal circuit. The one exception to this rule is the $F_1$ (or $\kappa=1$) cost function. This last describes an unusual geometry,\footnote{Let us add here that the $F_1$ measure also exhibits some unusual properties under a change of basis, as discussed in appendix \ref{base8}.} which is sometimes called the ``Manhattan metric.'' The key feature of this geometry is that the length is the same for all paths as long as they do not back-track at any point, and hence the straight-line circuit and the cMERA circuit have identical costs for this measure. 

This question certainly deserves further study. It appears that there are two possible approaches: the first would be to study more exotic cost functions in order to identify those which favour the cMERA circuit. This may be useful since given the AdS/MERA duality, it may provide better insight into the properties of the cost function that appears in holographic complexity. We also mention that this is likely not a straightforward approach since we found that introducing penalty functions (in position space) seems to drive the optimal circuit out of the normal-mode subspace, whereas the cMERA circuit is confined to this particular slice of the full circuit geometry by construction. A second option might be to introduce new physics in the selection of the ``optimal'' circuit. That is, while the straight-line and cMERA circuits have equivalent costs according to the $F_1$ cost function, there may be additional physics considerations, \eg some relation to renormalization group flows, which lead holography to favour a cMERA-like circuit.

\subsection*{Future directions:}

This paper provides only a preliminary investigation towards understanding circuit complexity in quantum field theory. We already mentioned a number of future directions that we expect will be fruitful. Some examples include extending the present calculations to evaluate the complexity of excited states, producing a more concrete connection between the ambiguities arising in our QFT calculations and those in holographic calculations of the complexity, and studying in detail the effect of penalty factors on the complexity and the structure of the optimal circuit for a lattice of oscillators. Other obvious extensions of the present work would include evaluating the complexity in fermionic theories or in interacting QFTs. 

In closing, we would like to draw a comparison with entanglement entropy in QFT. Entanglement entropy has a simple textbook definition: first one must construct the reduced density matrix $\rho_\mt{A}$ of the particular subsystem under study, and then one evaluates the von Neumann entropy of this density matrix as $S_\mt{EE}=-\sum \lambda_i\log\lambda_i$, where $\lambda_i$ are the eigenvalues of $\rho_\mt{A}$. However, much of the progress in understanding the properties and role of entanglement entropy came from the replica trick, introduced by Calabrese and Cardy \cite{Calabrese:2004eu,Calabrese:2005zw}. The latter applies familiar tools (\eg path integrals) in a novel setting (\eg the replicated background geometry) to evaluate the entanglement entropy. Returning to complexity, our present approach is to apply a more-or-less standard textbook definition to evaluating the complexity of states in a QFT, which is a useful preliminary step to gain an understanding of the properties of this new quantity. However, we would really like to develop a new approach, analagous to those developed for entanglement entropy above, which again uses familiar QFT techniques in a presumably novel setting to evaluate some quantity like the complexity. In other words, we are asking what is the new calculation of complexity which is the analog of Calabrese and Cardy's replica trick for entanglement entropy. Indeed, it may be that the first steps in this direction have already been taken in \cite{Cap1,Cap2}---see also \cite{Czechxx}.

\section*{Acknowledgments}
It is a pleasure to thank Micha Berkooz, Eugenio Bianchi, Shira Chapman, Adri\'an Franco-Rubio, Lucas Hackl, Markus Hauru, Michal Heller, Qi Hu, Steve Jordan, Hugo Marrochio, John Preskill, Djordje Radicevic, Grant Salton, Joan Simon, Guillaume Verdon-Akzam, Guifr\'e Vidal, Edward Witten, and Beni Yoshida for helpful conversations.  Research at Perimeter Institute is supported by the Government of Canada through the Department of Innovation, Science and Economic Development and by the Province of Ontario through the Ministry of Research \& Innovation. RCM is also supported by an NSERC Discovery grant, as well as research funding from the Canadian Institute for Advanced Research and from the Simons Foundation through the ``It from Qubit" Collaboration. RJ is supported by the $\Delta$-ITP consortium and the Foundation for Fundamental Research on Matter (FOM), both of which are parts of the Netherlands Organization for Scientific Research (NWO) funded by the Dutch Ministry of Education, Culture, and Science (OCW). RJ also gratefully acknowledges the support of the Perimeter Institute Visiting Graduate Fellows program. Finally, RCM would also like to thank the organizers of the It-From-Qubit ``Complexity and Black Holes'' workshop at Stanford University, the ``Tensor Networks for Quantum Field Theories II'' workshop at Perimeter Institute and the ``Strings 2017'' conference in Tel Aviv for the opportunity to present this work.

\vskip 1.5ex
\noindent While this paper was in preparation, we were informed of \cite{Chapman:2017rqy}, which seems to have significant overlap with the present work.

\begin{appendices}
\section{Example circuits}\label{sec:example}
In this section, we analyze the circuit depth of a few discrete circuits, as  introduced in section \ref{sec:gates}. In particular, let us consider the example given in eq.~\eqref{eq:eg1},
\beq
\psi_\mt{T}=U_1\,\psi_\mt{R}= Q_{22}^{\alpha_3}\,Q_{21}^{\alpha_2}\,Q_{11}^{\alpha_1}\,\psi_\mt{R}~,
\label{exam1}
\eeq
where the target wave function is given in eq.~\reef{eq:targetPhys} and the reference wave function, in eq.~\reef{eq:refPhys}. The question then is to determine the exponents $\alpha_i$ in this equation, \ie the number of times each type of gate is applied in the circuit.

Intuitively, $U_1$ is a string of gates running from the right to the left. The first gate applied is $Q_{11}$, which rescales the coefficient of $x_1^2$ in the exponent of the Gaussian wave function. Next, by applying $Q_{21}$, the two oscillators become entangled. Finally, the application of $Q_{22}$ rescales $x_2$ to ensure that $x_2^2$ appears with the correct coefficient. Hence let us begin the quantitative analysis by considering:
\beqn
Q_{11}^{\alpha_1}\,\psi_0(x_1,x_2)
=e^{\eps\alpha_1/2}\psi\lp e^{\eps\alpha_1}x_1,x_2\rp
=\sqrt{\frac{\omega_0}{\pi}}e^{\eps\alpha_1/2}\mathrm{exp}\left[-\frac{\omega_1}{2}x_1^2-\frac{\omega_0}{2}x_2^2\right]~,
\eeqn
where
\be
\omega_1\equiv e^{2\eps\alpha_1}\omega_0\implies\alpha_1=\frac{1}{2\eps}\log\!\lp\frac{\omega_1}{\omega_0}\rp~.
\label{exp1}
\ee
Next, applying the $Q_{21}$ gates yields
\beqn
\bal
Q_{21}^{\alpha_2}\,Q_{11}^{\alpha_1}\,\psi_0(x_1,x_2)
&=\sqrt{\frac{\omega_0}{\pi}}e^{\eps\alpha_1/2}e^{i\eps x_2p_1}\mathrm{exp}\left[-\frac{\omega_1}{2}x_1^2-\frac{\omega_0}{2}x_2^2\right]\\
&=\sqrt{\frac{\omega_0}{\pi}}e^{\eps\alpha_1/2}\mathrm{exp}\left[-\frac{\omega_1}{2}\lp x_1+\eps\alpha_2 x_2\rp^2-\frac{\omega_0}{2}x_2^2\right]\\
&=\sqrt{\frac{\omega_0}{\pi}}e^{\eps\alpha_1/2}\mathrm{exp}\left[-\frac{\omega_1}{2}x_1^2-\frac{1}{2}\lp\omega_0+\eps^2\alpha_2^2\omega_1\rp x_2^2-\eps\alpha_2\omega_1 x_1x_2\right]\,.
\eal
\eeqn
Note that the $x_1x_2$ cross-term will be rescaled in the next step, so we cannot fix any of the coefficients quite yet. Finally, we rescale $x_2$ with the $Q_{22}$ gates:
\be
Q_{22}^{\alpha_3}\,Q_{21}^{\alpha_2}\,Q_{11}^{\alpha_1}\,\psi_0(x_1,x_2)
=\sqrt{\frac{\omega_0}{\pi}}e^{\eps(\alpha_1+\alpha_3)/2}\mathrm{exp}\left[-\frac{\omega_1}{2}x_1^2-\frac{\omega_2}{2}x_2^2-\beta x_1x_2\right]~,\label{eq:eg1fin}
\ee
where $\alpha_2$ and $\alpha_3$ are determined by matching the second and third coefficients in the exponent, \ie
\be
\omega_2=\lp\omega_0+\eps^2\alpha_2^2\omega_1\rp e^{2\eps\alpha_3}~,
\;\;\;
 \beta\equiv\eps\alpha_2\omega_1  e^{\eps\alpha_3}~.
\ee
Solving the above constraints then yields
\be
\alpha_2=\frac{1}{\eps}\sqrt{\frac{\omega_0}{\omega_1}}\frac{\beta}{\sqrt{\omega_1\omega_2-\beta^2}}~,\qquad
\alpha_3=\frac{1}{2\eps}\log\!\lp\frac{\omega_1\omega_2-\beta^2}{\omega_0\,\omega_1}\rp~.
\label{exp2}
\ee
As a consistency check, note that with these identifications, the normalization factor of the final wave function becomes
\be
\sqrt{\frac{\omega_0}{\pi}}e^{\eps(\alpha_1+\alpha_3)/2}=\frac{\lp\omega_1\omega_2-\beta^2\rp^{1/4}}{\sqrt{\pi}}~,
\ee
which correctly preserves the unit norm. Of course, this was expected since, as discussed in the main text, the entangling and scaling gates, $Q_{ij}$ and $Q_{ii}$, preserve the norm when acting on Gaussian wave functions.

Hence the total number of gates in the circuit $U_1$ in eq.~\reef{exam1} is given by 
\be
\mathcal{D}(U_1)=|\alpha_1|+|\alpha_2|+|\alpha_3|
=\frac{1}{2\eps}\log\!\lp\frac{\omega_1\omega_2-\beta^2}{\omega_0^2}\rp
+\frac{1}{\eps}\sqrt{\frac{\omega_0}{\omega_1}} \frac{|\beta|}{\sqrt{\omega_1\omega_2-\beta^2}}~,\label{eq:Deg1Phys}
\ee
where we have assumed here that $\omega_1>\omega_0$ and $\omega_1\omega_2-\beta^2>\omega_0^2$.\footnote{See further comments at the end of this appendix.} As in the main text, we refer to $\mathcal{D}(U_1)$ as the {circuit depth}, rather than the complexity, since while it counts the total number of gates in the circuit, we have no reason to expect that $U_1$ is the optimal circuit. Recall that we introduced 
 absolute values in eq.~\reef{eq:Deg1Phys} in order to give an equal complexity cost for the inverse gates $Q_{ij}^{-1}$ as for the original gates $Q_{ij}$, \ie we count the appearance of $Q_{ij}^{-1}$ as one gate in a circuit. At a pragmatic level, this is required because $\alpha_2$ is negative in our example, \ie $\beta=(\omega_+-\omega_-)/2<0$. 

Note that in evaluating the exponents $\alpha_i$ in eqs.~\reef{exp1} and \reef{exp2}, we are implicitly treating them as real numbers. If we insisted on having integer exponents, then we would would need to round these results up or down to the nearest integer. In this case, we would define a measure of success for our transformation and choose the integer exponents to maximize this measure. For example, we could consider the overlap
\beq
\Big|\int d^2x\,\psi_\mt{T}^\dagger\ Q_{22}^{\alpha_3}\,Q_{21}^{\alpha_2}\,Q_{11}^{\alpha_1}\,\psi_\mt{R}\Big|^2 =1-\chi~,
\label{success}
\eeq
and choose the precise integer values of $\alpha_i$ to minimize $\chi$. Of course, using real exponents $\alpha_i$ is very much in line with describing the circuits in terms of path-ordered exponentials \reef{eq:pathPsi}. This discussion is related to the choice of a tolerance $\veps$ in eq.~\reef{scotch}, \ie rather than minimizing $\chi$, one might demand that $\chi\le\veps$.

Now let us briefly present a few other examples of simple circuits to further familiarize the reader with the concepts discussed here. First, let us consider applying the entangling gate before either of the scaling gates:
\beq
\psi_\mt{T}=U_2\psi_\mt{R}= Q_{22}^{\tilde\alpha_3}\,Q_{11}^{\tilde\alpha_1}\,
Q_{21}^{\tilde\alpha_2}\,\psi_\mt{R}~.
\eeq
Note that for comparison purposes, our numbering of the exponents is such that they are associated with the same gates as appear in eq.~\reef{exam1}. The calculation proceeds essentially as above; in the end, we must match the coefficients 
\beq
\omega_1=\omega_0e^{2\eps\tilde\alpha_1}~,\qquad
\omega_2=\lp1+\eps^2\tilde\alpha_2^2\rp\omega_0 e^{2\eps\tilde\alpha_3}~,\qquad
\beta=\eps\tilde\alpha_2\omega_0 e^{\eps(\tilde\alpha_1+\tilde\alpha_3)}~.
\eeq
Solving for the exponents $\alpha_i$ then yields
\beq
\tilde\alpha_1=\frac{1}{2\eps}\log\!\lp\frac{\omega_1}{\omega_0}\rp~,\qquad
\tilde\alpha_2=\frac{1}{\eps}\frac{\beta}{\sqrt{\omega_1\omega_2-\beta^2}}~,\qquad
\tilde\alpha_3=\frac{1}{2\eps}\log\!\lp\frac{\omega_1\omega_2-\beta^2}{\omega_0\omega_1}\rp~,
\label{exp3}
\eeq
and hence the circuit depth becomes
\beq
\mathcal{D}(U_2)=\sum|\tilde\alpha_i|=
\frac{1}{2\eps}\log\!\lp\frac{\omega_1\omega_2-\beta^2}{\omega_0^2}\rp
+\frac{1}{\eps}\frac{|\beta|}{\sqrt{\omega_1\omega_2-\beta^2}}~.
\label{horse0}
\eeq
Comparing the results in eqs.~\reef{exp1} and \reef{exp2} with those in eq.~\reef{exp3}, we see that the exponents for the scaling gates are identical, \ie $\alpha_1=\tilde\alpha_1$ and $\alpha_3=\tilde\alpha_3$, and only the exponent for the entangling gate has changed. Hence the circuit depth is almost identical to \eqref{eq:Deg1Phys}, except that the second term lacks the factor $\sqrt{\omega_0/\omega_1}$. If we assume $\omega_1>\omega_0$ as before, this implies that the present circuit will be slightly longer, \ie $\mathcal{D}(U_2)>\mathcal{D}(U_1)$.

As a third simple example, let us consider instead applying the entangling gate after both of the scaling gates:
\beq
\psi_\mt{T}=U_3\psi_\mt{R}= Q_{21}^{\hat\alpha_2}\,Q_{22}^{\hat\alpha_3}\,
Q_{11}^{\hat\alpha_1}\,
\psi_\mt{R}~.
\label{reed3}
\eeq
Again we skip over the details of the calculation; we find that we must match the coefficients 
\beq
\omega_1=\omega_0e^{2\eps\hat\alpha_1}~,\qquad
\omega_2=\lp e^{2\eps\hat\alpha_3}+\eps^2\hat\alpha_2^2 e^{2\eps\hat\alpha_1}\rp\omega_0 ~,\qquad
\beta=\eps\hat\alpha_2\omega_0 e^{2\eps\hat\alpha_1}~.
\eeq
Solving for the exponents $\alpha_i$ then yields
\beq
\hat\alpha_1=\frac{1}{2\eps}\log\!\lp\frac{\omega_1}{\omega_0}\rp~,\qquad
\hat\alpha_2=\frac{1}{\eps}\frac{\beta}{\omega_1}~,
\qquad
\hat\alpha_3=\frac{1}{2\eps}\log\!\lp\frac{\omega_1\omega_2-\beta^2}{\omega_0\omega_1}\rp~,
\label{exp4}
\eeq
and hence the circuit depth becomes
\beq
\mathcal{D}(U_3)=\sum|\tilde\alpha_i|=
\frac{1}{2\eps}\log\!\lp\frac{\omega_1\omega_2-\beta^2}{\omega_0^2}\rp
+\frac{1}{\eps}\frac{|\beta|}{\omega_1}~.
\label{horse}
\eeq
Again, comparing with the exponents in eqs.~\reef{exp1} and \reef{exp2} or in eq.~\reef{exp3}, we see that only the exponent for the entangling gate has changed. Hence the circuit depth here is similar to those for the previous two circuits, and whether the present circuit is longer or shorter depends on the values of the parameters $\omega_0$, $\omega_1$, $\omega_2$ and $\beta$

Let us consider one more general example. Another interesting circuit would be
\beq
\psi_\mt{T}=U_4\psi_\mt{R}= Q_{22}^{\bar\alpha_3}\,Q_{21}^{\bar\alpha_2}\,
\left(Q_{21}^{-1}\,Q_{11}\right)^{\bar\alpha_1}\,
\psi_\mt{R}~.
\label{reed4}
\eeq
Note that
\beq
\left(Q_{21}^{-1}\,Q_{11}\right)^n\psi(x_1,x_2)
=e^{n \eps/2}\psi\lp e^{n\eps}x_1-\eps e^\eps\frac{1-e^{n\eps}}{1-e^\eps}\,x_2,x_2\rp~,\label{eq:gatecombo}
\eeq
the derivation of which is as follows: first, consider
\beq
Q_{11}\psi=e^{\eps/2}\psi\lp e^\eps x_1,x_2\rp\;\implies
Q_{21}^{-1}Q_{11}\psi=e^{\eps/2}\psi\lp e^\eps x_1-\eps e^\eps x_2,x_2\rp~.
\eeq
Then acting with this combination twice yields
\beq
\bal
\lp Q_{21}^{-1}Q_{11}\rp^2\psi&=e^{2\eps/2}Q_{21}^{-1}\psi\lp e^{2\eps} x_1-\eps e^{\eps} x_2,x_2\rp
=e^{2\eps/2}\psi\lp e^{2\eps}\lp x_1-\eps x_2\rp-\eps e^{\eps} x_2,x_2\rp\\
&=e^{2\eps/2}\psi\lp e^{2\eps} x_1-e^\eps\lp e^{\eps}+1\rp\eps x_2,x_2\rp~.
\eal
\eeq
And a third time:
\beq
\lp Q_{21}^{-1}Q_{11}\rp^3\psi=e^{3\eps/2}Q_{21}^{-1}\psi\lp e^{3\eps} x_1-e^{\eps}\lp e^\eps+1\rp\eps x_2,x_2\rp
=e^{3\eps/2}\psi\lp e^{3\eps} x_1-e^\eps\lp e^{2\eps}+e^\eps+1\rp\eps x_2,x_2\rp~.
\eeq
Now the pattern is clear, and we deduce
\beq
\lp Q_{21}^{-1}Q_{11}\rp^n\psi
=e^{n\eps/2}\psi\lp e^{n\eps} x_1-\eps e^\eps\sum_{k=0}^{n-1} e^{k\eps}\,x_2,x_2\rp~.
\eeq
Since
\beq
\sum_{k=0}^{n-1}e^{k\eps}=\frac{1-e^{n\eps}}{1-e^\eps}~,
\eeq
this becomes \eqref{eq:gatecombo}, as claimed.

Now, acting with the circuit $U_4$ and matching coefficients as before, we find
\beq
\bal
\bar\alpha_1&=\frac{1}{2\eps}\log\lp\frac{\omega_1}{\omega_0}\rp~,\\
\bar\alpha_2&=\frac{1}{\eps}\sqrt{\frac{\omega_0}{\omega_1}}\frac{\beta}{\sqrt{\omega_1\omega_2-\beta^2}}+
\frac{1}{e^{-\eps}-1}\lp1-\sqrt{\frac{\omega_0}{\omega_1}}\rp~,\\
\bar\alpha_3&=\frac{1}{2\eps}\log\lp\frac{\omega_1\omega_2-\beta^2}{\omega_0\omega_1}\rp~.
\eal
\eeq
Expanding $\bar\alpha_2$ near $\eps\approx0$, we have
\beq
\bal
\bar\alpha_2&=\frac{1}{\eps}\sqrt{\frac{\omega_0}{\omega_1}}\frac{\beta}{\sqrt{\omega_1\omega_2-\beta^2}}-\lp\frac{1}{\eps}+\frac{1}{2}+\op\lp\eps\rp\rp\lp1-\sqrt{\frac{\omega_0}{\omega_1}}\rp+\\
&=\frac{1}{\eps}\left[\sqrt{\frac{\omega_0}{\omega_1}}\lp1+\frac{\beta}{\sqrt{\omega_1\omega_2-\beta^2}}\rp-1\right]-\frac{1}{2}\lp1-\sqrt{\frac{\omega_0}{\omega_1}}\rp+\op(\eps)~.
\eal
\eeq
We therefore find that the circuit depth for $U_4$ is
\beq
\mathcal{D}\lp U_4\rp=
\frac{1}{2\eps}\log\lp\frac{\omega_1\omega_2-\beta^2}{\omega_0^2}\rp
+\left|\frac{1}{\eps}\left[\sqrt{\frac{\omega_0}{\omega_1}}\lp1+\frac{\beta}{\sqrt{\omega_1\omega_2-\beta^2}}\rp-1\right]
-\frac{1}{2}\lp1-\sqrt{\frac{\omega_0}{\omega_1}}\rp+\op(\eps)\right|~.
\eeq
Here, as above, we assume $\omega_0<\omega_1$.

In general, we can describe the form of the circuit depth as being an overall factor of $1/\epsilon$ followed by a coefficient determined by the various physical parameters characterizing the target state and the reference state. More generally, the circuit depth might be given by an expansion in $\epsilon$, beginning with a $1/\epsilon$ term followed by a finite term and then potentially terms involving positive powers of $\epsilon$. However, since $\epsilon\ll1$, 
determining the complexity essentially requires finding the circuit which minimizes the coefficient of the leading $1/\epsilon$ term. 

For comparison to the results of the geometric approach in the main text, it is useful to express the present results in terms of the normal-mode frequencies via eq.~\reef{eq:omega12pm}. If we focus our attention on the first circuit $U_1$ in eq.~\reef{exam1}, the exponents given in eqs.~\reef{exp1} and \reef{exp2} become
\beq
\alpha_1=\frac{1}{2\eps}\log\!\lp\frac{\tom_++\tom_-}{2\omega_0}\rp~,
\quad\alpha_2=-\frac{1}{\eps}\sqrt{\frac{\omega_0}{\tom_++\tom_-}}\frac{\tom_--\tom_+}{\sqrt{2\tom_+\tom_-}}~,\quad
\alpha_3=\frac{1}{2\eps}\log\!\lp\frac{2\,\tom_+\tom_-}{\omega_0\,(\tom_++\tom_-)}\rp~.
\label{exam1a}
\eeq
As was alluded to above, to proceed further we must decide on the value of the reference frequency $\omega_0$ relative to the normal-mode frequencies. Given the discussion in section \ref{sec:N}, there are two natural hierarchies to consider: (i)  $\tom_+<\tom_-<\omega_0$ or (ii) $\omega_0<\tom_+<\tom_-$.\footnote{Implicitly, we chose the second hierarchy above in presenting our results in eqs.~\reef{eq:Deg1Phys}, \reef{horse0}, and \reef{horse}.} Of course, the ordering of the normal-mode frequencies is fixed and we are really only choosing $\omega_0$ here. In particular, in the first (second) hierarchy, $\omega_0$ is a UV (IR) frequency  larger (smaller) than any physical frequency in the coupled oscillator problem. Note that in the first case, all three exponents are negative, while in the second case, $\alpha_1,\alpha_3>0$ and $\alpha_2<0$. Evaluating $\mathcal{D}(U_1)=|\alpha_1|+|\alpha_2|+|\alpha_3|$ in these two cases yields:
\beq
\mathcal{D}(U_1)=\frac{1}{\eps}\sqrt{\frac{\omega_0}{2 \tom_+}+ \frac{\omega_0}{2 \tom_-}}\,\frac{\tom_-\!-\tom_+}{{\tom_+\!+\tom_-}}+\frac{1}{2\,\eps}\times\left\lbrace
\begin{matrix}
&
\log\frac{\omega_0^2}{\tom_+\tom_-}\quad{\rm for}\ \ \omega_0>\tom_+,\tom_-~,\\
&
\log\frac{\tom_+\tom_-}{\omega_0^2} \quad{\rm for}\ \ \omega_0<\tom_+,\tom_-~.\\
\end{matrix} 
\right.
\label{jjj}
\eeq
Recall that $\tom_->\tom_+$ from eq.~\reef{qm3}. We may now compare this result with those derived using the geometric approach of section \ref{sec:geometry}. In particular, if we recall the $F_1$ measure given in eq.~\eqref{eq:Fmetrics}, the complexity would be given by
\beq
{\cal C}=\frac12\left|\log\frac{\tom_+}{\omega_0}\right|+\frac12\left|\log\frac{\tom_-}{\omega_0}\right|=\left\lbrace
\begin{matrix}
&
\frac12\log\frac{\omega_0^2}{\tom_+\tom_-}\quad{\rm for}\ \ \omega_0>\tom_+,\tom_-~,\\
&
\frac12\log\frac{\tom_+\tom_-}{\omega_0^2} \quad{\rm for}\ \ \omega_0<\tom_+,\tom_-~.\\
\end{matrix} 
\right.
\label{formula1}
\eeq
Furthermore, recall that we should compare this with the coefficient of the $1/\epsilon$ factor in the discrete calculations (see footnote \ref{foot55}). Hence we see the second contribution in eq.~\reef{jjj} precisely matches the complexity above. However, there is an additional positive term in $\mathcal{D}(U_1)$, and therefore we see that -- at least by the $F_1$ measure -- $U_1$ is not the optimal circuit. 

We can also describe this circuit as a trajectory in the language of the path-ordered exponentials \reef{eq:pathPsi}. In this case, $U_1$ as given in eq.~\reef{exam1} becomes
\beqa
0\le s\le \frac{|\alpha_1|}{\mathcal{D}(U_1)}\qquad :&&\quad Y^{11}= \mathcal{D}(U_1)\,,\ Y^{22}=Y^{12}=Y^{21}=0~,
\nonumber\\
\frac{|\alpha_1|}{\mathcal{D}(U_1)}\le s\le \frac{|\alpha_1|+|\alpha_2|}{\mathcal{D}(U_1)}\ :&&\quad Y^{11}= Y^{22}= Y^{12}=0\,,\ Y^{21}=\mathcal{D}(U_1)~,
\label{lines3}\\
\frac{|\alpha_1|+|\alpha_2|}{\mathcal{D}(U_1)}\le s\le 1\qquad:&&
\quad Y^{11}= 0\,,\ Y^{22}=\mathcal{D}(U_1)\,,\ Y^{12}=Y^{21}=0~.
\nonumber
\eeqa
This form makes clear that the circuit consists of three separate ``straight'' segments, and so $U_1$ does not correspond to a geodesic path or an optimal circuit.

\section{Killing vectors and more geometry}\label{sec:appxKilling}

Inspecting the metric in eq.~\reef{metric1}, we can see three obvious Killing coordinates: $y,\, \tau,\, \theta$. When the penalty factors were introduced in section \ref{sec:penalty}, we found that this is reduced to two Killing coordinates, $y$ and $z=(\theta-\tau)/2$, in the geometry described by eqs.~\reef{penalty1} or \reef{penalty2}. However, by construction, all of these metrics are right-invariant, and hence the corresponding geometries must have one Killing vector for each generator \reef{Msimple}, namely, \textit{four}.\footnote{We thank Lucas Hackl for discussions on this point.} Furthermore, as we will see below, the structure of these Killing vectors will be completely independent of the particular choice of $G_{IJ}$ appearing in the metric (assuming it is a constant matrix), but rather is determined by the structure of eq.~\reef{eq:metricRightInv}.

One way to think of a Killing vector $k^i$ is as providing a coordinate transformation
\beq
x^i \to x^i+\veps\,k^i
\label{kill1}
\eeq
which leaves the geometry or line element invariant. (Note $\veps$ is just an infinitesimal parameter.) For example, eq.~\reef{metric1} is certainly invariant under $\delta \tau = \veps$ and so we write the corresponding Killing vector as
$k^i\partial_i = \partial_\tau$ or $k^i =\delta^i_\tau$. 

So let us identify the coordinate transformations which generally leave eq.~\reef{eq:metricRightInv} invariant. For a general coordinate shift in eq.~\reef{Umatrix}, we have
\beq
\delta U = \partial_i U\,\delta x^i \,.
\label{shift1}
\eeq
As long as $G_{IJ}$ is a constant matrix, all of the coordinate dependence is hidden in the one-forms $\tr\lp \dd U(s)\,U^{-1}(s)M^T_I\rp$, \cf eq.~\reef{penalty1}. However, it is clear that these expressions are invariant if we right-multiply $U$ by a global $\GLtwo$ transformation. Hence let us make the infinitesimal transformation: $U\to U\, \exp[\veps^I M_I]$, where the $\veps^I$ are (infinitesimal) constants. To leading order in these parameters, this reduces to
\beq
\delta U = U\,  M_I \veps^I~.
\label{shift2}
\eeq
Equating eqs.~\reef{shift1} and \reef{shift2}, we have
\beq
U\,M_I\veps^I=\pd_iU\delta x^i\,
\implies
\veps^I=\tr\lp U^{-1}\pd_i U\,M_I^T\rp\,\delta x^i~,
\label{shift3}
\eeq
where we have assumed that we are working with the orthogonal basis of generators satisfying $\tr\lp M_I M_J^T\rp=\delta_{IJ}$, \cf eq.~\reef{Msimple}. We now observe that, since the argument of the trace contains two free indices, we may view this object as a $4\times4$ matrix, which we can then invert to obtain
\beq
\delta x^i=\left[\tr\lp U^{-1}\pd_i U\,M_I^T\rp\right]^{-1}\veps^I
=\lp k_I\rp^i\veps^I~.
\eeq
Thus we obtain four independent Killing vectors $k_I=\lp k_I\rp^i\partial_i$.

Given our basis of generators in eq.~\reef{Msimple} and our parametrization of the circuit space in eq.~\eqref{Umatrix}, we can easily compute $\lp k_I\rp^i$.  We can then identify the Killing vectors by simply reading off this matrix row-by-row:
\beq
\bal
k_1=&\frac{1}{2}\pd_y-\frac12\sin (2z)\pd_\rho-\frac{\cos (2 z)}{2\sinh(2 \rho )}  \pd_x-\frac{\cosh(2\rho)}{2\sinh(2 \rho )}\cos (2 z) \pd_z~,\\
k_2=&\frac{1}{2} \cos (2 z)\pd_\rho-\frac{\sin (2 z)}{2\sinh(2 \rho )} \pd_x+\frac{1}{2} \left(1-\frac{\cosh (2\rho )}{\sinh(2\rho)}\, \sin (2z) \right)\pd_z~,\\
k_3=&\frac{1}{2} \cos (2 z)\pd_\rho-\frac{\sin (2 z)}{2\sinh(2 \rho )}\pd_x-\frac{1}{2} \left(1+\frac{\cosh (2\rho )}{\sinh(2\rho)}\, \sin (2z) \right)\pd_z~,\\
k_4=&\frac{1}{2}\pd_y+\frac12\sin (2z)\pd_\rho+\frac{\cos (2 z)}{2\sinh(2 \rho )}  \pd_x+\frac{\cosh(2\rho)}{2\sinh(2 \rho )}\cos (2 z) \pd_z~,
\eal
\label{Killed}
\eeq
where we are using the pseudo-lightcone coordinates of eq.~\eqref{lcCoords}, with $\theta=x+z$, $\tau=x-z$.
One can explicitly verify that these indeed satisfy the Killing equations, 
\beq
0=\nabla_i \lp k_I\rp_j+\nabla_j\lp k_I\rp_i
=\lp g_{j\ell}\nabla_i+g_{i\ell}\nabla_j\rp k_I^\ell~,
\eeq
for either of the metrics in eqs.~\reef{metric2} or \reef{penalty2}. However, it is clear that eq.~\reef{Killed} does not organize the Killing vectors in the simplest way, so we define:
\beq
\bal
\hat k_1\equiv&\quad \ k_1+k_4 = \pd_y~,\\
\hat k_2\equiv&-k_1+k_4=\sin (2z)\pd_\rho+\frac{\cos (2 z)}{\sinh(2 \rho )}  \pd_x+\frac{\cosh(2\rho)}{\sinh(2 \rho )}\cos (2 z) \pd_z~,\\
\hat k_3\equiv&\quad\ k_2+k_3 = \cos (2 z)\pd_\rho-\frac{\sin (2 z)}{\sinh(2 \rho )}\pd_x-\frac{\cosh (2\rho )}{\sinh(2\rho)}\, \sin (2z) \pd_z~,\\
\hat k_4\equiv&\quad \ k_2-k_3 =\pd_z~.\\
\eal
\label{Killed2}
\eeq
However, a simple inspection of the first metric \eqref{metric2} reveals that $\pd_x$ is also an independent Killing vector, hence:
\beq
\hat k_5\equiv\pd_x~.\label{Killed5}
\eeq
This is an accidental symmetry that emerges with the choice $G_{IJ}=\delta_{IJ}$.\footnote{We thank Lucas Hackl for discussions on the Killing symmetries.} However, as noted above, the four Killing vectors in eq.~\reef{Killed2} apply for any (constant) choice of $G_{IJ}$.

Of course, the existence of the above Killing vectors implies that there are an equal number of conserved momenta or charges which distinguish the geodesics, $c_I\equiv (\hat k_I)^i\,g_{ij}\,\dot x^j$. We make use of these momenta in solving for the optimal circuits in sections \ref{geod} and \ref{sec:penalty}.

\subsection*{AdS$_3$ geometry:} 

In section \ref{geod}, we noted the appearance of a three-dimensional anti-de Sitter geometry in discussing the parametrization of $U\in\GLtwo= \mathbb{R}\times\mathrm{SL}(2,\mathbb{R})$.  Of course, the appearance of AdS$_3$ is natural since it is the universal cover of the $\mathrm{SL}(2,\mathbb{R})$ subgroup. Here we would like to show how the AdS$_3$ geometry can be realized using the formalism introduced in section \ref{sec:geometry}. In particular, we consider the geometry that results from the choice 
\beq
G_{IJ}=\begin{pmatrix}1 & 0&0&0\\
0& 1&0&0\\
0 & 0&0&1\\
0 & 0&1&0\end{pmatrix}
=\eta_{IJ}\,.
\label{etaij}
\eeq
We have designated $G_{IJ}=\eta_{IJ}$ because the $I=3=\{12\}$ and $I=4=\{21\}$ directions are null and hence the metric has a Minkowski signature. With this choice, eq.~\reef{metric1} is replaced with
\beq
\bal
\dd s^2&=\eta_{IJ} \, \tr\lp \dd U(s)\,U^{-1}(s)M^T_I\rp\,\tr\lp \dd U(s)\,U^{-1}(s)M^T_J\rp\\
&=2\dd y^2+2\dd\rho^2-2\cosh^2\!\rho\ \dd\tau^2+2\sinh^2\!\rho\ \dd\theta^2~.
\eal\label{adds}
\eeq
Hence we have produced precisely the AdS$_{3}\times\mathbb{R}$ geometry anticipated in section \ref{geod}. Eq.~\reef{adds} describes the natural group invariant metric for $\GLtwo$, \ie the left- and right-invariant metric, whereas the Euclidean metric \reef{metric1} is a less symmetric metric with only right-invariance. However, the Lorentzian signature is undesirable for the problem of circuit complexity, since pieces of the circuit that correspond to null-geodesics have zero length, \ie zero cost. This would allow the construction of arbitrarily low-complexity circuits simply by deforming the circuit along the null directions.\footnote{Of course, moving in a timelike direction also yields a negative cost.}

\subsection*{Alternate basis of generators:} 

The basis of matrix generators in eq.~\reef{Msimple} is natural in the sense that it straightforwardly extends from the problem of complexity in the case of two coupled harmonic oscillators to the case of $N$ coupled oscillators. However, this is not the most convenient basis for certain calculations in section \ref{sec:geometry}. Hence for the interested reader, we describe here an alternate basis of generators which simplifies some of the calculations. In particular, consider the Pauli-like basis:
\beq
\bal
\widehat M_{1}&=\frac{1}{\sqrt{2}}\begin{pmatrix}1 & 0 \\ 0 & 1\end{pmatrix}=\frac{1}{\sqrt{2}}\,\mathbb{1}~,\qquad
\ \ \widehat M_{2}=\frac{1}{\sqrt{2}}\begin{pmatrix}1 & 0 \\ 0 & -1\end{pmatrix}=\frac{1}{\sqrt{2}}\,\sigma_3~, 
\\
\widehat M_{3}&=\frac{1}{\sqrt{2}}\begin{pmatrix}0 & 1 \\ 1 & 0\end{pmatrix}=\frac{1}{\sqrt{2}}\,\sigma_1~,\qquad
\widehat M_{4}=\frac{1}{\sqrt{2}}\begin{pmatrix}0 & 1 \\ -1 & 0\end{pmatrix}=-\frac{i}{\sqrt{2}}\,\sigma_2
~.
\eal\label{eq:Pauli}
\eeq
The normalization of the generators is still given by $\tr\!\left(\widehat M_I \widehat M_J^T\right)=\delta_{IJ}$. In fact, the new generators are easily related to the original generators in eq.~\reef{Msimple} by an orthogonal transformation: $\widehat M_I = R_I{}^J\, M_J$ with $R_I{}^J\in O(2)\times O(2)\in O(4)$. In this new basis, the $\widehat M_{2,3,4}$ generators naturally form the $\frak{sl}(2,\mathbb{R})$ subalgebra, with
\beq
[\wM_2,\wM_3]=\sqrt{2}\,\wM_4~,\;\;\;
[\wM_2,\wM_4]=\sqrt{2}\,\wM_3~,\;\;\;
[\wM_3,\wM_4]=-\sqrt{2}\,\wM_2~,
\eeq
while $\widehat M_1$ describes the remaining fibre over $\mathbb{R}$ in the $\GLtwo$ group.

With this new basis, the Killing vectors which emerge from the right-invariance of the metric naturally appear in the form given in eq.~\reef{Killed2}. 
One can easily show that working with these new generators, the metrics appearing in eqs.~\reef{metric1} and \reef{penalty1} are unchanged, \ie the corresponding $G_{IJ}$ are left unchanged by the rotation $R_I{}^J$ introduced above. Additionally, the AdS$_3$ geometry in eq.~\reef{adds} now results from the choice $G_{IJ}=\eta_{IJ}=\mathrm{diag}(1,1,1,-1)$.

\section{Normal-mode frequencies $\tom_k$} \label{sec:omegaDeriv}

The derivation of the normal-mode frequencies in eq.~\eqref{eq:eigenfreq} -- or eq.~\reef{freakd} for a lattice of coupled oscillators -- is straightforward, and can be found in a number of different sources, \eg any elementary condensed matter textbook. For completeness, we briefly review the result \eqref{eq:eigenfreq} for the periodic one-dimensional lattice discussed in section \ref{sec:N}. Essentially, we need only apply the inverse Fourier transform 
\beq
 x_a\equiv \frac{1}{\sqrt{N}}\sum_{k=0}^{N-1}\mathrm{exp}\lp \frac{2\pi i\,k}{N}\,a\rp\tilde x_k~,
\label{invFourier}
\eeq
to re-express the Hamiltonian \reef{qm88} in terms of the normal modes, \cf \reef{qm288}. In particular, we focus on the potential
\beq
V=\frac12\sum_{a=0}^{N-1}\Big[\omega^2x_a^2+\Omega^2\lp x_a-x_{a+1}\rp^2\Big]\,.
\label{eq:Vphys}
\eeq
Considering the second term involving the coupling between the oscillators, we find
\begin{equation*}
\bal
\Omega^2\sum_{a=0}^{N-1}\lp x_a-x_{a+1}\rp^2
&=\Omega^2\sum_{a=0}^{N-1}\left[\frac{1}{\sqrt{N}}\sum_{k=0}^{N-1}\mathrm{exp}\lp\frac{2\pi ik\,a}{N}
\rp \tilde x_k\lp1-\exp\lp\frac{2\pi ik}{N}\rp\rp\right]^2\\
&=\frac{\Omega^2}{N}\sum_{a,k,k'}\exp\lp\frac{2\pi i(k+k')a}{N}\rp \tilde x_k\tilde x_{k'}\lp1-\exp\lp\frac{2\pi ik}{N}\rp\rp \lp1-\exp\lp\frac{2\pi ik'}{N}\rp\rp\\
&=\Omega^2\sum_k\tilde x_k\tilde x_{-k}\lp1-\mathrm{exp}\lp\frac{2\pi ik}{N}\rp\rp\lp1-\mathrm{exp}\lp\frac{-2\pi ik}{N}\rp\rp\\
&=2\Omega^2\sum_k\tilde x_k\tilde x_{-k}\lp1-\cos\lp\frac{2\pi k}{N}\rp\rp
=4\Omega^2\sum_k|\tilde x_k|^2\,\sin^2\!\frac{\pi k}{N}~,
\eal
\end{equation*}
where in going to the third line we applied the normalization condition \eqref{eq:FourierNorm}, and in the last step we used $\tilde x_k\tilde x_{-k}=\tilde x_k\tilde x^\dagger{}_{\!k}$. Here all sums run from $0$ to $N\!-\!1$. The Fourier transform of the first term in the potential \eqref{eq:Vphys} is trivial, and thus we find
\beq
V=\frac12\sum_{k=0}^{N-1}\left[ \omega^2+4\Omega^2\,\sin^2\!\frac{\pi k}{N}\right]\, |\tilde x_k|^2
=\frac12 \sum_{k=0}^{N-1}\tom_k^2\,|\tilde x_k|^2\,.
\eeq
Hence we have identified the desired normal-mode frequencies,
\beq
\tom_{k}^2=\omega^2+4\Omega^2\,\sin^2\!\frac{\pi k}{N}~,
\label{eigenfreq2}
\eeq
\cf \eqref{eq:eigenfreq}. If instead we were examining a $d$-dimensional free scalar field, the lattice would be extended to $d\!-\!1$ (spatial) dimensions, whereupon the corresponding normal-mode frequencies become
\beq
\tom_{\vec k}^2=\omega^2+4\Omega^2\,\sum_{i=1}^{d-1}\sin^2\!\frac{\pi k_i}{N}~,
\label{eigenfreq3}
\eeq
where $k_i$ are the components of the momentum vector $\vec k=(k_1,k_2,\cdots,k_{d-1})$. Implicitly, we have assumed here that the lattice is square with periodic boundary conditions in each direction.

\section{Change of basis} \label{nmsub}

In this appendix, we would like to extend the discussion around eq.~\reef{bark2} describing the change of bases for the case of two coupled oscillators to the analogous transformation for a lattice of oscillators. In particular, we will focus on the case of a one-dimensional lattice of $N$ oscillators, although it is straightforward to extend the discussion to a lattice extending in $d\!-\!1$ (spatial) dimensions.

This transformation is particularly relevant in section \ref{sec:N}, where we presented a tentative argument that the metric on the normal-mode subspace is flat, \ie
\beq
\dd s^2_\mt{n-m}=|\dd \tilde y_0|^2+|\dd \tilde y_1|^2+\cdots+|\dd \tilde y_{\ssc N-1}|^2~,
\label{metric6xx}
\eeq
\cf \eqref{metric6}. However, we also noted that, at the time, this conclusion was somewhat premature, since implicitly we applied eq.~\reef{metric5}, which defines the metric in the position basis, to a calculation with the diagonal circuit \reef{raffle4} written in the normal-mode basis. That is, in eq.~\reef{metric5}, the indices $I,\,J$ run over pairs of position labels $(ab)$, and implicitly the generators act on Gaussian wave functions written in terms of coordinates $x_a$. In contrast, in eq.~\reef{raffle4}, we would write $\tilde M_\mt{n-m}=\tilde Y^{\tilde I} \tilde M_{\tilde I}$ where the tilde on the index $\tilde I$ indicates that it runs over pairs of momentum labels $(k\ell)$, and the tilde on $M$ indicates that these generators act on Gaussian wave functions written in terms of the normal coordinates $\tilde x_k$. In other words, in eq.~\reef{raffle4}, where we are restricting our attention to the normal-mode subspace, we are considering the diagonal generators $\tilde Y^{k\ell}=\delta^{k\ell}\, \tilde y_k$. 

Hence to show that the result \reef{metric6xx} is correct, we must take care to translate between the two bases of generators discussed above. As in eqs.~\reef{nmU} and \reef{straight9}, we can transform from generators acting in the normal-mode basis to those in the position basis via\footnote{Note that this transformation removes the tilde from $M$ but not from the index. For example, the new generator $M_{kk}$ acts on Gaussian wave functions written in terms of the oscillator position coordinates $x_a$, but still has the effect of scaling the $k$th normal mode $\tilde x_k$.} 
\beq
M_{\tilde I} = \RN\!{}^\dagger\, \tilde M_{\tilde I} \,\RN\,.
\label{lot0}
\eeq
Implicitly, the normal-mode generators $\tilde M_{\tilde I}$ have the same form as that given in eq.~\reef{matrix}, namely
\beq
\big[\tilde M_{k\ell}\big]{}_{pq}=\delta_{kp}\delta_{\ell q}\,,
\label{lot1}
\eeq
where we have denoted $\tilde I=(k\ell)$ with momentum labels $k,\,\ell$. Similarly $p,\,q$ are the row and column indices, respectively, of the $N\times N$ matrix, which also take values as momentum labels (since the generator acts in the normal-mode space). Now let us combine these two equations to write\footnote{Note that the complex conjugation appears on the first factor in $\wRN=\RN^*\otimes \RN$ because our convention is that written in terms of the normal modes, the Gaussian wave functions involve both $\tilde x_k$ and $\tilde x_k^\dagger$, \eg the appearance of $|\tilde x_k|^2$ in eq.~\reef{targetk}.}
\beqa
\left[M_{k\ell}\right]{}_{ab} &=& \big[\RN\!{}^\dagger \big]{}_{ap} \big[\tilde M_{k\ell}\big]{}_{pq} \big[\RN\big]{}_{qb}
=\big[\RN\!{}^\dagger \big]{}_{a k} \, \big[\RN\big]{}_{\ell b}
\nonumber\\
&=& \big[\RN\!{}^\dagger \big]{}_{c k} \, \big[\RN\big]{}_{\ell d}\ \big[M_{cd}\big]{}_{ab}
\nonumber\\[.9ex]
\implies&& M_{\tilde I} = [\wRN]{}_{\tilde I  J}\, M_{J} \qquad
{\rm with}\ \ \wRN=\RN^*\otimes \RN\,.
\label{lot2}
\eeqa
In going from the second to third line, we used eq.~\reef{matrix} and identified $\tilde I=(k\ell)$ and $J=(cd)$. This equation generalizes eq.~\reef{bark2} from $\GLtwo$ in the previous section to the case of $\GLN$ studied here. Furthermore, given the properties of $\RN$, one can easily see that the matrix $\wRN$ is a unitary matrix. Hence we can invert the transformation in eq.~\reef{lot2} to write $M_{I} =[\wRN\!{}^\dagger]_{I\tilde J}\, M_{\tilde J}$. 

Similarly, we can invert the transformation in eq.~\reef{lot0}, \ie transform from generators acting in the position basis to the normal-mode basis via
\beq
\tilde M_{ I}=\RN\,M_{I}\,\RN\!{}^\dagger   \,,
\label{lot0a}
\eeq
and combine this expression with eq.~\reef{matrix}, $[ M_{ab}]{}_{cd} =\delta_{ac}\delta_{bd}$, to write
\beqa
\left[\tilde M_{ab}\right]{}_{k\ell} &=& \big[\RN\big]{}_{kc} \big[M_{ab}\big]{}_{cd} \big[\RN\!{}^\dagger \big]{}_{d\ell}
=\big[\RN \big]{}_{ ka} \, \big[\RN\!{}^\dagger\big]{}_{b\ell }
\nonumber\\
&=& \big[\RN\big]{}_{pa} \, \big[\RN\!{}^\dagger \big]{}_{bq}\, \big[\tilde M_{pq}\big]{}_{k\ell}
\nonumber\\[.9ex]
\implies&& \tilde M_{I} = [\wRN\!{}^\dagger]{}_{I  \tilde J}\, \tilde M_{\tilde J} \qquad
{\rm with}\ \ \wRN\!{}^\dagger=\RN^T\otimes \RN\!{}^\dagger\,.
\label{lot2a}
\eeqa
In going from the second to third line, we used eq.~\reef{lot1} and identified $I=(ab)$ and $\tilde J=(pq)$. As before, we can easily invert the transformation in eq.~\reef{lot2a} to write $\tilde M_{\tilde I} =[\wRN]_{\tilde I J}\, \tilde M_{J}$. As our notation indicates, $\wRN$ is precisely the unitary matrix appearing in eq.~\reef{lot2}, and hence it also plays a role in transforming the generators acting in the normal-mode space. 

Hence by using the special structure of the generators in eqs.~\reef{matrix} and \reef{lot1}, we have re-organized the  transformation acting on the matrix indices in eqs.~\reef{lot0} and \reef{lot0a} to a transformation acting on the generator labels in eqs.~\reef{lot2} and \reef{lot2a}, respectively.

With these tools in hand, let us consider re-expressing the cost function \reef{cost5} or the metric \reef{metric5} in terms of the normal-mode basis, using eq.~\reef{lot0}. Here, we show the calculation for the metric; the transformation of the cost function follows in a similar manner. Beginning with the differential $\dd Y^I=\tr\lp \dd U\,U^{-1}\, M^\dagger_I\rp$ defined in eq.~\reef{metric5}, we transform the circuit to the normal-mode space via $U=\RN\!{}^\dagger \,\tilde U\RN$, which yields 
\beq
\dd Y^I=\tr\lp \dd\tilde U \,\tilde U^{-1}\,\RN\, M^\dagger_I \,\RN\!{}^\dagger\rp=\tr\lp \dd\tilde U \,\tilde U^{-1}\,\tilde M^\dagger_I \rp=[\wRN]_{\tilde I\, I}\ \dd\tilde Y^{\tilde I}
\label{tran1}
\eeq
where $\dd\tilde Y^{\tilde I}=\tr\lp \dd\tilde U \,\tilde U^{-1}\,\tilde M^\dagger_{\tilde I} \rp$,
and we have employed eqs.~\reef{lot0a} and \reef{lot2a} in the second and third equalities. Hence the metric \reef{metric5} transforms as
\beq
\dd s^2=[\wRN]_{\tilde I\, I}\,\delta_{IJ}\, [\wRN\!{}^\dagger]_{J\,\tilde J}\, \dd\tilde Y^{\tilde I}\,
(\dd\tilde Y^{\tilde J})^* =\delta_{\tilde I \tilde J}\, \dd\tilde Y^{\tilde I}\,
(\dd\tilde Y^{\tilde J})^* \,.
\label{tran2}
\eeq
Note that here we are using the fact that $\wRN$ is a unitary matrix. Thus we have found that the metric takes precisely the same form whether expressed in terms of the oscillator position space or the normal-mode space.\footnote{The fact that this transformation preserves the cost function essentially follows from the Plancherel theorem, which states that the Fourier transform preserves the $L^2$ norm. We thank Adri\'an Franco-Rubio for a discussion on this point.} Of course, the same is true of the cost function \reef{cost5}, \ie it can also be written as
\beq
\mathcal{D}(U)=\int_0^1\dd s\sqrt{\delta_{\tilde I\tilde J}\,Y^{\tilde I}(s)\,( Y^{\tilde J}(s))^*}\,, \qquad{\rm where}\ \ Y^{\tilde I}(s)=\tr\lp\pd_s\tilde U(s)\,\tilde U^{-1}(s)\tilde M^\dagger_{\tilde I} \rp\,.
\label{cost5a}
\eeq

Note that this transformation is slightly different than that expressed in eq.~\reef{metric1a} for the metric for two coupled oscillators. In the latter case, we are considering the metric to still be in the position basis but evaluated with a different basis of generators. The same invariance holds here for a lattice of oscillators, as can be seen by applying eq.~\reef{lot2} directly to the metric \reef{metric5} to produce
\beq
\dd s^2=\delta_{\tilde I \tilde J}\ \dd Y^{\tilde I}\,
\big(\dd Y^{\tilde J}\big)^*~, \qquad{\rm where}\qquad \dd Y^{\tilde I}=\tr\lp \dd U\,U^{-1}\, M^\dagger_{\tilde I}\rp\,.
\label{tran3}
\eeq
Of course, the same change of basis could also be performed with eq.~\reef{lot2a} when working in the normal-mode space.

\subsection{General cost functions} \label{base8}

In eq.~\reef{Dalpha}, the $\kappa$ cost functions were defined with a sum over the components of the velocity $Y^{\tilde I}$ in the normal-mode basis. Here we would like to apply the techniques developed above to examine the differences that arise from using the original oscillator position basis. That is, we could equally well define cost functions with
\beq
\mathcal{D}_\kappa=\int_0^1\dd s\sum
\left|Y^{I}(s)\right|^\kappa~.\label{DalphaP}
\eeq
In the discussion of the $F_2$ cost function in the previous section, we found that this change of basis had no effect on the complexity; but here we will find that, in fact, the complexity is not basis independent. As a simple example, let us consider the case of two coupled oscillators for which the optimal circuit $U_0(s)$ appears in eq.~\reef{solver5}, for which the velocity components in the position basis become
\beq
Y^{11}=Y^{22}=y_1\,,\qquad
Y^{12}=Y^{21}=-\rho_1\,.
\label{veloc}
\eeq
These two factors are written in terms of the normal-mode frequencies in eq.~\reef{fin2}, but we can re-express these results as
\beq
y_1=\frac14\left(\log\frac{\tom_-}{\omega_0}+\log\frac{\tom_+}{\omega_0}\right)~,\qquad
\rho_1=\frac14\left(\log\frac{\tom_-}{\omega_0}-\log\frac{\tom_+}{\omega_0}\right)~.
\label{fin2a}
\eeq
Recall that $\tom_->\tom_+$, but in the following, we also assume that $\tom_\pm>\omega_0$, which ensures that both $y_1$ and $\rho_1$ are positive quantities. Now we evaluate the cost of $U_0$ using eq.~\reef{DalphaP} for a few values of $\kappa$,\footnote{In the case that $\omega_0>\tom_\pm$, one should replace
$\tom_\pm/\omega_0\to\omega_0/\tom_\mp$ in these formulae. Note this substitution only really changes the results for odd $\kappa$.}
\beq
\bal
\mathcal{D}_\kappa(U_0)&=2y_1^\kappa+2\rho_1^\kappa
=\begin{cases}
\log\frac{\tom_-}{\omega_0}&\mathrm{for }\;\;\kappa=1~,\\[1ex]
\frac14\left(\log^2\frac{\tom_-}{\omega_0}+\log^2\frac{\tom_+}{\omega_0}\right)&\mathrm{for }\;\;\kappa=2~,\\[1ex]
\frac1{16}\left(\log^3\frac{\tom_-}{\omega_0}+3\log\frac{\tom_-}{\omega_0}\,\log^2\frac{\tom_+}{\omega_0}\right)&\mathrm{for }\;\;\kappa=3~,\\[1ex]
\frac1{64}\left(\log^4\frac{\tom_-}{\omega_0}+\log^4\frac{\tom_+}{\omega_0}+6\log^2\frac{\tom_-}{\omega_0}\,\log^2\frac{\tom_+}{\omega_0}\right)&\mathrm{for }\;\;\kappa=4~.
\end{cases}
\eal\label{trials}
\eeq
Hence we see that it is only for $\kappa=2$ that we reproduce the cost found using eq.~\reef{Dalpha} in the normal-mode basis,
$\widetilde{\mathcal{D}}_\kappa(U_0)\simeq \log^\kappa(\tom_-/\omega_0)+ \log^\kappa(\tom_+/\omega_0)$.

These differences in the cost can be understood using the approach developed to implement a change of basis for a lattice of oscillators in section \ref{sec:N}. In particular, transforming from the position basis to the normal-mode basis can be described in terms of the unitary matrix $\wRN$ defined in eq.~\reef{lot2}. Given the definition of the velocity components in eq.~\reef{cost5}, we then have
\beq
Y^{\tilde I}=\big[\wRN^*\big]{}_{\tilde I J}\,Y^J\,,
\label{chan0}
\eeq
or, inverting this expression, $Y^{I}=\big[\wRN^T\big]_{I\tilde  J}\,Y^{\tilde J}$. Furthermore, recall that the quadratic construction $\delta_{IJ}\,Y^I (Y^J)^*= \delta_{\tilde I\tilde J}\,Y^{\tilde I} (Y^{\tilde J})^*$ is invariant under this change of basis. Therefore the cost evaluated with the $F_2$ or $\kappa=2$ cost functions are invariant as well. 

However, this discussion also makes clear that if we include penalty factors, then these quadratic cost functions are no longer invariant. That is, the penalty factors introduce a more general metric $G_{IJ}$, which transforms nontrivially under the change of basis, \ie 
\beq
G_{\tilde I\tilde J}= \big[\wRN\big]{}_{\tilde J J}\,G_{IJ}\,\big[\wRN^\dagger\big]{}_{I\tilde I}\,,
\label{chan1}
\eeq
where we assumed symmetry of the metric $G_{IJ}=G_{JI}$.

This also suggests how we should treat the more general $\kappa$ cost functions. We should generalize eq.~\reef{DalphaP} to allow for general penalty factors by writing
\beq
\mathcal{D}_\kappa=\int_0^1\dd s\!\!\sum_{I_1,I_2,\cdots,I_\kappa}
G_{I_1I_2\cdots I_\kappa}\,
|Y^{I_1}(s)|\, |Y^{I_2}(s)|\cdots |Y^{I_\kappa}(s)|~,\label{DalphaPP}
\eeq
where $G_{I_1I_2\cdots I_\kappa}$ is a symmetric tensor with $\kappa$ indices.  In eq.~\reef{Dalpha}, we are implicitly considering simple ``penalty'' tensors of the form
\beqa
G_{I_1I_2\cdots I_\kappa}&=&\delta_{I_1I_2}\,\delta_{I_2I_3}\cdots
\delta_{I_{\kappa-1}\,I_\kappa} \qquad{\rm for}\ \ \kappa\ge2\,,
\nonumber\\
G_I&=&1\qquad\qquad\qquad\qquad\qquad{\rm for}\ \ \kappa=1\,.
\label{tens2}
\eeqa
In general, it is clear that the unitary transformation will not leave these penalty tensors (or more general choices) invariant. This simply reflects the fact that in choosing different gates, we are treating different gates as fundamental and that in general, we expect the results for the complexity to depend on the choice of the elementary gate set.

Of course, this does not mean that the complexity must be evaluated in one particular basis. However, if the cost function is fixed with a certain choice of basis, then changing the basis requires that we properly transform the cost function to the new basis. To gain a better understanding of this situation, let us investigate the case of $\kappa=1$ in more detail. In addition to the simplicity of this case, recall that this was also the cost function favoured in the comparison to holographic complexity in section \ref{sec:continuum}.

Let us begin with the case of $N=2$, in which case the transformation matrix $\wR=\wR_2$ takes the simple form given in eq.~\reef{bark2}. For $\kappa=1$, the penalty tensor \reef{tens2} becomes the four component vector
\beq
G_I=(1,1,1,1)\,,
\label{tens3}
\eeq
which is actually an eigenvector of $\wR$. Hence if we transform as in eq.~\reef{chan1}, we find the rather surprising result that
\beq
G_{\tilde I} = \wR_{\tilde I J}\, G_J= (2,0,0,0)\,.
\label{tens4}
\eeq
That is, expressing our $\kappa=1$ cost function \reef{DalphaP} in terms of the normal-mode basis, we are only penalizing the scaling gate  associated with $\tilde x_+$! The other (normal-mode) gates can be inserted in the circuit at zero cost. However, we must add that the transformation in eq.~\reef{tens3} is slightly naive since it assumes that the absolute values in the cost function \reef{DalphaP} play no role, \ie we are assuming that all $Y^{\tilde I}\ge0$
(or all $Y^{\tilde I}\le0$). However, one finds that, depending on the signs of the various velocity components, only one of the normal-mode gates is penalized at a time. For example, with $Y^{\tilde I}\ge0$ for ${\tilde I}=++,--$ and $Y^{\tilde I}\le0$ for ${\tilde I}=+-,-+$,\footnote{This is the case in eq.~\reef{veloc} for the optimal circuit with $\tom_\pm>\omega_0$. Hence in eq.~\reef{trials}, the $\kappa=1$ cost function only depends on $\tom_-$.} one finds $G_{\tilde I} = (0,0,0,2)$, \ie only the scaling gate associated with $\tilde x_-$ is penalized. 

Similar results arise if we begin with the $\kappa$ cost functions \reef{Dalpha} expressed in terms of the normal-mode basis and examine their structure in the position basis. In this case for $\kappa=1$, the original and transformed penalty tensors become 
\beq
G_{\tilde I}=(1,1,1,1)\quad\longrightarrow\quad
G_{I} = \wR^T_{I\tilde J}\, G_{\tilde J}= (2,0,0,0)\,.
\label{tens5}
\eeq
Hence we have the rather curious result that this cost function is only penalizing the scaling gate associated with $x_1$, the position of the first oscillator. Of course, we must again remind the reader that eq.~\reef{tens5} assumes that the absolute values in the cost function \reef{Dalpha} play no role. This assumption is more natural in this case, as with a natural choice of $\omega_0$ we find that all $Y^{\tilde I}\ge0$ (or all $Y^{\tilde I}\le0$) for the optimal circuit, \ie all of the scaling components have a definite sign and all components in the entangling directions vanish. 

Furthermore, using eq.~\reef{Dalpha}, we might note that the cost of our straight-line circuit is simply
\beq
\widetilde{\mathcal{D}}_{\kappa=1}(U_0) =\tilde y_{+}+\tilde y_{-}={\textstyle \frac{1}{2}\log\frac{\tom_+}{\omega_0} +
\frac{1}{2}\log\frac{\tom_-}{\omega_0}}\,,
\label{ckao}
\eeq
again assuming $\tom_\pm>\omega_0$. Here we emphasize that since only two of the velocity components were non-vanishing, namely $Y^{++}$ and $Y^{--}$, we would arrive at the same cost for a family of penalty tensors of the form
\beq
G_{\tilde I}=(1,\pen_1^2,\pen_2^2,1)~.
\label{tens6}
\eeq
In this case, transforming to the position basis as in eq.~\reef{tens5} yields
\beq
G_{I} = \wR^T_{I\tilde J}\, G_{\tilde J}=\frac12 (2+\pen_1^2+\pen_2^2,\ \pen_2^2-\pen_1^2,  \ \pen_1^2-\pen_2^2,\ 2-\pen_1^2-\pen_2^2)\,.
\label{tens7}
\eeq
At first sight, this result seems to yield a more reasonable penalty tensor relative to eq.~\reef{tens5}. However, upon closer examination, we see that $G_{12}=-G_{21}$, and hence one of these penalty factors will be negative. That is, the cost of the circuit will be reduced by including more of one type of the entangling gates in the normal-mode basis! The only resolution of this unsatisfactory situation is to set the two penalty factors equal, \ie $\pen_1=\pen_2=\pen$, whereupon eq.~\reef{tens7} becomes
\beq
G_{I} = (1+\pen^2,\ 0,  \ 0,\ 1-\pen^2)\,,
\label{tens8}
\eeq
which still requires that $\pen^2\le1$ in order that $G_{22}\ge0$.

The above results are somewhat unsatisfying, in that a perfectly reasonable penalty tensor in one basis yields an undesirable or even inconsistent (in the case of negative penalty factors) cost function in another basis. We return to this point in the discussion in section \ref{discuss}; however, we should say that some of these issues arise because we focused on the simple case of $\kappa=1$. For example, if we consider instead the $\kappa=2$ cost function \reef{DalphaP} with our penalized metric \reef{pen99}, then transforming to the normal-mode basis yields eq.~\reef{pen99a}. While the resulting metric has negative entries, we know that this in itself is not worrisome. Rather, one must examine the eigenvalues of the new metric, and since these have not been changed by the transformation, all remain positive. 

To close this section, let us comment on extending this discussion to a lattice of oscillators.  In particular, we observe that the essential features of the complexity noted in section \ref{sec:continuum} using the $\kappa$ cost functions \reef{Dalpha} constructed in the normal-mode basis remain unchanged when working with eq.~\reef{DalphaP} in the position basis. For a $(d\!-\!1)$-dimensional spatial lattice of oscillators, the $\kappa=1$ penalty tensor in eq.~\reef{tens2} becomes
\beq
G_{\tilde I}=(N^{d-1},0,0,\cdots,0)
\label{weird}
\eeq
in terms of the normal modes. Hence as in eq.~\reef{tens4}, only the scaling gate of lowest normal mode is penalized, but the cost of that single gate has been increased to $N^{d-1}$, the total number of oscillators in the lattice.\footnote{Note that in eq.~\reef{tens4}, the penalty associated to the $M_{++}$ was increased to $2$, the number of oscillators.} The cost for the straight-line circuit then becomes
\beq
{\cal D}_{\kappa=1}(U_0)=\frac12\,N^{d-1}\,\left|\log\frac{m}{\omega_0}\right|=\frac{V}{2\,\delta^{d-1}}\,\left|\log\frac{m}{\omega_0}\right|\,.
\label{weird2}
\eeq
Hence the cost is still proportional to $V/\delta^{d-1}$, as desired to emulate the holographic complexity. This factor is again multiplied by a logarithmic factor whose argument depends on the reference frequency $\omega_0$. However, since only the lowest eigenfrequency $\tilde\omega_{\vec k=0}=m$ appears in the (single) logarithmic factor, the cut-off scale can only appear in this result through the reference frequency $\omega_0$. Hence $\delta$ appears if $\omega_0$ is chosen as a UV frequency, \eg $\omega_0=e^{-\sigma}/\delta$, but it does not appear if the reference frequency is chosen as an IR frequency. We observe that this is the opposite of the situation discussed in section \ref{sec:continuum}.

\section{Approximating the complexity}\label{sec:approx}

In section \ref{sec:continuum}, we compared our result \reef{complexityNd} for the complexity of the ground state of a $(d\!-\!1)$-dimensional spatial lattice of $N^{d-1}$ oscillators to the analogous results for holographic complexity. We could easily identify the leading contribution in the limit of large $N$ and a small UV cut-off distance, \ie $m\delta\ll1$. In particular, this led us to consider the generalized family of $\kappa$ cost functions given in eq.~\reef{Dalpha}, which yields
\beq
\CC=\frac{1}{2^\kappa}\sum_{\{k_i\}=0}^{N-1}\left|\log\!\lp{\tilde\omega_{\vec k}}/{\omega_0}\rp\right|^\kappa\,.
\label{nothing}
\eeq
To identify the leading contribution to either eq.~\reef{complexityNd} or \reef{nothing}, we made the crude approximation of replacing $\tom_{\vec k}\sim1/\delta$ for all momenta. In the following, we would like to avoid this approximation and examine the complexity \reef{nothing} in more detail. Our result \eqref{exit} is still an approximation, but it allows us to consider the subleading contributions to eq.~\reef{calpha}. In particular, we will determine the leading corrections involving the mass. 

First we substitute the normal-mode frequencies \reef{freakd} into the above expression for general $\kappa$ to find
\beq
\CC
=\frac{1}{4^\kappa}\sum_{\{k_i\}=0}^{N-1}\left|\log\!\lp\frac{m^2}{\omega_0^2}+\lp\frac{2}{\omega_0\delta}\rp^{\!2}\,\sum_{j=1}^{d-1}\sin^2\!\lp\frac{\pi k_j}{N}\rp\rp\right|^\kappa
\label{atall}
\eeq
Now, it is certainly true that the second term in the argument of the logarithm dominates for most of the terms in the sum over all momenta. But we would like to be more careful in retaining the leading corrections arising from the mass term. To simplify the analysis below, we will assume that the reference frequency is an IR frequency with $\omega_0<m$.

As a first step, we should isolate the infrared contributions which come from terms in the sum where the mass actually dominates or is comparable to the momentum term in the argument. For large $N$, we can take the usual continuum limit for these contributions with $p_i=2\pi k_i/(N\delta)$. The IR contribution in eq.~\reef{atall} then becomes
\beq
\CC_\mt{IR}=\frac{V}{4^\kappa}\int_0^{\Lambda_\mt{IR}}
\frac{\dd^{d-1}p}{(2\pi)^{d-1}}\ \left[\log\!\lp\frac{m^2+p^2}{\omega_0^2}\rp\right]^\kappa\,.
\label{ir2}
\eeq
Note that we have dropped the absolute value symbol here since we are assuming that $\omega_0<m$. The cut-off $\Lambda_\mt{IR}$ in this integral is an IR scale which delineates the boundary of the IR contributions to the momentum sum in eq.~\reef{atall}. Implicitly, we are also letting $k_i$ and $p_i$ range over positive and negative values so that all of the IR contributions come in the vicinity of $\vec k=0$---see footnote \ref{crap}. Choosing this cut-off to be $\Lambda_\mt{IR}\propto m$, the IR contribution takes the general form
\beq
\CC_\mt{IR}=V m^{d-1} \sum_{a=0}^\kappa
c_{a} \left[\log m/\omega_0\right]^a\,,
\label{ir3}
\eeq
where the numerical coefficients $c_{a}$ are independent of $m$ and $\omega_0$, but will depend on the spacetime dimension $d$. The leading contribution then takes the form
$\CC_\mt{IR}\simeq c_{\kappa}\,V m^{d-1} 
 \left[\log m/\omega_0\right]^\kappa$.

Having isolated the IR contribution, we return to the UV contributions to eq.~\reef{atall}. In these remaining terms, we can consider $m^2/\omega_0^2$ to be a small correction to the argument of the logarithm, and so we perform a Taylor series expansion and keep only the first correction in $m\delta$:
\beqa
\CC_\mt{UV}&\simeq&\frac{1}{4^\kappa}\sum_{\{k_i\}>\mathrm{IR}}
\log\!\left[\lp\frac{2}{\omega_0\delta}\rp^{\!2}\,\sum_{j=1}^{d-1}\sin^2\!\lp\frac{\pi k_j}{N}\rp\right]^\kappa
\label{eq:Cgeneral}\\
&&\qquad\qquad\quad\times\ \left\lbrace1+\kappa\,\frac{(m\delta)^2}{4}\left(\sum_{j=1}^{d-1}\sin^2\lp\frac{\pi k_j}{N}\rp\ 
\log\!\left[\lp\frac{2}{\omega_0\delta}\rp^{\!2}\,\sum_{j=1}^{d-1}\sin^2\!\lp\frac{\pi k_j}{N}\rp\right]
 \right)^{\!-1}\right\rbrace~,
\nonumber
\eeqa
where the notation $\sum_{\{k_i\}>\mathrm{IR}}$ indicates that the summation begins at the IR cut-off, \ie $|\vec k|\ge \Lambda_\mt{IR}\,N\delta/(2\pi)$. The leading term will produce the expected leading contribution in eq.~\reef{calpha} with some overall numerical coefficient which depends on $\kappa$ and $d$. Of course, there will also be a subleading dependence on our IR cut-off $\Lambda_\mt{IR}\propto m$.

To proceed further, we focus on the case $\kappa=1$, which simplifies the calculations slightly and was also the case that we found best emulated the holographic complexity. The following analysis is essentially unchanged for larger values of $\kappa$. Substituting $\kappa=1$ into eq.~\eqref{eq:Cgeneral} yields
\beq
\CC_\mt{UV}\simeq \frac{1}{4}\sum_{\{k_i\}>\mathrm{IR}}\left\lbrace\log\!\left[\lp\frac{2}{\omega_0\delta}\rp^{\!2}\,\sum_{j=1}^{d-1}\sin^2\!\lp\frac{\pi k_j}{N}\rp\right]
+\frac{(m\delta)^2}{4}\left[\sum_{j=1}^{d-1}\sin^2\lp\frac{\pi k_j}{N}\rp\right]^{\!-1}\right\rbrace~.
\label{eq:Ck1}
\eeq
We now examine the two sums separately. First, we break the leading sum into two:
\beq
\log\!\lp\frac{4}{\omega_0^2\delta^2}\sum_{j=1}^{d-1}\sin^2\!\lp\frac{\pi k_j}{N}\rp\rp
=2\log\frac{2}{\omega_0\delta}\\
+\log\sum_{j=1}^{d-1}\sin^2\!\lp\frac{\pi k_j}{N}\rp~.
\eeq
Since the first term is independent of $\vec k$, the corresponding sum over the UV modes yields a factor of $N^{d-1}$ (up to corrections proportional to $\Lambda_\mt{IR}^{d-1}$). Summing over the second term  is more complicated, but numerical fits for a range of $N$ and $d$ suggest that the sum takes the form  
\beq
\frac{1}{4}\sum_{\{k_i\}>\mathrm{IR}}\log\sum_{j=1}^{d-1}\sin^2\!\lp\frac{\pi k_j}{N}\rp
=a_{d-1}N^{d-1}+a_{d-3}N^{d-3}+\cdots+a_0~,
\eeq
where the $a_i$ are fixed numerical coefficients. Note that the constant term $a_0$ appears for both odd and even $d$, and the former may also have a logarithmic correction (\ie $\log N$). Turning now to the second sum in eq.~\eqref{eq:Ck1}, we again carried out numerical fits to find
\beq
\frac14\sum_{\{k_i\}>\mathrm{IR}}\left[\sum_{j=1}^{d-1}\sin^2\!\lp\frac{\pi k_j}{N}\rp\right]^{-1}\simeq
b_{d-1}N^{d-1}+b_{d-3}N^{d-3}+\cdots+b_0~.
\eeq

Collecting results, our approximation to the total complexity for $\kappa=1$ (and assuming $\omega_0<m$) is therefore
\beqa
\CC&\simeq&V m^{d-1} \left(c_{\mt{IR}\,1} \log m/\omega_0+ c_{\mt{IR}\,0} \right)+\frac{N^{d-1}}2\log\frac{2}{\omega_0\delta}
+\lp a_{d-1}N^{d-1}+a_{d-3}N^{d-3}+\cdots+a_0\rp
\nonumber\\
&&\qquad\qquad+\lp m\delta\rp^2\lp b_{d-1}N^{d-1}+b_{d-3}N^{d-3}+\cdots+b_0\rp~.
\label{exit}
\eeqa
To make a comparison with holographic complexity, as in section \ref{sec:continuum}, we substitute $N^{d-1}=V/\delta^{d-1}$, and introduce $L=V^{1/(d-1)}$ as the linear size of the lattice. The complexity \reef{exit} then becomes
\beqa
\CC&\simeq&
\frac{1}{2}\frac{V}{\delta^{d-1}}\log\frac{2}{\omega_0\delta}
+\frac{V}{\delta^{d-1}}\lp a_{d-1}+a_{d-3}\frac{\delta^2}{L^2}+\cdots\rp+\frac{m^2V}{\delta^{d-3}}\lp b_{d-1}+b_{d-3}\frac{\delta^2}{L^2}+\cdots\rp 
\nonumber\\
&&\qquad\quad+V m^{d-1} \left(c_{1} \log m/\omega_0+ c_{0} \right)\,.
\label{exit2}
\eeqa
Hence we find the expected leading term, which corresponds to the result in eq.~\reef{calpha} with $\kappa=1$. We also find a subleading term proportional to ${V}/{\delta^{d-1}}$, as is found in holographic complexity in the CA proposal \cite{Carmi:2016wjl}. Additionally, we would highlight the correction proportional to ${m^2V}/\delta^{d-3}$, for which analogous results can again be found in holographic calculations---see section \ref{sec:continuum} for further comments. It is interesting that we also see corrections, \eg of the form ${V}/(L^2\delta^{d-3})$. Of course, this term is far more suppressed that the previous one, but it also involves a fractional power of the volume, $V^{\frac{d-3}{d-1}}$. Such fractional powers would never arise in holographic complexity.

To make the above formulae more concrete, consider the case of a one-dimensional lattice ($d=2$), in which case eq.~\reef{exit2} reduces to
\beq
\CC=\frac{1}{2}\frac{L}{\delta}\log\frac{2}{\omega_0\delta}
+a_{1}\frac{L}{\delta}+a_0+L m \left(c_{1} \log m/\omega_0+ c_{0} \right)~,
\label{eq:Cd1}
\eeq
where we have replaced $V=L$ to emphasize that the volume is only a linear length here.

\section{Optimal geodesic for penalized geometry}\label{sec:subspace}

We would like to find the optimal geodesic in the penalized geometry \reef{penalty2}, but as commented below eq.~\reef{eq:kPenFull}, finding the general solution for geodesics satisfying the desired boundary conditions seems out of reach. Recall that we were able to show that the simple straight-line geodesic describing the optimal circuit \reef{solver5} in the unpenalized geometry remains a geodesic in our new penalized geometry. However, it was also easy to show that the segmented path described by eq.~\reef{segments} was shorter than this geodesic when the penalty factor was large, \ie  $\pen\gg\rho_1,y_1$; \cf eq.~\reef{compar2}.
 
To make progress towards finding the optimal geodesic in the new geometry, we make a simplifying assumption. To begin, we examine the penalized metric \reef{penalty2} and observe that as the radius $\rho$ increases, the fastest growing component of the metric is $g_{zz}\sim \pen^2 e^{4\rho}$ (for generic $x$, but this component still grows as $g_{zz}\sim e^{4\rho}$ for $x=\pi/2$). This suggests that motion in the $z$ direction will quickly be suppressed as the geodesics move out from the origin. Therefore, we simplify our problem by considering motion on a constant-$z$ submanifold:
\beq
\dd s^2=2\dd y^2+2\left[\pen^2-\lp \pen^2-1\rp\sin^2\!\lp2x\rp\right]\dd\rho^2+2 \pen^2\dd x^2~.\label{eq:metricSimpleX}
\eeq
The particular value of $z$ in question will be fixed below by the boundary condition that $z=x$ at $s=0$. We will return to justify our assumption of no (or little) motion in the $z$ direction at the end of this appendix. 

Working with this simpler geometry \eqref{eq:metricSimpleX}, the analysis of the geodesics becomes much more tractable. First, we observe that both $\pd_y$ and $\pd_\rho$ are now Killing vectors, for which the associated conserved quantities are
\beq
2\,\bar c_1\equiv 2\dot y~,\qquad
2a\, \bar c_2\equiv 2\left[ \pen^2-\lp \pen^2-1\rp\sin^2\!\lp2 x\rp\right]\dot\rho~,
\eeq
where the factors of 2 and $\pen$ were chosen to simplify expressions below, and we have used the notation $\bar c_i$ to avoid confusion with the $\hat c_i$ in eq.~\reef{revel0}. As before, the first constraint gives the usual solution \reef{yrun} for $y$, \ie $y(s)=y_1\,s$ with $\bar c_1=y_1$. The second constraint yields
\beq
\dot\rho =\frac{\pen\,\bar c_2}{\pen^2-\lp \pen^2-1\rp\sin^2\!\lp2 x\rp}\,.
\label{revel2}
\eeq
The normalization of the tangent vector then becomes
\beq
\bal
k^2&=2\dot y^2+2\left[ \pen^2-\lp \pen^2-1\rp\sin^2\!\lp2 x\rp\right]\dot\rho^2+2\pen^2\dot x^2\\
&=2y_1^2+\frac{2\pen^2\,\bar c_2^2}{\pen^2-\lp \pen^2-1\rp\sin^2\!\lp2 x\rp}+2\pen^2\dot x^2~.\label{eq:ktemp}
\eal
\eeq
It is possible to integrate this equation to find $s(x)$. To simplify the subsequent equations, we shall define $\bar k$ via
\beq
2\bar k^2\equiv k^2-2y_1^2\ .
\label{newerX}
\eeq
Isolating $\dot x$ in eq.~\reef{eq:ktemp} then yields
\beq
\frac{\dd s}{\dd x}=\sqrt{2}a\left[ 2\bar k^2-\frac{2\pen^2\, \bar c_2^2}{\pen^2-\lp \pen^2-1\rp\sin^2\!\lp2 x\rp}\right]^{-1/2}~.
\label{revel3}
\eeq
 Upon integrating and choosing the constant of integration such that $s(x_1\!=\!\pi/2)=1$, the result can be simplified to
\beq
s(x)=1-\frac{\pen^2}{2\sqrt{\bar k^2-\bar c_2^2}}\ \Pi\!\lp-(\pen^2-1);-f(x)\,\Bigg|\,\frac{\lp \pen^2-1\rp \bar c_2^2}{\bar k^2-\bar c_2^2}\rp~,\label{eq:sX}
\eeq
where
\beq
f(x)\equiv\frac{1}{\sin^{-1}\!\lp\cot(2x)\sqrt{\pen^2+\tan^2\!(2x)}\rp}~,
\eeq
and $\Pi$ is the incomplete elliptic integral of the third kind, which we write here as
\beq
\Pi(-n;-z|m)=-\int_0^z\frac{\dd t}{\lp1+n\sin^2t\rp\sqrt{1-m\sin^2t}}~.
\eeq
We now combine the expressions for $\dot \rho$ and $1/\dot x$ in eqs.~\reef{revel2} and \reef{revel3} to find 
\beq
\frac{\dd\rho}{\dd x}=\frac{\dot\rho}{\dot x}=\frac{\sqrt{2}\pen^2\,\bar c_2}{\pen^2-\lp \pen^2-1\rp\sin^2\!\lp2 x\rp}\left[\bar k^2-\frac{2\pen^2\, \bar c_2^2}{\pen^2-\lp \pen^2-1\rp\sin^2\!\lp2 x\rp}\right]^{-1/2}\,.
\label{revel4}
\eeq
This expression can likewise be integrated to obtain
\beq
\rho(x)=\rho_1-\frac{i\pen\bar c_2}{2\sqrt{\bar k^2-\bar c_2^2}}\ F\!\lp i\,g(x)\Bigg|\,1-\frac{\lp \pen^2-1\rp\bar c_2^2}{\bar k^2-\bar c_2^2}\rp~,\label{eq:rX}
\eeq
where
\beq
g(x)\equiv\frac{1}{\sinh^{-1}\!\lp\pen\cot(2x)\rp}~,
\label{flip0}
\eeq
and $F$ is the incomplete elliptic integral of the first kind,
\beq
F(z=ix|m)=i\,\int_0^x\frac{\dd \tau}{\sqrt{1+m\sinh^2\tau}}~,\qquad x\in\mathbb{R}~.
\eeq
Note that in eq.~\reef{eq:rX}, we have fixed the integration constant via the boundary condition $\rho(x_1=\pi/2)=\rho_1$. 

Now, unfortunately \eqref{eq:sX} cannot be inverted to find an analytical expression for $x(s)$, which we could then use to obtain $\rho(s)$ via eq.~\eqref{eq:rX}. However, we can still study the behaviour of these geodesics numerically. Before doing so, it remains to relate the parameters $\bar c_2$ and $k$ (or $\bar k$) to the boundary values $\rho_1$ and $x_0$ (as well as $x_1=\pi/2$). Let us first examine the parameter range for which we obtain a real result. It turns out that $F$ in eq.~\eqref{eq:rX} is always complex; hence in order for $\rho(x)$ to be real, the coefficient must also be imaginary, which requires
\beq
\bar k^2>\bar c_2^2\;\implies\;k^2>2\lp y_1^2+\bar c_2^2\rp~.\label{eq:real}
\eeq
Turning now to $s(x)$ in eq.~\eqref{eq:sX}, the elliptic integral $\Pi$ in this case is always real, and the coefficient will also be real in precisely the same regime \eqref{eq:real}. Therefore this is the only restriction on our parameters required to ensure a real result. 

We have fixed the integration constants in both $s(x)$ and $\rho(x)$ via the boundary conditions at the end-point of the geodesic, namely $s(x=\pi/2)=1$ and $\rho(x=\pi/2)=\rho_1$. For the optimal geodesic, we further choose the boundary condition $x(s=0)=\pi/4$, which minimizes the cost of motion in the $\rho$ direction, \cf eq.~\reef{rats6}.\footnote{Alternatively, we could choose $x(s\!=\!0)=3\pi/4$, but the resulting trajectory is simply of a copy of the present geodesic rotated 180$^o$ around the $(\theta,\tau)=(\pi,0)$ axis---see figure \ref{fig:spiralPi}.} However, we must be careful in evaluating eqs.~\reef{eq:sX} and \reef{eq:rX} at this value of $x$; in particular, we must consider the limits
\beq
\bal
\lim_{x\rightarrow\pi/4^+}s(x)&=1-\frac{\pen^2}{2\sqrt{\bar k^2-\bar c_2^2}}\ \Pi\!\lp-(\pen^2-1)\bigg|\,\frac{\lp \pen^2-1\rp \bar c_2^2}{\bar k^2-\bar c_2^2}\rp~,\\
\lim_{x\rightarrow\pi/4^+}\rho(x)&=
\rho_1-\frac{\pen\,\bar c_2}{2\sqrt{\bar k^2-\bar c_2^2}}\ K\!\lp\frac{\lp \pen^2-1\rp \bar c_2^2}{\bar k^2-\bar c_2^2}\rp~,\label{eq:limits}
\eal
\eeq
where $F$ and $K$ are the complete elliptic integrals of the first and third kind, respectively, defined via
\beq
K(z)=F\lp\pi/2\,|\,z\rp~,\qquad
\Pi(n\,|\,m)=\Pi\lp n;\pi/2\,|\,m\rp~.
\eeq
The parameters $\bar c_2$ and $\bar k$ must be chosen so that both these limits vanish, since initially we must have $s=0$ and $\rho_0=0$. In principle, we have two equations and two unknowns, but in practice the elliptic integrals are intractable. Fortunately, for our purposes a general solution is not required: we seek only a valid case to compare with the length  \eqref{eq:straight} of the simple straight-line geodesic.

To that end, observe that the elliptic integral $K$ is of order 1 almost everywhere, except when the argument approaches 1 in where it diverges, $\lim_{w\rightarrow1}K(w)=\infty$. Since we want $\rho$ to be large, let us choose
\beq
\frac{\lp \pen^2-1\rp\bar c_2^2}{\bar k^2-\bar c_2^2}=1-\eps\quad\implies\quad
\bar k^2=\bar c_2^2\,\frac{\pen^2-\eps}{1-\eps}~,\label{eq:kEps}
\eeq
where $0<\eps\ll1$. Note that this is within the reality domain \eqref{eq:real} since $\pen>1$. The boundary condition that the limits \eqref{eq:limits} should vanish then allows us to solve for $\rho_1$ and $\bar c_2$; one finds:
\beq
\rho_1=\frac{\pen}{2}\sqrt{\frac{1-\eps}{\pen^2-1}}\,K\lp1-\eps\rp~,\qquad
\bar c_2=\frac{\pen^2}{2}\sqrt{\frac{1-\eps}{\pen^2-1}}\,\Pi\lp1-\pen^2\,\big|\,1-\eps\rp~.\label{eq:params}
\eeq
One can then make $\rho_1$ arbitrarily large by taking $\eps\rightarrow0$; note that $\bar c_2$ becomes arbitrarily large in the same limit. In fact, the divergence in both cases is logarithmic:
\beq
\bal
\rho_1&=\frac{\pen}{\sqrt{\pen^2-1}}\lp\frac{1}{4}\log\!\lp\frac{1}{\eps}\rp+\log2\rp+\op(\eps)~,\\
\bar c_2&=\frac{1}{4\sqrt{\pen^2-1}}\log\!\lp\frac{1}{\eps}\rp
+\lp\frac{1}{2}\tan^{-1}\!\sqrt{\pen^2-1}+\frac{\log2}{\sqrt{\pen^2-1}}\rp
+\op(\eps)~,
\eal\label{hous2x}
\eeq
where higher-order terms vanish in the limit $\eps\rightarrow0$. 

We may now numerically compare the length of this geodesic to the proposed minimum \eqref{eq:straight} associated with the straight-line circuit. Substituting $\bar c_2$ and $\rho_1$ from eq.~\eqref{eq:params} into $\bar k$ given eq.~\eqref{eq:kEps} and the analogous quantity $\bar k_0$ from eq.~\eqref{eq:straight}, we find
\beq
\bar k=\frac{\pen^2}{2}\sqrt{\frac{\pen^2-\eps}{\pen^2-1}}\,\Pi\lp1-\pen^2\,\big|\,1-\eps\rp
\qquad{\rm and}\qquad
\bar k_0=\frac{\pen^2}{2}\sqrt{\frac{1-\eps}{\pen^2-1}}\,K\lp1-\eps\rp~,\label{eq:vs}
\eeq
where $2\bar k_0^2\equiv k_0^2-2y_1^2$. 

Of course, while these expressions are well-suited to numerics, we would also like to express $\bar k$ in terms of the coordinates $\rho_1$, $y_1$, so as to compare with \eqref{eq:straight} on more physical footing. We can obtain an approximation of this form by first replacing $\bar c_2$ in \eqref{eq:kEps} by its expression in \eqref{eq:params}, and then expanding for $\eps\rightarrow0$:
\beq
\bar k=\frac{\pen}{\sqrt{\pen^2-1}}\lp\frac{1}{4}\log\!\lp\frac{1}{\eps}\rp+\log2\rp
+\frac{\pen}{2}\tan^{-1}\!\sqrt{\pen^2-1}+\op(\eps)~, 
\eeq
where as above the $\op(\eps)$ terms vanish as $\eps\rightarrow0$, and we shall drop them henceforth. Comparing this expression to eq.~\eqref{hous2x}, we observe that we can equivalently write this as
\beq
\bar k\simeq\frac{\pen}{2}\tan^{-1}\!\sqrt{\pen^2-1}+\rho_1
\,\simeq\,\frac{\pi}{4}\pen+\rho_1-\frac{1}{2}-\frac{1}{12\pen^2}+\op\!\lp\frac1{\pen^4}\rp~,\label{eq:kbarCoords}
\eeq
where in the second approximation we have performed an expansion in the limit $\pen\rightarrow\infty$.

Additionally, it will be interesting to compare these geodesics against the segmented trajectory described in eq.~\reef{segments}. The length of this path is given by eq.~\reef{eq:ks}, and so as in eq.~\reef{newerX}, we define $2\bar k_s^2 =k_s^2 -2 y_1^2$, 
\beq
\bal
\bar k_s&=\frac{1}{4}\left[\pi^2\pen^2+8\rho_1\lp2\rho_1+\sqrt{\pi^2\pen^2+ (4y_1)^2}\rp\right]^{\!\frac12}\\
&=\frac{\pen}{4}\left[\pi^2+4\lp\frac{1-\eps}{\pen^2-1}\rp^{\!\frac12}K\!\lp1-\eps\rp\,\lp\lp\frac{1-\eps}{\pen^2-1}\rp^{\!\frac12} K\!\lp1-\eps\rp+\pi\lp1+\frac{16y_1^2}{\pi^2\pen^2} \rp^{\!\frac12}\rp\right]^{\!\frac12}~,
\eal\label{eq:kStraight}
\eeq
where in the second line we have replaced $\rho_1$ using eq.~\eqref{eq:params}. Note that unlike $\bar k$ and $\bar k_0$ in eq.~\eqref{eq:vs}, the parameter $y_1$ still appears in this expression---although this contribution is suppressed for $\pen\gg y_1$. Again however, the second line above is more suited to numerics than physical inspection; to compare with \eqref{eq:kbarCoords}, we shall expand with $\pen\gg\rho_1,y_1$ (as well as $\rho_1,y_1\gg1$, and assuming $\rho_1$ and $y_1$ are roughly the same order of magnitude). Hence:
\beq
\bar k_s=\frac{\pi \pen}{4}\left[1+\frac{8\rho_1}{\pi\pen}\lp\frac{2\rho_1}{\pi\pen}+\sqrt{1+\lp\frac{4y_1}{\pi\pen}\rp^2}\rp\right]^{\!\frac12}
\simeq\frac{\pi}{4}\pen+\rho_1+\frac{8}{\pi^2}\frac{y_1^2}{\pen^2}\,\rho_1+\ldots~.\label{eq:ksCoords}
\eeq
We mentioned above that the segmented path constitutes a remarkably good approximation to the geodesic. Comparing $\bar k$ in \eqref{eq:kbarCoords} and $\bar k_s$ in \eqref{eq:ksCoords}, one can see evidence for this claim in that the leading-order behaviours are precisely the same; deviations arise only in the subleading terms, which are increasingly negligible for large values of $\pen$. We discuss this point further in the main text---see eq.~\reef{eq:overhead}. We also explicitly confirm that the two paths are very close to one another in the large $\pen$ regime by examining $x(s)$ and $\rho(s)$ numerically, as shown in figure \ref{fig:funcS}.

In closing this appendix, we remind the reader that in order to make progress, we confined our attention to motion in the constant-$z$ subspace given by the simpler metric \eqref{eq:metricSimpleX}. Hence for completeness, we should go back and examine whether or not this was a reasonable assumption.  In particular, we wish to argue that, at least in the limit $\pen\gg1$, the particular class of geodesics with $x_0=\pi/4$ and $x_1=\pi/2$ obtained for the constant-$z$ subspace are a good approximation to the corresponding geodesics in the full geometry \eqref{penalty2}. Intuitively, we motivated this restriction by the observation that movement in the $z$-direction is relatively costly. We can quantify this by considering the behaviour of $\dot z$ given in eq.~\eqref{revel1}. Recall that $\tau_0=0$, and hence $z_0=x_0=\pi/4$. Then the finiteness condition \eqref{lov2} requires that we set $\hat c_3=0$.\footnote{Note that $\hat c_2$ is still free, since we can rewrite eq.~\eqref{lov2} as $\hat c_3=\pen^2\hat c_2\cot(2z_0)|_{z_0=\pi/4}=0$.} Along the initial segment, where $x=\pi/4$, the derivatives \eqref{revel1} then reduce to
\beq
\dot x|_{x=\pi/4}=-\frac{\hat c_4\cosh\lp2\rho\rp}{\pen^2}~,\qquad
\dot z|_{x=\pi/4}=-\frac{\hat c_4}{2\pen^2}~,\qquad
\dot\rho|_{x=\pi/4}=\frac{\hat c_2}{2}~.\label{eq:suppressed}
\eeq
Therefore, in the large $\pen$ limit under consideration, motion in both the $x$- and $z$-directions is highly suppressed, while only motion along $\rho$ is inexpensive. Along the second segment, where we rotate around to $x=\pi/2$, both $\dot x$ and $\dot z$ pick up terms which are $\op(1)$ in $\pen$, but $\dot z$ is still exponentially suppressed in $\rho_1$ relative to $\dot x$. (Meanwhile $\dot \rho$ decreases sharply to 0 on this segment in the limit $\pen\gg1$.) Thus geodesics in the full spacetime \eqref{penalty2} can indeed be approximated by those in the constant-$z$ subspace \eqref{eq:metricSimpleX}, at least in the limits that we are considering. 

\end{appendices}

\bibliographystyle{ytphys}
\bibliography{drafty}
\end{document}